\documentclass[aps,pre,twocolumn,groupedaddress,longbibliography]{revtex4-1}
\usepackage{amsmath,amssymb,yhmath,graphicx,tipa,color,textcomp,url,fancyvrb}
\usepackage{booktabs}
\usepackage[T1]{fontenc}
\usepackage[utf8]{inputenc}
\usepackage{longtable}

\newtheorem{theorem}{Theorem}

\newcommand{\bv}[1]{\boldsymbol{\mathbf{#1}}}
\newcommand{\PP}{ \mathbb{P} }

\newcommand{\EE}{ \mathbb{E} }
\newcommand{\RR}{ \mathbb{R} }

\newcommand{\cl}{\mathfrak{C}}
\newcommand{\rar}{\rightarrow}
\newcommand{\ra}{\rangle}
\newcommand{\la}{\langle}

\newcommand{\dd}{\delta}

\begin{document}

\title{Spatial evidence that language change is not neutral}

\author{James Burridge}
\email[]{james.burridge@port.ac.uk}
\affiliation{School of Mathematics and Physics, University of Portsmouth, Portsmouth PO1 3HF, United Kingdom }

\author{Tamsin Blaxter}
\email[]{ttb26@cam.ac.uk}
\affiliation{Gonville \& Caius College, Cambridge, CB2 1TA, United Kingdom}

\begin{abstract}
The neutral theory of genetic and linguistic evolution holds that the relative frequencies of variants evolve by random drift. Neutral evolution remains a plausible null model of language change. In this paper we provide evidence against the neutral hypothesis by considering the geographical patterns observed in language surveys. We model speakers as neurons in a Hopfield network embedded in space, analogous to one of the classical two dimensional lattice models of statistical physics. The universality class of the model depends on the form of the activation function of the neurons, which encodes learning behaviour of speakers. We view maps generated by the Survey of English Dialects as samples from our network. Maximum likelihood analysis, and comparison of spatial auto-correlations between real and simulated maps, indicates that the maps are more likely to belong to the conformity-driven Ising class, where interfaces are driven by surface tension, rather than the neutral Voter class, where they are driven by noise. 
\end{abstract}

\maketitle

\section{Introduction}

Languages are complex, constantly evolving structures which take a wide variety of forms \cite{fas14,lab01}. Language change involves evolutionary processes which can vary substantially between different parts of the linguistic system (phonology, morphology, syntax, lexicon), and changes may be driven by purely linguistic effects, social phenomena, migration, geography, technology and changes in wider society \cite{cha98,tru00,blo33,mil85}. Nevertheless, every language is generated and maintained by a large number of interacting speakers with similar properties (vocal apparatus, a need to communicate, to display status and cooperate). It is therefore natural for statistical physicists to construct models which capture how languages arise and evolve, based on the interactions of such agents \cite{lor11, cas09,oli08,bur17,bur19,abr03,bax06,bar06,bar06_2}. A popular and simple model, which also serves as a model of genetic evolution \cite{mor58,fis30}, is neutral evolution \cite{bax06,cro00,bly07,bly12_2,kau17}, wherein linguistic variants survive with a probability equal to their current frequency within the population. Despite the simplicity of this assumption, and the complexity of real languages, neutral evolution remains a surprisingly robust \textit{null} model of language change \cite{bly07,kau17}. 

Recent work on the spatial evolution of language (in both birds \cite{bur16} and humans \cite{tes06,bur17,bur18}) suggests that geographical boundaries between language features, known to linguists as \textit{isoglosses}, may be analogous to the domain walls seen in classical lattice models of statistical physics \cite{kra10,car96} undergoing surface tension-driven coarsening \cite{bra94}. The best known example of this is the Ising model evolving according to Glauber dynamics \cite{glau63}.  The surface tension effect requires some non-linearity in the local copying rule \cite{dor01,dro99}, which in the social context implies a form of social conformity or majority rule \cite{che05,asc56,mor12}, violating the assumption of neutrality. This would appear to provide evidence against the neutral hypothesis, but surface tension is not the only copying mechanism capable of generating distinct spatial domains. For example,  the voter model \cite{lig85}, in which agents select their state by copying a randomly selected neighbour \cite{kra10}, is the lattice analogue of neutral evolution, because states are reproduced with a probability equal to their local frequency. Although the voter model lacks surface tension \cite{cha98} it still evolves towards increased spatial order, characterised by logarithmically decaying correlations, and interfaces driven by noise. This raises the question: if local speaker to speaker copying is responsible for geographical variations in language use, then is the neutral hypothesis sufficient to explain observed patterns, or is a non-linear copying rule a more likely explanation?

The primary purpose of this paper is to address the above question, and to do this we employ a classical model of associative memory, the Hopfield Neural Network \cite{hop82, hop84}. This provides a  model of a language community if we view each speaker as a single neuron, responding to her surroundings via an \textit{activation} function. The form of this function can be as complex as we wish, allowing us, in principle, to model a wide variety of language change processes. However, in this work we investigate two simple cases which, if the network is embedded in two dimensional space, produce behaviour analogous to classical voter (neutral) and Ising (non-neutral) lattice models. After formulating coarse grained equations describing the large scale evolution of our spatial network, we analyse its ordering behaviour, and construct an approximate probabilistic model of a language survey, carried out on a small subset of the population. We can then infer which form of the activation function is more likely,  using data from the Survey of English Dialects (SED) \cite{ort62}, a large scale language survey of disappearing traditional English rural ``folk-speech'', carried out in the 1950s. The relative lack of mobility, compared to modern people, of the communities within which the observed language features evolved over the preceding centuries allows us to ignore the migration of speakers in our coarse grained dynamics, which is then driven only by local copying. This allows the neutral hypothesis to be tested under controlled conditions, analogous to those found in classical lattice models. While modern surveys have generated vastly more linguistic data \cite{lee16,vau17}, the spatial patterns of language features have been mixed and diluted by population movement and connectivity. By addressing the simple but fundamental question of whether neutral evolution provides an adequate description of language evolution in a simpler age, we can inform the construction of more sophisticated models of the modern world, where the factors affecting language evolution are more diverse and difficult to model.

\section{Hopfield  networks as language communities}

\subsection{Hopfield Networks}

In Hopfield's original model of a neural network \cite{hop82}, each neuron, $i$ is in one of two \textit{activation states}: firing ($V_i=1$) or not ($V_i=0$). These states are updated randomly in time, based on the total input to neuron - nerve impulses travelling down the axons (outgoing trunks) of the neurons which connect it. The total input is 
\begin{equation}
H_i = \sum_{j \neq i} \omega_{ij} V_j
\end{equation}
where $\omega_{ij}$ is the synaptic interconnection strength from $j$ to $i$. When the neuron updates it state, it uses the following threshold rule
\begin{equation}
V_i = \begin{cases}
1 & \text{ if } H_i \geq U_i \\
0 & \text{ if } H_i < U_i,
\end{cases}
\label{eqn:thresh}
\end{equation}
where $U_i$ is the activation threshold. According to (\ref{eqn:thresh}), neuron $i$ fires if enough of the other neurons that connect to it, weighted by the connection strengths, are also firing. Hopfield later generalized  this discrete model to a continuous state version by introducing an \textit{internal activation} $u_i$ which lags behind its instantaneous inputs \cite{hop84}. The relation between this internal state and the output of the neuron is given by the \textit{activation function}
\begin{equation}
v_i = g_i(u_i).
\end{equation}
The network is then described by the set of coupled ordinary differential equations
\begin{equation}
\tau_m \frac{du_i}{dt} =  \sum_{j} \omega_{ij} g_j(u_j(t)) - u_i(t),  
\label{eqn:hop}
\end{equation}
where $\tau_m$ is a time constant (the \textit{memory time}) which controls the extent of the lag between internal activation and output.

\subsection{Hopfield networks as language communities}

Equation (\ref{eqn:hop}) may also be viewed as a simple model for the state of a single speaker within a speech community. Writing (\ref{eqn:hop}) in integral form
\begin{equation}
u_i(t) = \frac{1}{\tau_m} \int_{-\infty}^t e^{\frac{(s-t)}{\tau_m}} \sum_{j} \omega_{ij} g_j(u_j(s)) ds 
\end{equation}
we see that $u_i(t)$ is an exponentially decaying time average (a memory) for the states of the other nodes of the network to which node $i$ is in contact. If $\sum_j \omega_j=1$ and $g_j: [0,1] \rightarrow [0,1] $ then we can interpret $g_i(u_i)$ as the relative frequency with which speaker $i$ uses one variant of a binary linguistic variable, and $u_i$ as their memory for the relative frequency with which others have used it. The activation function $g_i$ encodes how a speaker responds to or learns from the rest of the community. We can generalise (\ref{eqn:hop}) to the case of $q$ different linguistic states by promoting the memory (internal state) to a vector 
\begin{equation}
\bv{u}_i = (u_{i1},u_{i2}, \ldots, u_{iq})
\end{equation}
where $u_{ik}$ is the memory of speaker $i$ for the fraction of speakers in state $k$. In this case $\bv{u}_i$ belongs to the $q$-dimensional simplex $\Delta^q$, and the activation function becomes a vector valued mapping $\bv{g}_i: \Delta^q \rightarrow \Delta^q$. We write 
\begin{equation}
\bv{v}_i(t) = \bv{g}_i(\bv{u}_i(t)),
\end{equation} 
for the \textit{external state} of speaker $i$. 

\subsection{Stochastic model}

\label{sec:stoch}

To allow for the fact that language evolution is inherently stochastic, we generalize the Hopfield model by interpreting  $\bv{v}_i$ as a probability mass function over the set of possible language states. We introduce a second time constant $\tau_s$, the \textit{switching time}, which determines the time period between stochastic changes in state. At time $t$ the \textit{emitted} state $\bv{X}_i(t)$ of speaker $i$ is then a sample from the probability mass function $\bv{v}_i(t-\tau_s)$, so that
\begin{equation}
\PP(\bv{X}_i(t)=\bv{e}_k)=v_k(t-\tau_s)
\end{equation}
where $\bv{e}_k$ is a unit vector in the direction of the $k$th linguistic state. Defining $\delta \bv{u}_i(t) \triangleq \bv{u}_i(t)-\bv{u}_i(t-\tau_s)$ we introduce the following discrete dynamics, analogous to (\ref{eqn:hop})
\begin{align}
\delta \bv{u}_i(t) &= \frac{\tau_s}{\tau_m} \left( \sum_j \omega_{ij} \bv{X}_j(t) - \bv{u}_i(t-\tau_s) \right). 
\label{eqn:dhop}
\end{align}
Whereas $\tau_m$ measures the length of time that historical linguistic behaviour remains relevant to present behaviour, the constant $\tau_s$ controls the rate of stochastic switching. For given $\tau_m$, as $\tau_s \rar 0$ the model becomes deterministic, because the internal state is an average over a very large historical random sample of other speakers' behaviour, which consequently has small sample error. For large $\tau_s$, speakers randomly select a state and then stick to it for longer, meaning that individuals have a noisier sample of each others' internal states. For given $\tau_s$, increasing the memory time constant slows down the deterministic component of the dynamics, and reduces stochasticity by averaging over a larger sample of random updates.

In this paper we wish to analyse the approximate spatial behaviour of this model. To do so we divide space into a grid of square cells with side $a$, each with population $N$. We write $\cl(\bv{r})$ for the set of speakers in the cell centred on $\bv{r}$. The cell average internal state is then
\begin{equation}
\bv{u}(\bv{r}) \triangleq \frac{1}{N}\sum_{i \in \cl(\bv{r})} \bv{u}_i,
\end{equation}
with $\bv{v}(\bv{r}) \triangleq \bv{g}(\bv{u}(\bv{r}))$. The symbol $\triangleq$ denotes a definition. We let $\la \bv{r} \ra$ denote the set of cell centres which are nearest neighbours to $\bv{r}$ and introduce the \textit{interaction range}, $\sigma$, in terms of which we approximate cell-aggregated interaction strengths as follows
\begin{equation}
 \sum_{i \in \cl(\bv{r})} \omega_{ij} \approx \begin{cases}
1-2 \left( \frac{\sigma}{a} \right)^2 & \text{ if } j \in \cl(\bv{r}) \\
\frac{1}{2} \left( \frac{\sigma}{a} \right)^2 & \text{ if } j \in \cl(\bv{r}'), \bv{r}' \in \la \bv{r} \ra \\
0 & \text{ otherwise.}
\end{cases}
\end{equation}
According to this, each speaker receives a total connection weight of $1-2(\sigma/a)^2$ from other speakers in their own cell, and $(\sigma/a)^2/2$ from speakers in each nearest neighbour cell. We will see below that this definition is consistent with a Gaussian spatial interaction kernel. We also define the cell-aggregated emitted state
\begin{equation}
\bv{Y}(\bv{r}) = \sum_{j \in \cl(\bv{r})} \bv{X}_j.
\end{equation}
Averaging the noise term of our discrete dynamics (\ref{eqn:dhop}) over one cell we obtain
\begin{align}
\frac{1}{N} & \sum_{i \in \cl(\bv{r})} \sum_j \omega_{ij} \bv{X}_j = \frac{1}{N}  \sum_j \left(\sum_{i \in \cl(\bv{r})}  \omega_{ij} \right) \bv{X}_j \\
& = \left[1-2 \left( \frac{\sigma}{a} \right)^2 \right] \frac{\bv{Y}(\bv{r})}{N} + \frac{1}{2} \left( \frac{\sigma}{a} \right)^2 \sum_{\bv{r}' \in \la \bv{r} \ra} \frac{\bv{Y}(\bv{r}')}{N} \\
& =\frac{1}{N} \left[ \bv{Y}(\bv{r}) + \frac{\sigma^2}{2} \nabla^2 \bv{Y}(\bv{r}) \right]
\label{eqn:saddle}
\end{align}
where $\nabla^2$ is the discrete spatial second derivative
\begin{equation}
\nabla^2 \bv{Y}(\bv{r}) \triangleq \frac{1}{a^2} \left( \sum_{\bv{r}' \in \la \bv{r} \ra} \bv{Y}(\bv{r}') - 4 \bv{Y}(\bv{r}) \right)
\end{equation}
making (\ref{eqn:saddle}) a saddle point approximation \cite{ma71} to a Gaussian spatial average. The cell aggregated emitted state is approximately multinomial $\bv{Y}(\bv{r}) \sim \text{multinomial}(\bv{v}(\bv{r}),N)$ allowing us to efficiently simulate the spatial dynamics as follows
\begin{align}
\delta \bv{u}(\bv{r}) = \frac{\tau_s}{\tau_m} \left( \frac{1}{N} \left[ \bv{Y}(\bv{r}) + \frac{\sigma^2}{2} \nabla^2 \bv{Y}(\bv{r}) \right] - \bv{u}(\bv{r}) \right)
\label{eqn:spat_sim}
\end{align}
where  time dependence has been suppressed for brevity.

In order to understand the behaviour of the model analytically, it is useful to write the cell noise as a sum of deterministic and stochastic terms
\begin{equation}
N^{-1} \bv{Y}(\bv{r}) = \bv{v}(\bv{r}) + \frac{\bv{\epsilon}(\bv{r})}{\sqrt{N}}
\end{equation}
where the statistical properties of the stochastic term may be understood using the normal approximation to the multinomial distribution \cite{geo12} (see appendix \ref{ap:nomu} for details). Let $\mathsf{O}_{\bv{v}}$ be an orthogonal matrix ($\mathsf{O}_{\bv{v}}^T = \mathsf{O}_{\bv{v}}^{-1}$) whose last column is $\bv{v}$, and define  $\bv{Z} = (Z_1, Z_2, \ldots, Z_{q-1},0)^T$ where $Z_i \sim \mathcal{N}(0,1)$, then
\begin{equation}
\bv{\epsilon} \approx \bv{v}^{\odot \frac{1}{2}} \odot \mathsf{O}_{\bv{v} } \bv{Z}
\label{eqn:eps}
\end{equation}
where $\odot$ denotes the Hadamard (element-wise) product and $\odot \tfrac{1}{2}$ is the Hadamard square root. For example, in the binomial case $q=2$, this yields
\begin{equation}
\bv{\epsilon} = \sqrt{v_1 (1-v_1)} Z_1 \begin{bmatrix}
-1 \\
+1
\end{bmatrix},
\end{equation}
where we made use of the fact that $v_1 = 1-v_2$. Approximating the noise terms from nearest neighbour cells with their mean values, we obtain
\begin{equation}
\delta \bv{u}(\bv{r}) = \frac{\tau_s}{\tau_m} \left( \underbrace{\bv{v}(\bv{r}) - \bv{u}(\bv{r})}_{\text{Nonlinearity}} + \underbrace{\frac{\sigma^2}{2} \nabla^2 \bv{v}(\bv{r})}_{\text{Diffusion}} + \underbrace{\frac{\bv{\epsilon}(\bv{r})}{\sqrt{N}}}_{\text{Noise}} \right).
\label{eqn:spat_evo2}
\end{equation}
For practical purposes it is useful to consider a continuous time approximation to (\ref{eqn:spat_evo2}). If $\bv{W} = (W_1, W_2, \ldots, W_{q-1}, 0)^T$ is a vector of standard Brownian motions then
\begin{equation}
\bv{\epsilon} \overset{d}{=} \frac{1}{\sqrt{\tau_s}} \bv{v}^{\odot \frac{1}{2}} \odot \mathsf{O}_{\bv{v} } \int_0^{\tau_s} d\bv{W}_t,
\end{equation}
where $\overset{d}{=}$ denotes equality in distribution. Defining the continuous time differential form of $\bv{\epsilon}$ via
\begin{equation}
\bv{\epsilon} \triangleq \int_0^{\tau_s} d\bv{\epsilon}_t,
\end{equation}
we have
\begin{equation}
d \bv{\epsilon} = \frac{1}{\sqrt{\tau_s}} \bv{v}^{\odot \frac{1}{2}} \odot \mathsf{O}_{\bv{v} }d\bv{W}.
\end{equation}
The differential form of our evolution equation is then
\begin{align}
\nonumber
d \bv{u}  =  \frac{1}{\tau_m} &  \left(  \bv{v} - \bv{u} + \frac{\sigma^2}{2} \nabla^2 \bv{v} \right) dt \\
& + \frac{\sqrt{\tau_s}}{\tau_m  \sqrt{N} }  \bv{v}^{\odot \frac{1}{2}} \odot \mathsf{O}_{\bv{v} }d\bv{W}.
\label{eq:acsde}
\end{align}
where $\bv{r}$ dependence has been omitted for brevity. Equation (\ref{eq:acsde}) is discrete in space and continuous in time. Assuming that changes in the continuous version of $\bv{u}$ are small over the interval $[t,t+\tau_s]$ then the discrete form (\ref{eqn:spat_evo2}) may be retrieved by integrating (\ref{eq:acsde}) from $t$ to $t+\tau_s$. In the limit $\tau_s \rar 0$ the noise term in (\ref{eq:acsde}) vanishes, as expected.

\begin{figure}
	\centering
	\includegraphics[width=0.75\linewidth]{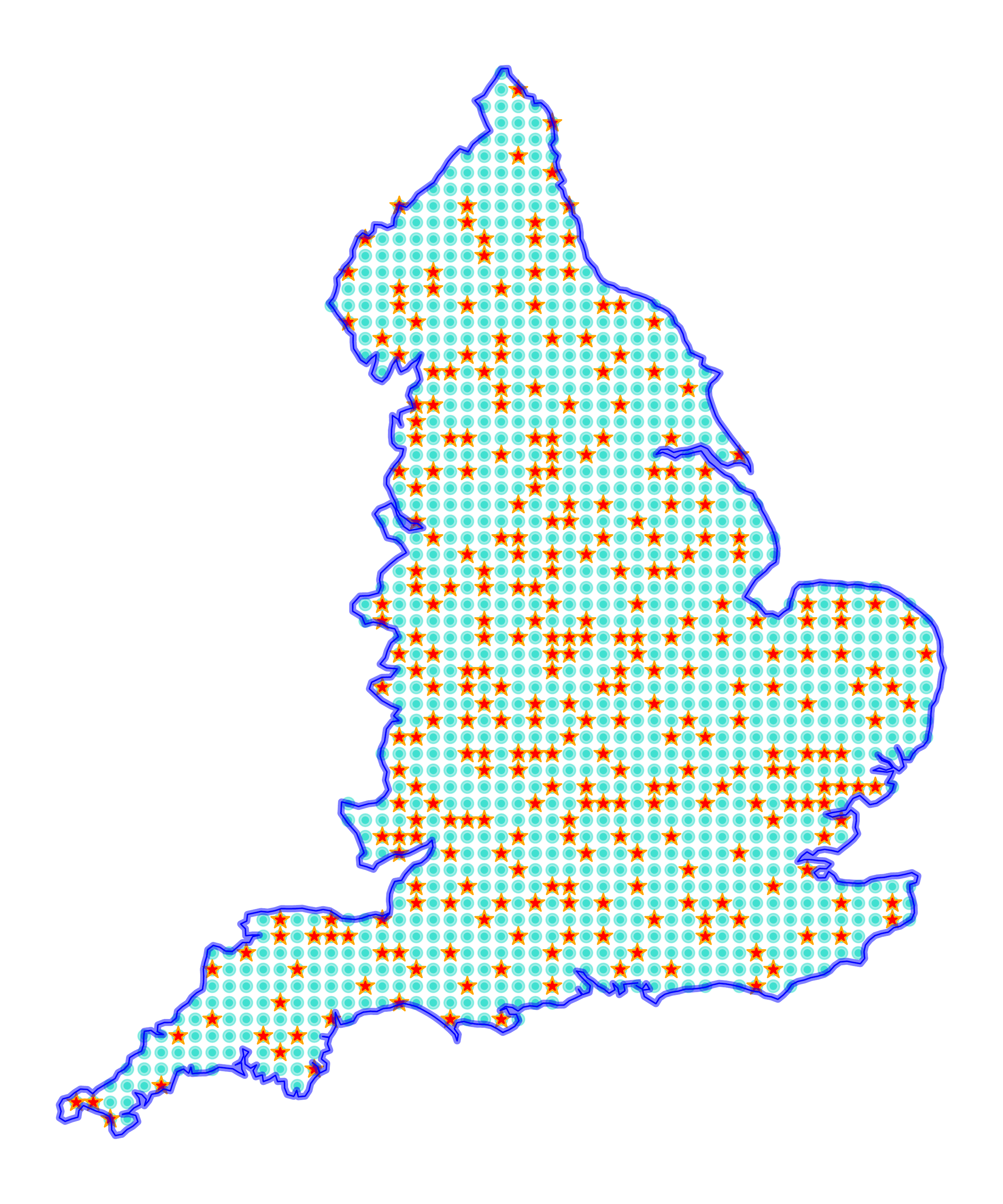}
	\caption{ Grid used for simulating English language features. Each light blue dot is the centre of a 10km $\times$ 10km grid square. Each red star is an SED survey location. There are 1329 grid squares, and 310 survey locations. }
	\label{fig:grid}
\end{figure}

Writing our evolution equation in this way allows us to explore the effects of rescaling the units of time by a factor of $c$, that is ($\dd t_{\text{new}} = c \dd t_{\text{old}}$). Such a rescaling yields the equation
\begin{align}
\nonumber
d \bv{u}  =  \frac{1}{c \tau_m} & \left(  \bv{v} - \bv{u} + \frac{\sigma^2}{2} \nabla^2 \bv{v} \right) dt \\
& + \frac{\sqrt{\tau_s}}{\tau_m  \sqrt{c N} } \bv{v}^{\odot \frac{1}{2}} \odot \mathsf{O}_{\bv{v} }d\bv{W}.
\label{eqn:rescaled}
\end{align}
Consider neutral evolution, where $\bv{u}=\bv{v}$, with population $N$ per cell and interaction range $\sigma$. If we reduce the population density to $N' = c N$ where $c<1$, then provided we also increase the interaction range to
\begin{equation}
\sigma' = \frac{\sigma}{\sqrt{c}}
\label{eqn:sig_sim}
\end{equation}
then the dynamics (\ref{eqn:rescaled}) is identical apart from a rescaling of time. Conversely, if we use simulated population $N_{\text{sim}} = cN$ and interaction range $\sigma_{\text{sim}}$ then we will obtain spatial distributions with approximately the same statistical properties as if we have simulated the model with the full population, $N$, and interaction range
\begin{equation}
\sigma_{\text{eff}} = \sqrt{\frac{N_{\text{sim}}}{N}} \sigma_{\text{sim}}.
\label{eqn:sigeff}
\end{equation}
This will allow us to explore neutral evolution by simulating smaller cell populations, which converge within a computationally feasible time frame. In this paper our focus will be on English folk speech, as recorded in the Survey of English dialects \cite{ort62}. For simulations we divide England into a grid of $10\text{km}\times10\text{km}$ squares, as shown in Figure \ref{fig:grid}. There are 1329 grid squares, each containing $10^4$ speakers giving a total population of $13.29$ million, a level reached in the mid 1830s \cite{wri89}.

\section{Interfaces and matching}

Our evolution equation (\ref{eqn:spat_evo2}) can generate spatial distributions of language use where domains emerge in which a particular linguistic feature dominates. The structure and dynamics of these domains depends strongly on the form of the activation function $\bv{g}$. In this paper we consider two alternatives, which we refer to as \textit{neutral}, $\bv{g}(\bv{u})=\bv{u}$, and \textit{conformity driven}. 

\subsection{Conformity driven case}

\label{sect:non_neut}

\begin{figure}
	\centering
	\includegraphics[width=\linewidth]{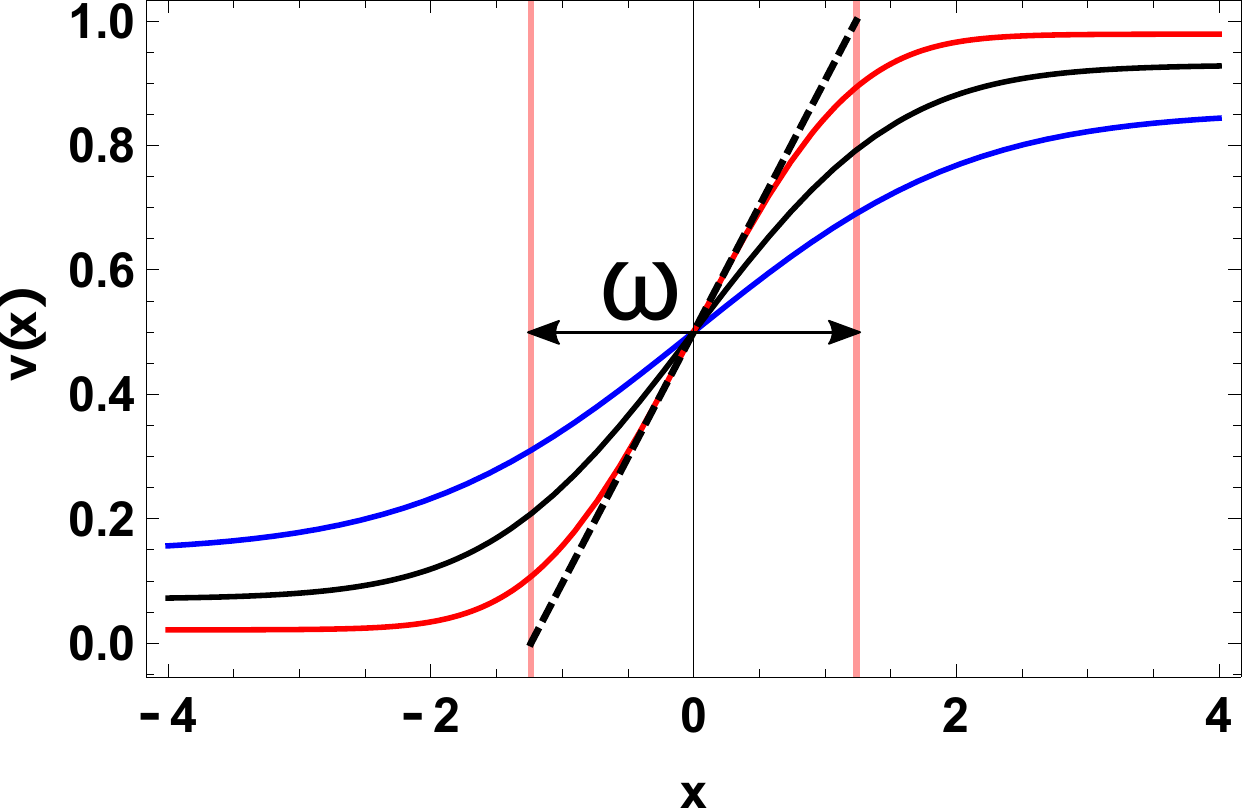}
	\caption{ Interface shape when $\sigma=1$ and $\beta \in \{2.5,3,4\}$ (blue, black, red). Dashed line shows gradient of interface for $\beta=4$, and red vertical lines show width of interface $\omega(\sigma,\beta)$ in this case. }
	\label{fig:interface}
\end{figure}

\begin{figure}
	\centering
	\includegraphics[width=\linewidth]{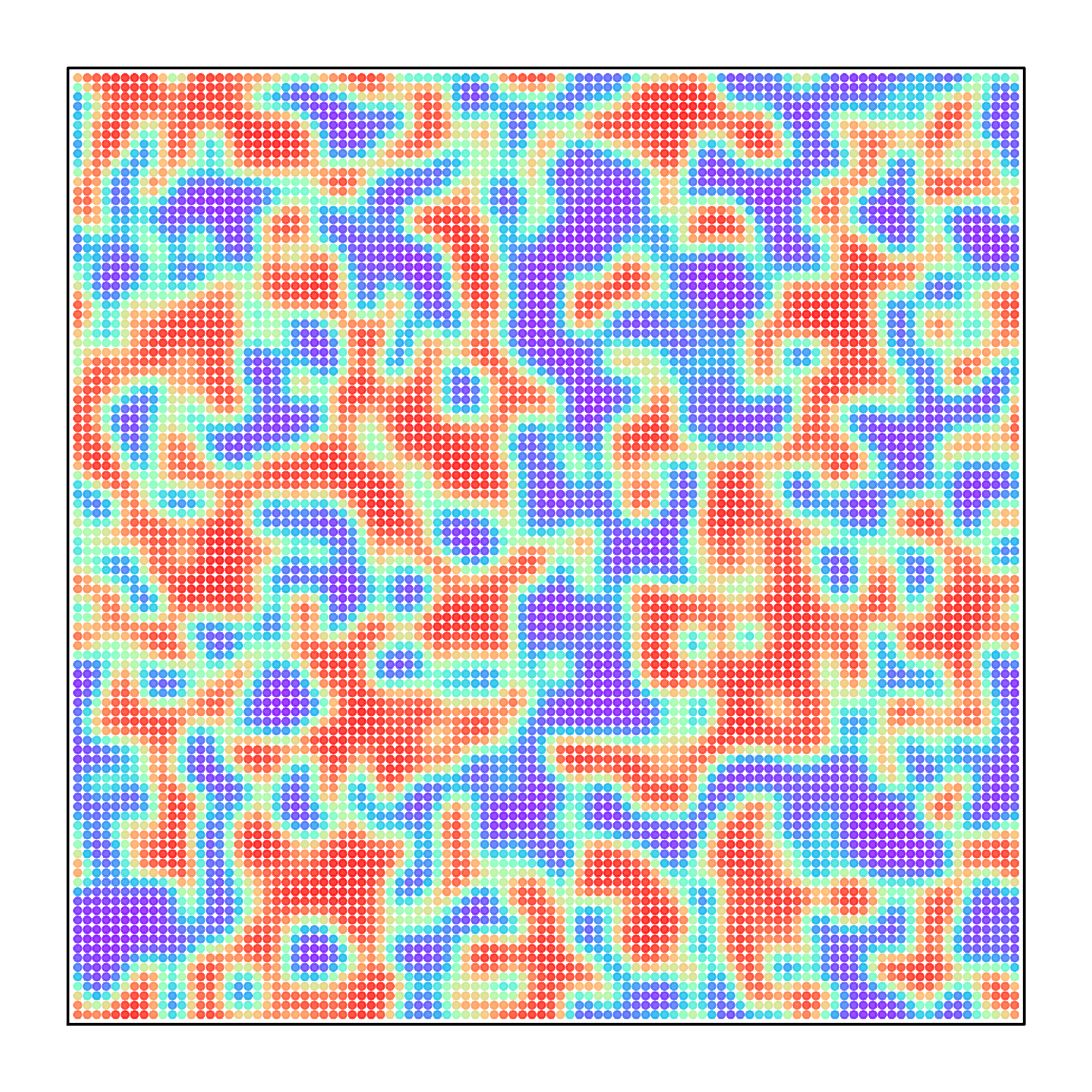}
	\caption{ Spatial distribution of the two state probability mass function $\bv{v}(\bv{r})$ over a $1000 \text{km} \times 1000 \text{km}$ toroidal system with $a=10$km. Parameter values $\sigma=5$km, $N=10^4$, $\beta=2.5$. $\tau_m=2, \tau_s=1$.  Evolution shown after 25 time steps starting from randomized initial conditions. Red and blue correspond to the two possible variants.   }
	\label{fig:doughnut}
\end{figure}

For simplicity we consider a binary variable, so $\bv{v}$ may be written
\begin{equation}
\bv{v} = (v,1-v).
\end{equation} 
 We define the following activation function, which gives the probability for selecting variant one, 
\begin{equation}
g(u) \triangleq \frac{e^{\beta u}}{e^{\beta u} + e^{\beta(1-u)}}.
\label{eqn:potts_g}
\end{equation}
The parameter $\beta$, which we call the \textit{conformity number}, is analogous to inverse temperature in physical systems \cite{kra10}. As $\beta \rar 0$, corresponding to a very noisy or ``hot'' system, $g(u) \rar \tfrac{1}{2}$ meaning that variants are selected entirely at random. As $\beta \rar \infty$ speakers select the variant which is most common in their memory, leading to spatially ordered states. The fact that speakers using activation function (\ref{eqn:potts_g}) tend to adopt the behaviour of the majority is the origin of the term \textit{conformity driven} \cite{mor12, che05}. The critical inverse temperature $\beta_c = 2$ marks the transition between the disordered case, where variants persist in approximately equal proportions, and the ordered case when one variant dominates. The inverse activation function is
\begin{equation}
g^{-1}(v) = \frac{1}{2} \left[1 + \frac{1}{\beta} \ln \left( \frac{v}{1-v} \right) \right].
\end{equation}    
The most important property of conformity driven dynamics is its ability to maintain spatial interfaces \cite{bra94, che05}. To see how this occurs, consider an interface aligned along the $y$ axis, so that $v = v(x)$. Assuming the population is large enough so that the system is well described by the deterministic component of its dynamics, then the steady state shape of the interface, a smoothed step function,  solves
\begin{equation}
\frac{\sigma^2}{2} v''(x) = \frac{1}{2} \left[1 + \frac{1}{\beta} \ln \left( \frac{v}{1-v} \right) \right] - v
\label{eqn:equv}
\end{equation} 
where $v''(x)$ denotes the lattice second derivative. We assume that $v(x)$ is sufficiently slowly varying so that $x$ may be treated as continuous and (\ref{eqn:equv}) treated as an ordinary differential equation. 
Without loss of generality we can assume the interface is centred on the origin, where $v(0)=\tfrac{1}{2}$, which is a fixed point of the right hand side of (\ref{eqn:equv}). As $x \rar \pm \infty$,  $v''(x) \rar 0$ and $v$ approaches one of the two other fixed points, which solve
\begin{equation}
\frac{v}{1-v} = \exp \left( \beta (2v-1)\right).
\end{equation} 
We write these solutions, which lie to the left and right of $v=\tfrac{1}{2}$ as  $v^\ast_-$ and $v^\ast_+$. We now define the potential function
\begin{equation}
V(v) \triangleq \frac{(1-2v)^2}{8} - \frac{\ln (2-2v)}{2 \beta} - \frac{v \ln \left( \frac{v}{1-v} \right)}{2 \beta},
\end{equation}
in terms of which we may write our equilibrium equation (\ref{eqn:equv})
\begin{equation}
\frac{\sigma^2}{2} v''(x) = -\frac{dV}{dv}.
\label{eqn:newt}
\end{equation}
Noting that $v''(x) = v'(x) \tfrac{d}{dv} v'(x)$, and integrating (\ref{eqn:newt}) with respect to $v$ we obtain the conservation law
\begin{equation}
E \triangleq \frac{\sigma^2}{4} (v')^2 + V(v) 
\end{equation}
where $E$ is a constant, which we may view as a conserved ``energy''. To see this, note that if we interpret $v,x$ as as position and time variables, then (\ref{eqn:equv}) describes the motion of a particle of mass $\sigma^2/2$ moving in a potential $V(v)$. Noting that $V(\tfrac{1}{2})=0$, then the gradient of the interface at the origin is given by
\begin{equation}
v'(0) = \sqrt{\frac{4E}{\sigma^2}}.
\end{equation}
To find $E$ we note that $\lim_{x \rar \pm \infty} v'(x) = 0$ so 
\begin{align}
E &= V(v^\ast_+)  = V(v^\ast_-)  \\
& \sim \frac{1}{8} +\frac{\ln 2}{2 \beta} \text{ as } \beta \rar \infty.
\end{align}
A simple measure of the width of the interface is the reciprocal of its gradient at the origin, which has asymptotic behaviour
\begin{equation}
\omega(\sigma,\beta) \sim \sqrt{2} \sigma \left( 1+ \frac{2 \ln 2}{\beta} \right)  \text{ as } \beta \rar \infty.
\label{eqn:omega}
\end{equation}
From this we see that the width of the interface scales linearly with the interaction range, and becomes wider at higher temperatures. As $\beta \rar 2^+$, the interface becomes infinitely wide as the system transitions to disorder. Figure \ref{fig:interface} shows some example interfaces, obtained by numercially solving equation (\ref{eqn:equv}). 
Figure \ref{fig:doughnut} shows how such interfaces can spontaneously from, starting from randomized initial conditions. This process, known as \textit{coarsening}, is widely observed in two dimensional physical models of phase ordering, which have been adapted many times to model social phenomena, including language \cite{bur16,bur17,bur18,tes06,cas09,bar06,bar06_2}. 

\begin{figure}
	\centering
	\includegraphics[width=0.9\linewidth]{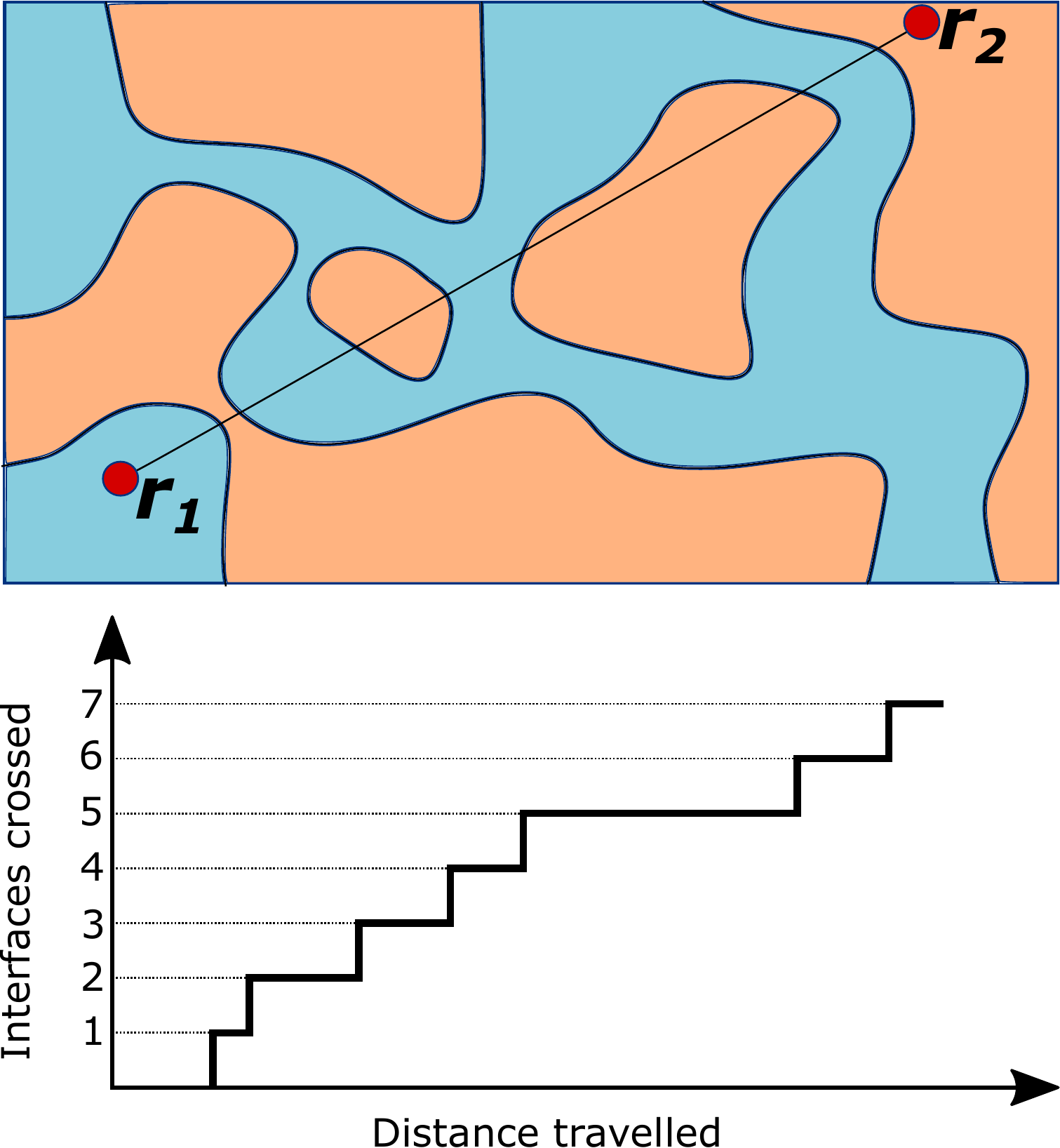}
	\caption{ Interface crossing count, travelling between $\bv{r}_1$ and $\bv{r}_2$ in a binary linguistic system. If an even number of interfaces are crossed, then the speakers at $\bv{r}_1$ and $\bv{r}_2$ will match with high probability. }
	\label{fig:renew}
\end{figure}

\begin{figure}
	\centering
	\includegraphics[width=0.9\linewidth]{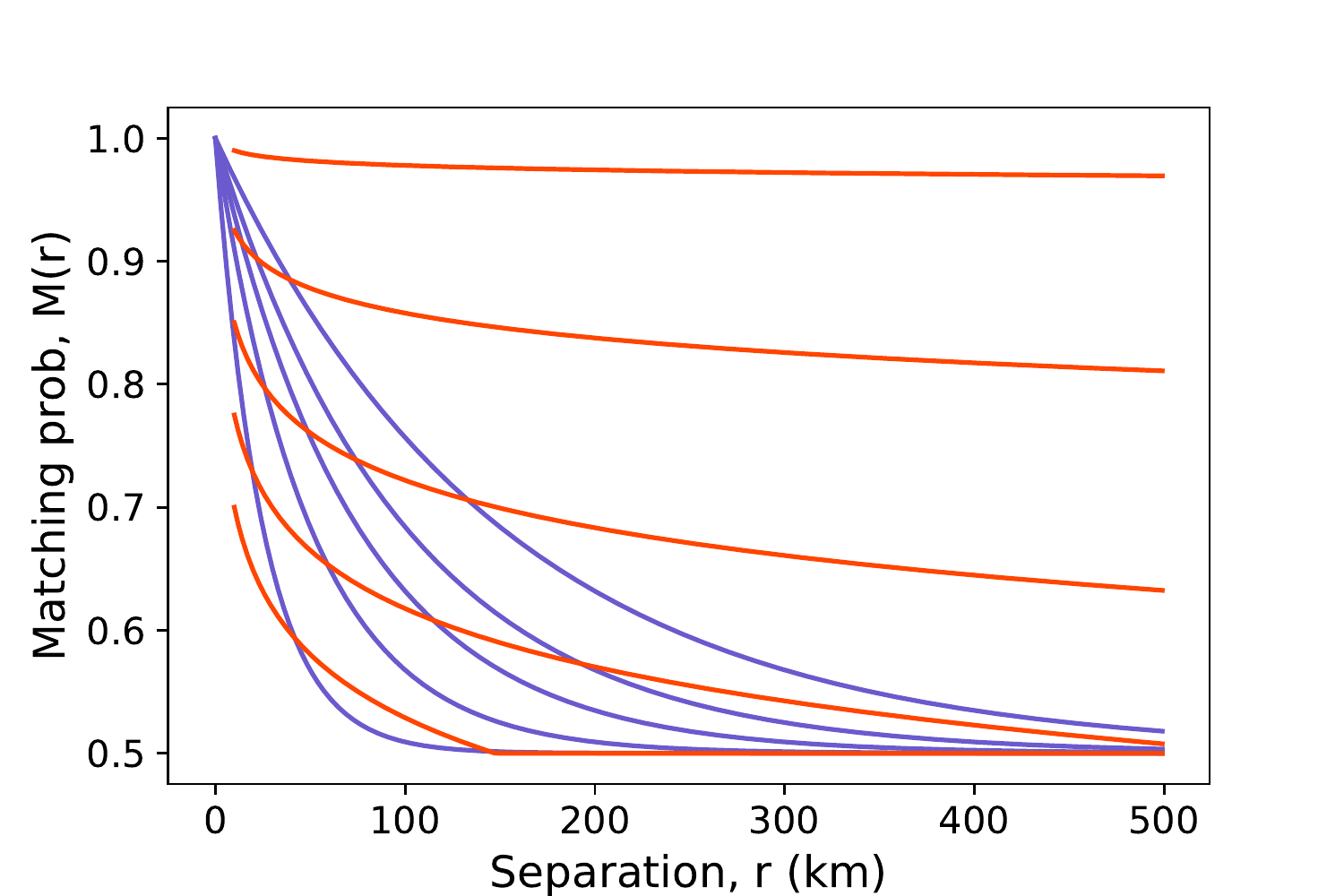}
	\caption{ Blue curves show conformity driven matching probability functions given by (\ref{eqn:expmat}) for the five $\lambda$ values given in table \ref{tab:param}. Red curves show neutral matching probabilities given by (\ref{eqn:msol}) for five $(b,c)$ values, also given in table \ref{tab:param}.  }
	\label{fig:mat}
\end{figure}

The central aim of our paper is to infer what forms of activation function realistically capture the true processes which drive language evolution. If linguistic conformity is a significant driver of change, then in systems where most interactions are short range, and speakers do not migrate too much, then we would expect to see the formation of interfaces. Our methods of inference are primarily based on the \textit{matching probability}, $M(\bv{r}_1,\bv{r}_2)$, between two locations. This gives the probability that the emitted states of two speakers in cells $\bv{r}_1$ and $\bv{r}_2$ are identical. In the case of a binary linguistic variable, we have 
\begin{align}
M(\bv{r}_1,\bv{r}_2) = 2v(\bv{r}_1) v(\bv{r}_2) - v(\bv{r}_1) -  v(\bv{r}_2) + 1. 
\end{align}
It is possible to estimate this matching probability by direct simulation, or by adapting analytical techniques developed to calculate correlations in physical systems which exhibit phase ordering, starting from randomized initial conditions \cite{bra94}. However, from a social-linguistic perspective these methods have some potential drawbacks. We know that the positions of interfaces can be influenced by  initial conditions (determined by history and migration, and by the locations of innovations \cite{cry03}), population distributions,  geographical features, and localized cultural identities \cite{lab01,bur17,cha98}, and these may affect the sizes of domains. For this reason, as well as direct simulations, we also use an alternative approach in which typical domain size is a free parameter. We imagine walking from $\bv{r}_1$ to $\bv{r}_2$, and counting the interfaces that we cross on the way (see Figure \ref{fig:renew}). 
The crossing points of any long straight line drawn across the system will form a point process \cite{grim01}. To facilitate calculations, we will assume that the intervals between crossing points are independent random variables drawn from some distribution $f(\Delta r)$ (the marginal of the joint interval distribution), so that the locations of crossing points form a renewal process \cite{grim01}. The simplest choice of marginal is exponential
\begin{equation}
f(\Delta r) = \frac{1}{\lambda} \exp\left( - \frac{\Delta r}{\lambda} \right),
\end{equation}
where $\lambda$ is the average distance between crossings - a measure of the typical size of a single domain. In this case the crossing points form a Poisson Point Process \cite{grim01} with intensity $\lambda^{-1}$, and the number, $N(r)$, of crossings on a line of length $r$ is a Poisson random variable with expectation $\EE[N(r)] = r/\lambda$, and mass function
\begin{equation}
\PP(N(r)=k) \triangleq p_k(r) = \frac{1}{k!} \left(\frac{r}{\lambda}\right)^k e^{-\frac{r}{\lambda}}.
\end{equation}
Assuming that $\beta$ is large, so that domains are linguistically \textit{pure} with narrow interfaces, then the matching probability for two points separated by a distance $r = | \bv{r}_1 - \bv{r}_2 |$ is the probability that an even number of interfaces are crossed on the journey between them
\begin{align}
M(\bv{r}_1,\bv{r}_2) &= \sum_{k=0}^\infty p_{2k}(r) \\
&= e^{-\frac{r}{\lambda}} \cosh \left( \frac{r}{\lambda} \right).
\label{eqn:expmat}
\end{align}
These matching probabilities are plotted in Figure \ref{fig:mat}.
We note that exponentially decaying match probabilities (or correlations) are generic in phase ordering systems driven by short range interactions \cite{bra94}.

\begin{table}
\begin{tabular}{|c|c|c|c|c|c|}
	\hline
	Curve & 1  & 2  & 3  & 4  & 5 \\
	\hline
	\hline
	Conformity: $\lambda$ (km) & 50 & 100 & 150  & 200 & 300  \\
	\hline
	\hline
	Neutral: $b$ & 0.700  & 0.775 & 0.850  & 0.925 & 0.990  \\
	\hline
	Neutral: $c$ & -0.0743 & -0.0684 & -0.0556 & -0.0292 & -0.0053  \\
	\hline
\end{tabular}
\caption{\label{tab:param} Parameter values for the five different conformity driven $(\lambda)$ and neutral $(b,c)$ matching curves. Neutral $(b,c)$ pairs estimated by fitting logarithmic function (\ref{eqn:msol}) with $\epsilon=10$km to simulated matching probabilities using model parameters $\sigma = 3.16$km and $N=5$ (approximately equivalent to $\sigma=70m$ when $N=10^4$), and $\tau_s=1, \tau_m=2, q=2$.  }
\end{table}

\subsection{Neutral case}

In the case of neutral evolution we have $\bv{v}=\bv{g}(\bv{u})=\bv{u}$ so our spatial system is governed by the noisy discrete diffusion equation
\begin{equation}
\delta \bv{v} =
\frac{\tau_s}{\tau_m} \left( \underbrace{\frac{\sigma^2}{2} \nabla^2 \bv{v}}_{\text{Diffusion}} + \underbrace{\frac{\bv{\epsilon}}{ \sqrt{N} }}_{\text{Noise}}\right). 
\label{eqn:neut}
\end{equation}
In the two variant case the continuous time approximation to this equation is
\begin{equation}
dv =  \frac{1}{\tau_m} \frac{\sigma^2}{2} (\nabla^2 v)  dt + \frac{\sqrt{\tau_s}}{\tau_m}\sqrt{\frac{1}{ N}  v(1-v)} dW.
\label{eqn:neut_sde}
\end{equation}
An equation of this form, but with different parameters, also approximates Kimura's stepping stone model of neutral genetic evolution \cite{kim53,cox02}, where the population is divided into a grid of cells (``demes'' or ``colonies'' in the language of genetics) each containing $N_K$ individuals. The model \cite{cox02} evolves in discrete time with each member of generation $n+1$ inheriting their type (one of two possible alleles A or B) from a member of generation $n$ who is selected from the same cell with probability $1-m$ or from a nearest neighbour cell with probability $m$. A continuous time approximation for this process, is 
\begin{equation}
dv = \frac{m}{4} a^2 (\nabla^2 v) dt + \sqrt{\frac{1}{ N_K}  v(1-v)} dW
\label{eqn:kim_sde}
\end{equation}
where $v$ is now the population fraction with allele A.  The approximation applies when the population per cell  is large enough so that the noise may be approximated with a Wright-Fisher diffusion \cite{eth09}.  The spatial patterns generated by the stepping stone model have been heavily studied in the equilibrium setting (see \cite{cox02, eth09} and references therein), where it is used to understand how genetic differences accumulate with distance. 

Spatial variations in our model and the stepping stone model are the result of a competition between diffusion and local noise. Whereas diffusion acts to equalise states between nearby sites, local noise generates spatial variations. For sufficiently large interaction range the diffusion term will equalise the linguistic/genetic state across the speakers in a system much faster than noise effects can create locally distinct variations. In this case the system behaves as if it were a single well-mixed group (we note the ``well mixed group'' condition has recently been corrected \cite{cox02} from its original form \cite{kim71}). After some time, noise effects will drive the population into one of two pure states. When this occurs the system is said to have \textit{fixed}. In order for spatially distinct domains to form, diffusion must act sufficiently slowly so that parts of the system can temporarily enter different pure or near-pure states. Even if this is the case, the entire system will eventually fix in one or other state.

\begin{figure}
	\centering
	\includegraphics[width=\linewidth]{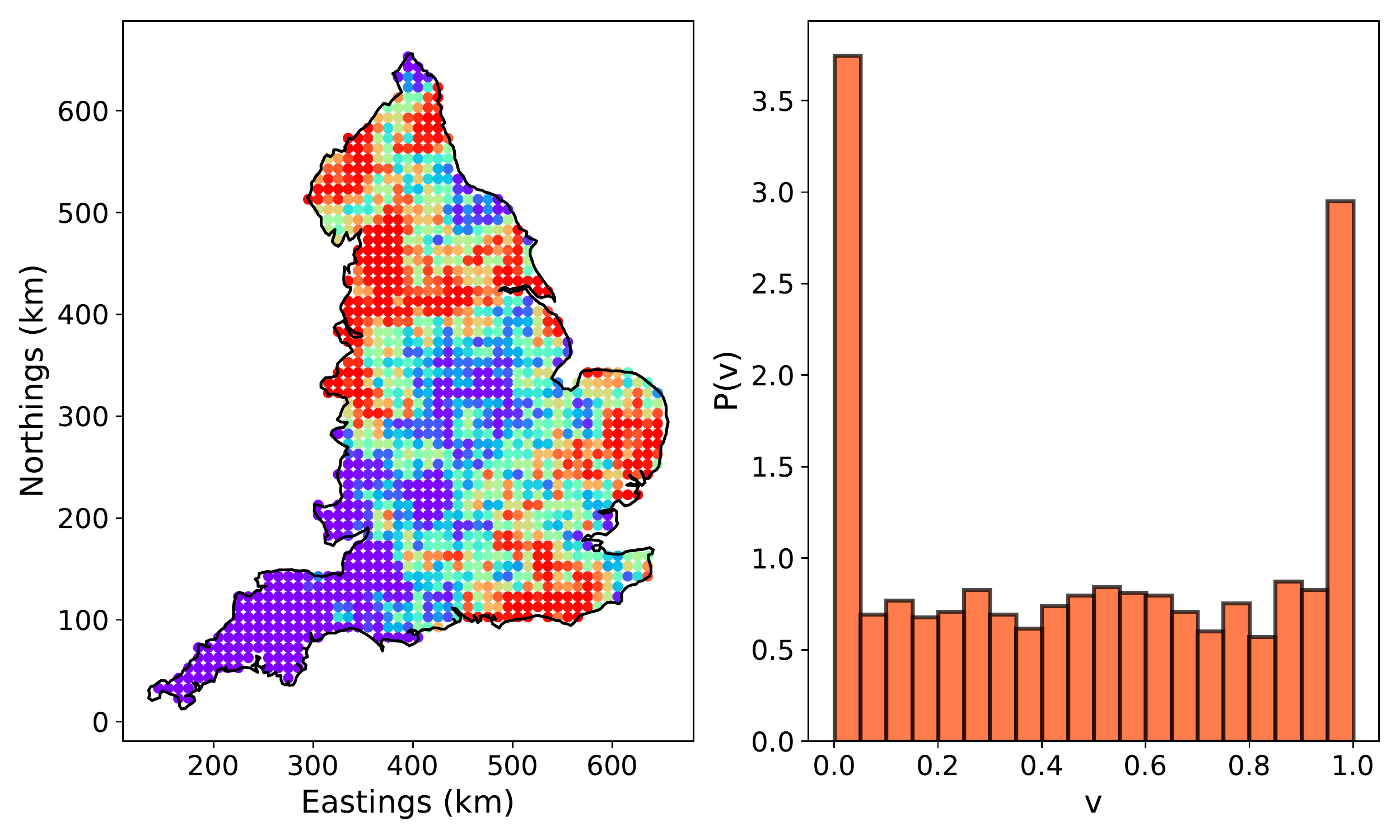}
	\caption{ External sates of neutrally evolving $q=2$ state system with $\tau_m=2, \tau_s=1$. The simulated population per cell and interaction range are $N_{\text{sim}}=5$ and $\sigma_{\text{sim}}=3.16$km  giving an effective interaction range of $\sigma_{\text{eff}} \approx 70$m when $N=10^4$ and $a=10$km.  Histogram shows distribution of external states. }
	\label{fig:neut_N_5}
\end{figure}

\begin{figure}
	\centering
	\includegraphics[width=\linewidth]{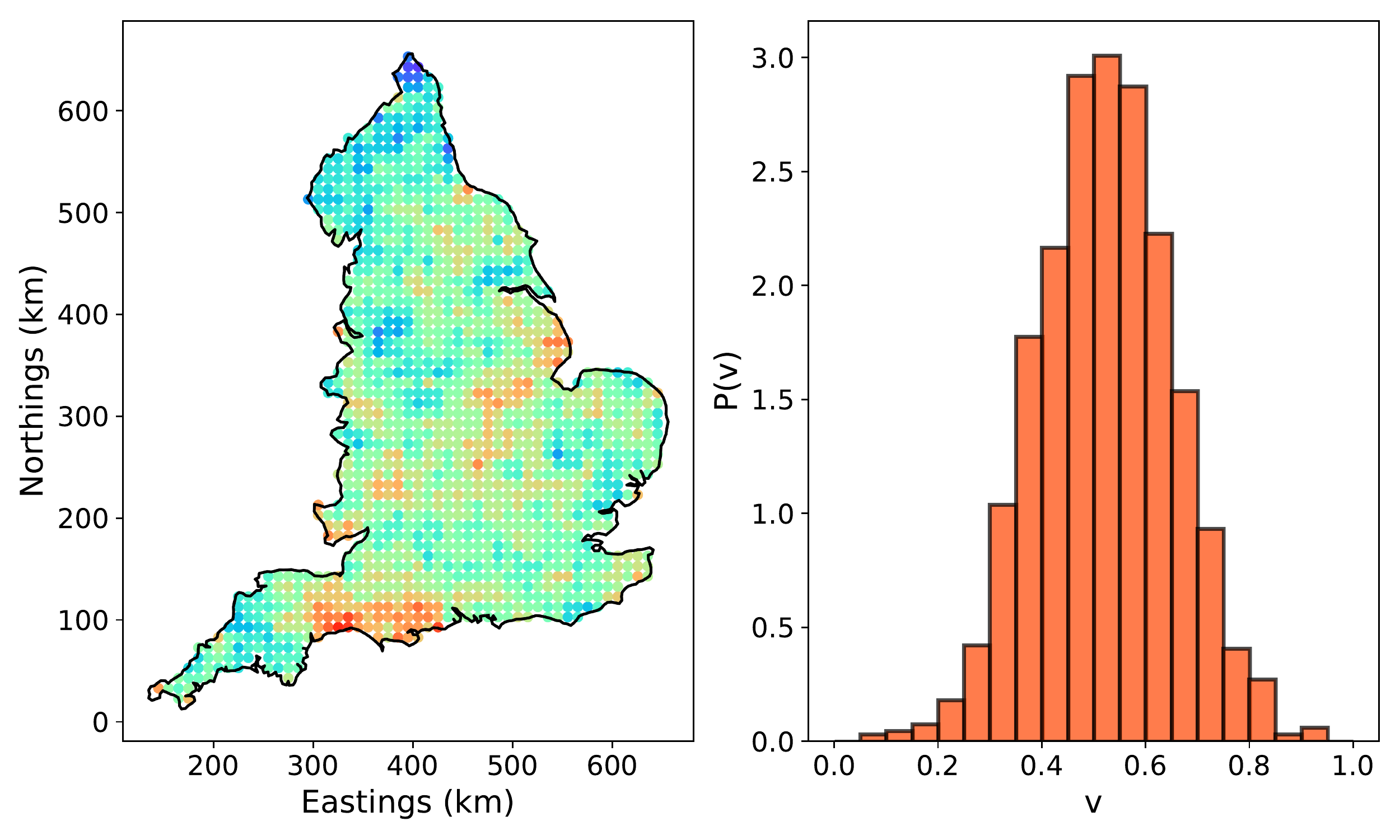}
	\caption{ External sates of neutrally evolving $q=2$ state system with $\tau_m=2, \tau_s=1$. The simulated population per cell and interaction range are $N_{\text{sim}}=50$ and $\sigma_{\text{sim}}=3.16$km  giving an effective interaction range of $\sigma_{\text{eff}} \approx 220$m when $N=10^4$ and $a=10$km.  Histogram shows distribution of external states. }
	\label{fig:neut_N_50}
\end{figure}

To see how interaction range affects this process, we give a heuristic derivation of the conditions under which distinct zones form.  Consider the behaviour of a purely diffusive ($N \rar \infty$) version of (\ref{eqn:neut_sde}), which describes the spatial diffusion of particles, genes or linguistic variants with diffusion coefficient $D=\tau_s \sigma^2/(2 \tau_m)$. The root mean squared displacement of a diffusing particle, evolving according to (\ref{eqn:neut_sde}) after time $t$, is  
\begin{align}
d(t) &= \sqrt{4 D t} \\
&= \sigma \sqrt{\frac{2t}{\tau_m}}.
\end{align}
This is the \textit{diffusion distance}.  In a system of linear size $L$, the time to diffuse across the system is then
\begin{equation}
t_{\text{mix}} \triangleq \frac{\tau_m}{2} \left(\frac{L}{\sigma}\right)^2.
\end{equation}
We call this the \textit{mixing time}. If diffusion acts sufficiently quickly, then the linguistic zone may be thought of as a single \textit{panmictic} (random mixing) group \cite{cox02,eth09} of size $M\approx N (L/a)^2$ obeying
\begin{equation}
dv = \frac{\sqrt{\tau_s}}{\tau_m \sqrt{M}} \sqrt{v(1-v)} dW_t.
\label{eqn:WF}
\end{equation}
The expectation of the time $T$ required for this group to fix, starting from state $v(0)=x$ is
\begin{equation}
\EE[T|v(0)=x] = -\frac{2 M \tau_m^2}{\tau_s} \ln \left( (1-x)^{1-x} x^x \right). 
\end{equation} 
The typical \textit{fixation time}, starting from an equal proportion of each variant ($x=1/2$) is then
\begin{equation}
t_{\text{fix}} \triangleq \frac{2 \ln 2 M \tau_m^2}{\tau_s}.
\end{equation}
Now suppose that the mixing and fixing times are comparable. Setting $t_{\text{mix}}=t_{\text{fix}}$ we obtain the condition
\begin{equation}
4 \ln 2 \frac{\tau_m}{\tau_s} N \left(\frac{\sigma}{a}\right)^2 = 1.
\end{equation}
In terms of population density $\rho = N/a^2$ this gives an approximate critical interaction range
\begin{equation}
\sigma_c \approx \sqrt{\frac{\tau_s}{\tau_m \rho}},
\end{equation}
where we have neglected the multiplicative constant $(4 \ln 2)^{-1/2} \approx 0.6 $. If $\sigma$ is of the order of $\sigma_c$ or smaller, then variants cannot mix fast enough to keep the system in an effectively fully connected state. Subregions may then form which, due to slow mixing, are isolated for long enough to allow temporary ``fixation'' into different pure states before system wide fixation occurs. In order for the population to be panmictic (not geographical), $\sigma$ must be substantially larger than $\sigma_c$ \cite{cox02}. We note that a condition for ``marked'' spatial variation in the stepping stone model is given in \cite{kim71} 
\begin{equation}
mN_K <1,
\label{eqn:kim_cond}
\end{equation}
which is equivalent (by comparing (\ref{eqn:neut_sde}) and (\ref{eqn:kim_sde})) to a critical interaction range
\begin{equation}
\hat{\sigma}_c = \frac{1}{\sqrt{2}} \sqrt{\frac{\tau_s}{\tau_m \rho}}
\end{equation}
matching our heuristically derived value up to a constant close to unity.  Notice that $\sigma_c$ does not depend on system size. As the local mixing rate increases, people become more connected and the size of group which can be considered to have the same internal state increases. This in turn reduces the noise and slows down the dynamics, meaning that a less rapid mixing rate is sufficient to keep even larger groups in the same internal state. Taking $\rho=104.7$, which was the population density in England in 1841  we find that
\begin{equation}
\sigma_c \approx \frac{1}{10} \sqrt{\frac{\tau_s}{ \tau_m}}. 
\end{equation}
Assuming that $\tau_s/\tau_m \approx 1$, then if the interaction range is substantially greater than a hundred meters, distinctive zones will not form. As noted in section \ref{sec:stoch}, using  realistic cell populations and varying the interaction range requires unfeasibly long simulation times. We therefore simulate using a fixed, moderate interaction range and reduce the cell population, using relation (\ref{eqn:sigeff}) to estimate the effective interaction range which would generate similar spatial distributions if the cell population took a realistic value. The effect of reducing cell population / effective interaction range is illustrated in Figures \ref{fig:neut_N_5} and \ref{fig:neut_N_50} which show the evolution of the neutral model with $\sigma_{\text{sim}}=3.16$km and cell populations $N_{\text{sim}}=5,50$ giving effective interaction ranges of 70m and 220m. In the shorter range case, distinctive localised linguistic zones appear, creating a bimodal probability distribution of $v(\bv{r})$ over the system as a whole. When $\sigma_{\text{eff}} \approx 220$m, although there are small spatial fluctuations, the distribution of $v(\bv{r})$ is clustered around a single value, meaning that the population as a whole are evolving as a single group. Interfaces and strong regional variations are therefore a feature of both neutral and non-neutral evolution, but in the neutral case, for realistic population densities, very low geographical connectivity is required for domains to form. In addition, from Figure \ref{fig:neut_N_5}, the interfaces are geographically wide, and of a much more complex shape than the non-neutral case, where their evolution is driven a by a surface tension effect \cite{bra94,kra10,bur17}. 

\begin{figure}
	\centering
	\includegraphics[width=\linewidth]{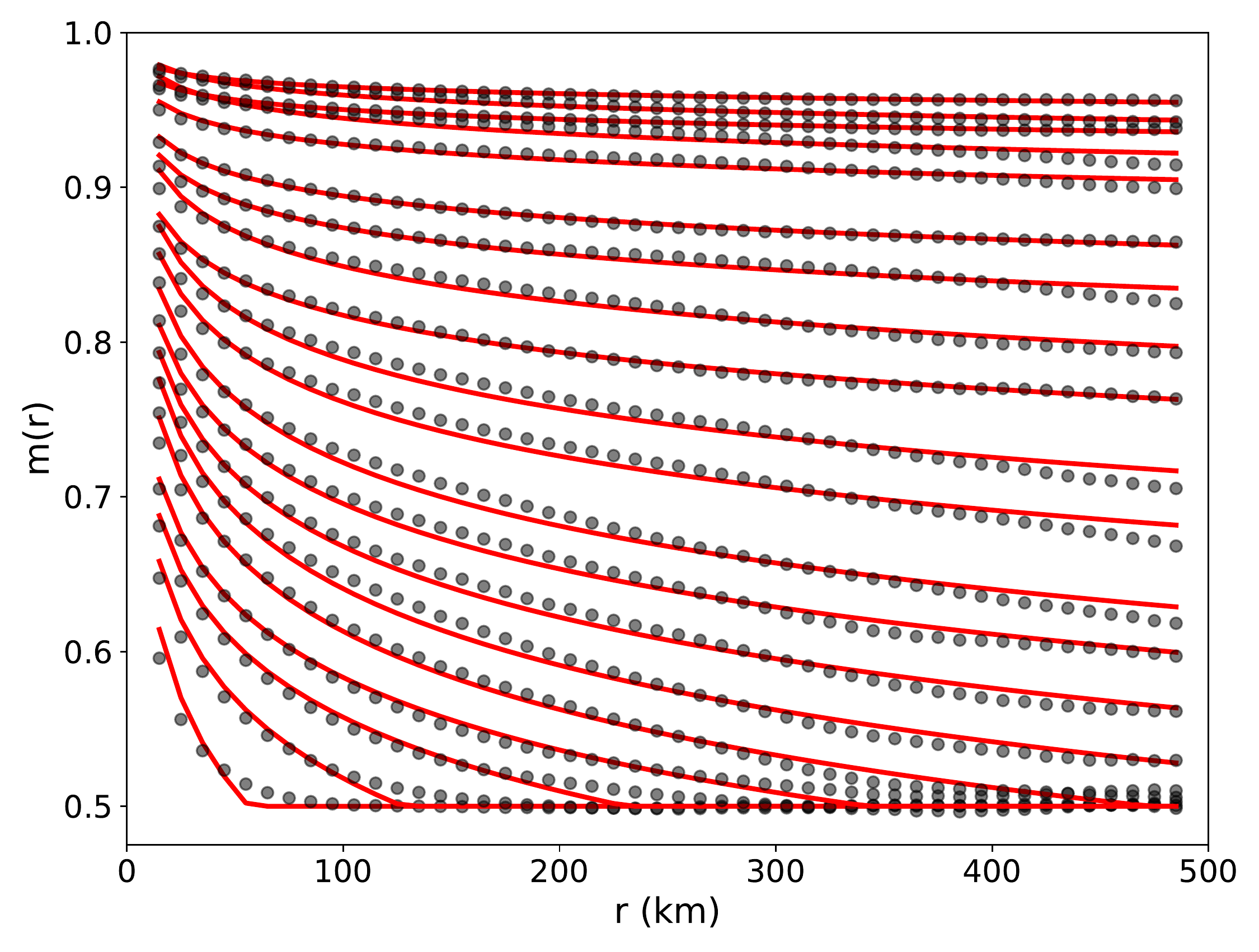}
	\caption{ Dots show equal time snapshots of the simulated correlation function $m(\bv{r})$ calculated from 500 realisations of  a neutrally evolving $q=2$ state model with $\tau_m=2, \tau_s=1$, simulated on England starting from randomized, spatially uncorrelated initial conditions. Snapshots of the ensemble of systems were taken at a (quadratically increasing) sequence of 8 times in the interval $T \in [1, 2 \times 10^4]$. The simulated population per cell and interaction range are $N_{\text{sim}}=5$ and $\sigma_{\text{sim}}=3.16$km  giving an effective interaction range of $\sigma_{\text{eff}} \approx 70$m when $N=10^4$ and $a=10$km.  Red curves show least squares fits to $\max(m(\bv{r}),1/2)$ where $m(\bv{r})$ defined by (\ref{eqn:msol}) and $\epsilon=10$km.  }
	\label{fig:m_Eng}
\end{figure}

\begin{figure}
	\centering
	\includegraphics[width=\linewidth]{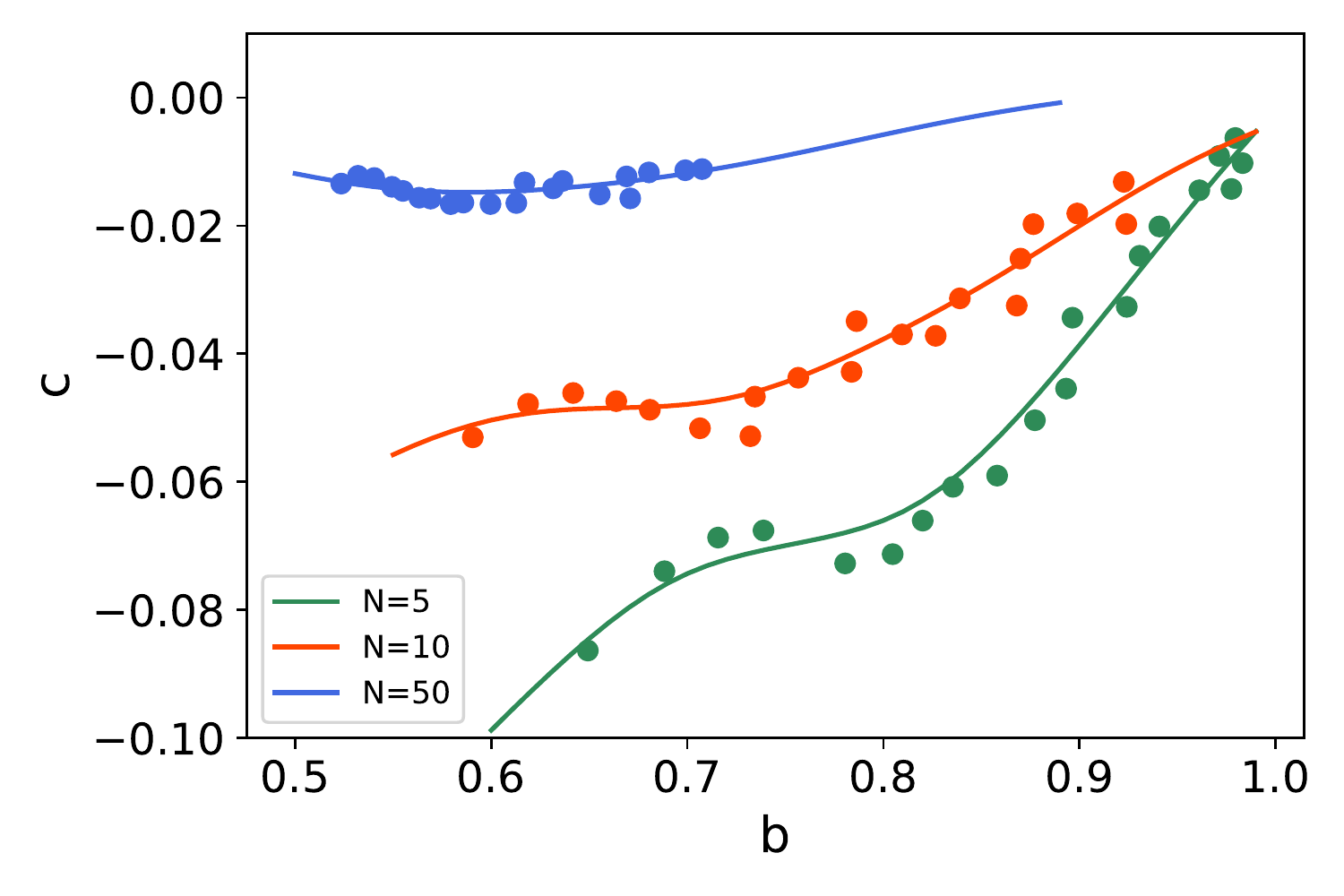}
	\caption{ Relationships between the parameters of the matching probability (\ref{eqn:msol}), with $\epsilon=10$km, obtained by fitting to snapshots at times in the interval $T \in [1, 2 \times 10^4]$ of the simulated correlation function $m(\bv{r})$ calculated from 500 realisations of the neutrally evolving model. In every case we have $\tau_m=2, \tau_s=1, \sigma_{\text{sim}}=3.16$km, $N=10^4$ and $a=10$km. The three curves correspond to different simulated cell populations $N_{\text{sim}} \in \{5,10,50\}$.  Solid lines show Kernel Density Regression on simulated data with bandwidth $h=0.05$.  }
	\label{fig:bc}
\end{figure}

To calculate matching probabilities, we consider a very large (spatially invariant) system and derive and equation for $M(\bv{R},\bv{R}+\bv{r})$ averaged over all locations $\bv{R}$ to give a matching probability which depends only on displacement
\begin{equation}
m(\bv{r}) \triangleq  \left \la M(\bv{R},\bv{R}+\bv{r}) \right \ra_{\bv{R}}
\end{equation}
where $\la \cdot \ra_{\bv{R}}$ denotes the average over $\bv{R}$. Assuming that terms of order $\delta v(\bv{r}_1) \delta v(\bv{r}_2)$ can be neglected then the change in $M(\bv{R},\bv{R}+\bv{r})$ per time step, when $\bv{r} \neq \bv{0}$, is
\begin{multline}
\delta M(\bv{R},\bv{R}+\bv{r}) = \\
\sigma^2 \frac{\tau_s }{\tau_m}  \Big[ v(\bv{R})  \nabla^2 v(\bv{R}+ \bv{r}) + v(\bv{R}+\bv{r}) \nabla^2 v(\bv{r}) \\ 
- \frac{1}{2} (\nabla^2 v(\bv{R}) + \nabla^2 v(\bv{R} + \bv{r}))\Big] + \text{noise} \\
= \frac{\sigma^2}{2} \frac{\tau_s }{\tau_m}  (\nabla^2_{\bv{R}} + \nabla^2_{\bv{R}+\bv{r}}) M(\bv{R},\bv{R}+\bv{r}) + \text{noise}
\end{multline} 
where in the final line we have made the position at which derivatives are taken explicit. Averaging over $\bv{R}$ we obtain, for $\bv{r} \neq \bv{r}$
\begin{align}
\la \nabla^2_{\bv{R}} M(\bv{R},\bv{R}+\bv{r}) \ra_{\bv{R}} &= \nabla^2 m(\bv{r}) \\
\la \nabla^2_{\bv{R}+\bv{r}} M(\bv{R},\bv{R}+\bv{r}) \ra_{\bv{R}} &=\nabla^2 m(\bv{r}) \\
\la \text{noise} \ra_{\bv{R}} &\approx 0
\end{align}
so
\begin{equation}
\delta m(\bv{r}) =  \sigma^2 \frac{\tau_s }{\tau_m} \nabla^2 m(\bv{r}),
\label{eqn:m}
\end{equation}
with continuous time approximation
\begin{equation}
 \dot{m}(\bv{r}) =  \frac{\sigma^2  }{\tau_m} \nabla^2 m(\bv{r}).
\label{eqn:mt}
\end{equation}
Let us suppose that the interaction range is sufficiently small so that locally pure spatial zones can form, and that the system starts from a spatially uncorrelated state. The typical size of pure zones will slowly grow, and a finite system will eventually consists of one single pure zone. However, we note that there will typically be large fluctuations in the sizes and patterns of zones before this occurs. Within each pure zone the matching probability is one, so when such zones exist $m(\bv{0})$ will be close to one. We note that until fixation it will not equal one, because cells on boundaries of pure zones will be in a mixed state. We obtain separation dependent matching probabilities from (\ref{eqn:m}) using a quasi-static approximation, originally devised to calculate correlations in the voter model \cite{kra10}. We let $\epsilon$ be a radius beyond which $m$ is sufficiently slowly varying so that our discrete equation (\ref{eqn:mt}) may be approximated by a continuous diffusion equation. We then note that  the continuous version of (\ref{eqn:mt}) describes the density of diffusing particles at a distance from a \textit{source} region $\mathcal{S}(\epsilon)=\{\bv{r} \text{ s.t. } |\bv{r}|<\epsilon\}$, which appears once locally pure zones have formed, and is held at constant density equal to the value of $m(\bv{r})$ when $|\bv{r}| \approx \epsilon$. The density of particles larger distances is initially $1/2$ (the matching probability between distant sites). Over time, particles from the source region will diffuse outwards, increasing the density away from the the origin, corresponding to increased matching probabilities for larger $\bv{r}$ values. At a time $t$ after the initial formation of the source region, these extra particles have little effect on $m(\bv{r})$ beyond the diffusion distance $\sigma \sqrt{t/\tau_m}$, which divides the region around the source into a \textit{near} and a \textit{far} zone.  If we assume that within the near zone the particle distribution has reached equilibrium then the solution within this zone is
\begin{equation}
m(\bv{r}) = b + c \ln \left( \frac{|\bv{r}|}{\epsilon}\right)
\label{eqn:msol}
\end{equation} 
where $b$ is the matching probability at short range, and $c$ is fixed by the value of $m(\bv{r})$ at the edge of the near zone. Figure \ref{fig:m_Eng} shows least squares fits of this function to simulated matching probabilities averaged over all points in England, using an effective interaction range $\sigma_{\text{eff}}=70m$, from which we see that (\ref{eqn:msol}) provides a good approximation to the empirical matching probabilities in the English dialect domain. 
By fitting a large number of such curves we are able to find a relationship, using kernel regression, between $b$ and $c$, giving a one parameter family of correlation functions for the cases $N_{\text{sim}} \in \{5,10,50\}$ (Figure \ref{fig:bc}). As $T \rar \infty$, the system moves toward fixation, corresponding to $b \rar 1$ and $c \rar 0^{-}$. At earlier times, larger values of $|c|$ mean faster decaying matching probabilities, more spatial variations and higher spatial autocorrelation (see section \ref{sec:auto}). We will see in section \ref{sec:inf} that the lowest cell population $N_{\text{sim}}=5$ is best able to generate distributions consistent with the SED, so for the remainder of the paper we work with the family of matching curves generated for this case. The corresponding $(b,c)$ values are listed in Table \ref{tab:param}, and the curves plotted in Figure \ref{fig:mat}. 

It may be of interest to linguists to note that our matching function is closely related to S\'{e}guy's curve \cite{seg71,seg73}, which gives the relationship between geographical and linguistic distance. The linguistic distance between two locations may be defined as the number of variables that differ between them \cite{ner03}. If all variables evolve according to the same dynamics then this is just $1-M(\bv{r}_1,\bv{r}_2)$. There is some debate as to what family of functions S\'{e}guy's curve belongs \cite{ner03}, and this question has recently been addressed from the point of view of Statistical Physics \cite{bur17}. In his original work S\'{e}guy fitted families of curves based on logarithmic increase, generating ``fat tailed'' distance relationships more consistent with neutral evolution than conformity driven dynamics.

\section{Inference}

\label{sec:inf}

\subsection{The survey of English Dialects}

We wish to use data from the Survey of English Dialects (SED) to infer the most likey class of activation function. The SED contains 310 survey locations within the British mainland, excluding the Isle of Wight and the Isle Mann (see Figure \ref{fig:grid}). The geographical distribution of language features recorded in the SED are the result of local copying processes with biases which depend on the linguistic feature in question, and on social factors. We view language change as a branching process which generates a single new variant at a time. In order to become established some mechanism must exist which, at least temporarily, biases speakers in favour of new variants. Such mechanisms might be socially conditioned (used to signal social group or generation) or linguistic (the new variant is innately more preferable). However, the existence of long lived stable interfaces between variants suggests that such biases are in many cases weak, short lived or contextual. Geographical distributions will also be influenced by migration events and political changes. The focus of our work is only to understand the fundamental class of the copying process, assuming that variants are all approximately equivalent. We wish to know if, in the absence of differences in intrinsic ``fitness'', variants survive with a probability equal to their current frequency, or whether it is more probable that speakers preferentially select variants which are already more common.

\begin{figure}
	\centering
	\includegraphics[width=\linewidth]{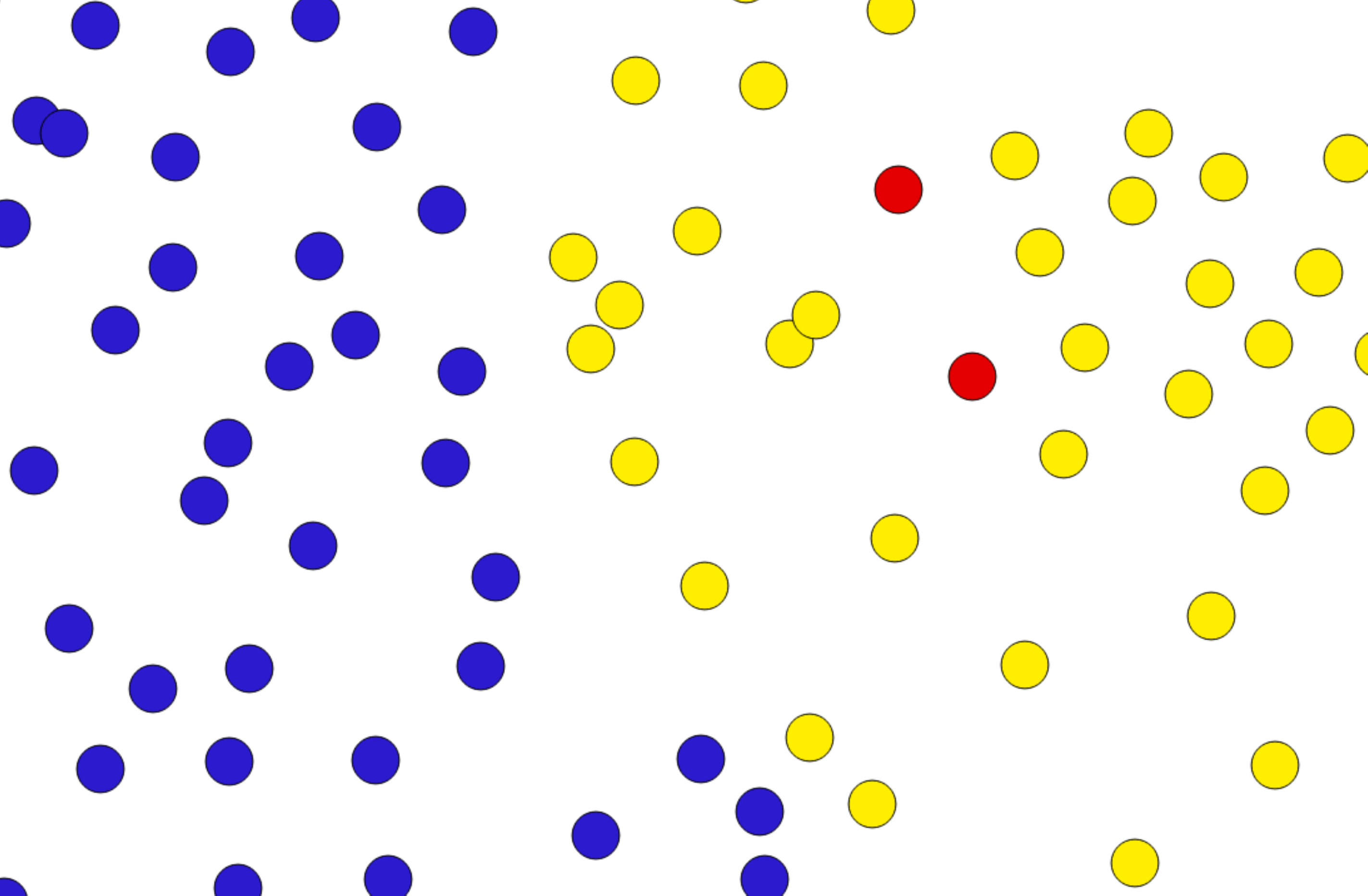}
	\caption{ A third variant embedded within a clearly defined domain.  }
	\label{fig:merge1}
\end{figure}

\begin{figure}
	\centering
	\includegraphics[width=\linewidth]{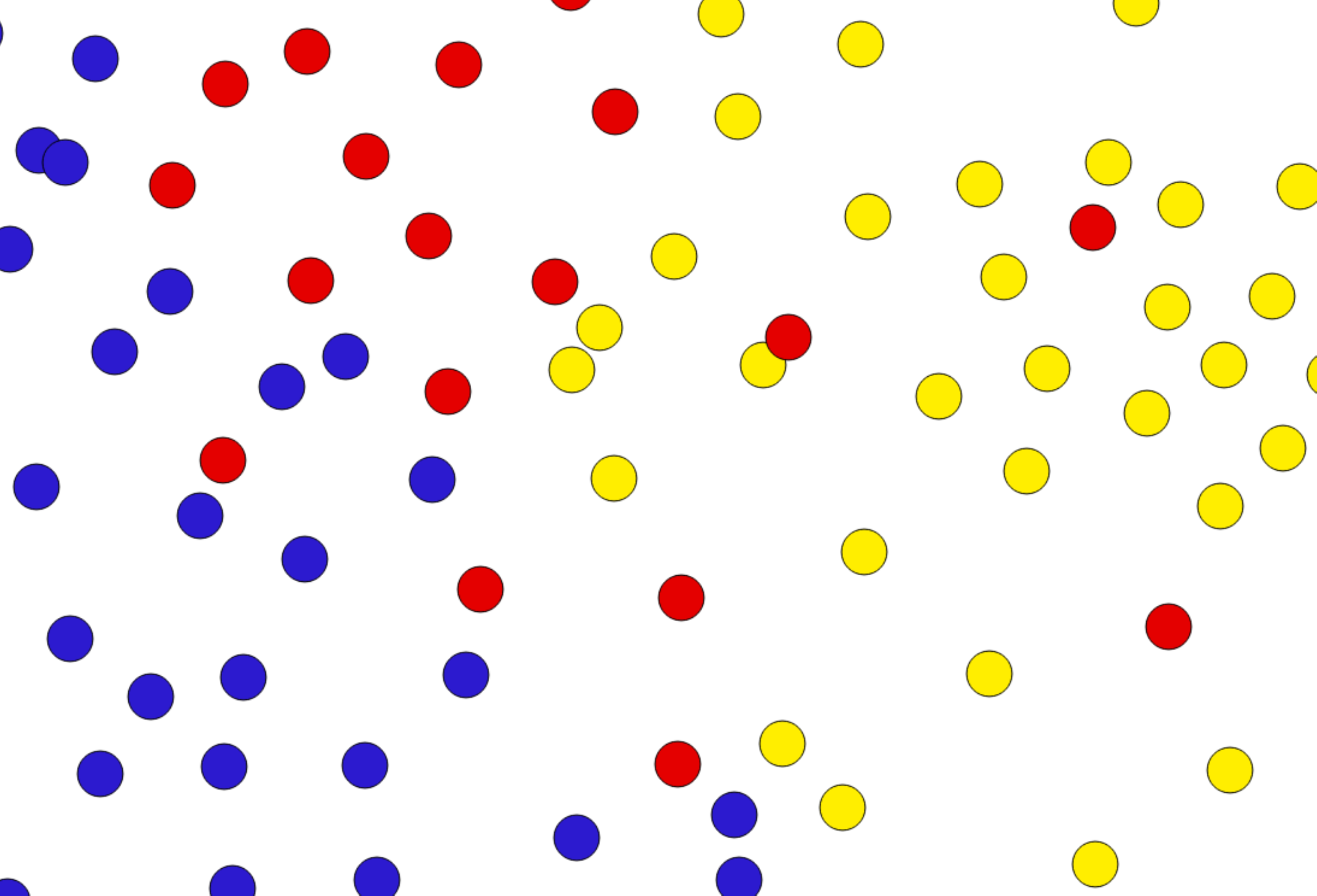}
	\caption{ A third variant not well-embedded within a clearly defined domain.  }
	\label{fig:merge2}
\end{figure}

We consider variables for which it is possible to identify a single branch in their evolution. In some cases this involves reducing multi-variant maps to the bi-variant case by merging or excluding variants from the system. Where one variant transparently reflects an additional change modifying the output of an earlier change, we can merge these so that the dataset reflects just the distribution of the earlier change. Where the relationships among variants is non-transparent because a later change obscures an earlier one, or where a variant is formally unrelated to the others, there is no justification to merge it with any other: even if we know the chronology of innovations, we cannot reconstruct with any surety which variant would have occurred at localities with the most recent innovation. With such datasets our only option is to exclude the problematic variant altogether. If the variant occurs only within an otherwise well-defined domain (Figure \ref{fig:merge1}), 
then we can exclude it and impute the missing data without a problem for two reasons: firstly, we are modelling the changes in distribution of the other two variants and the boundary between the domains in which these are dominant remains clear; secondly, a distribution of this type suggests that the third variant is a later innovation within one of the existing regions, even if we do not have specific historical evidence to support this. If, on the other hand, a problem variant is not embedded within a clearly defined domain but occurs along another domain boundary (an ``isogloss'') (Figure \ref{fig:merge2}),
there is no way to treat the variable as binary since we cannot reconstruct the distribution of the two variants we are interested in at a crucial point where they interact, and we lack evidence for the chronology of innovations. Variables falling into this class are excluded from our study. Figures \ref{fig:high_moran} and \ref{fig:low_moran} show binary distribution maps for 40 of the 68 variables which we consider. A detailed description of the linguistic variables used in our analysis and, where relevant, their reduction to binary form, is given in appendix \ref{ap:SED}.

\subsection{Spatial Autocorrelation}

\label{sec:auto}

\begin{figure}
	\centering
	\includegraphics[width=\linewidth]{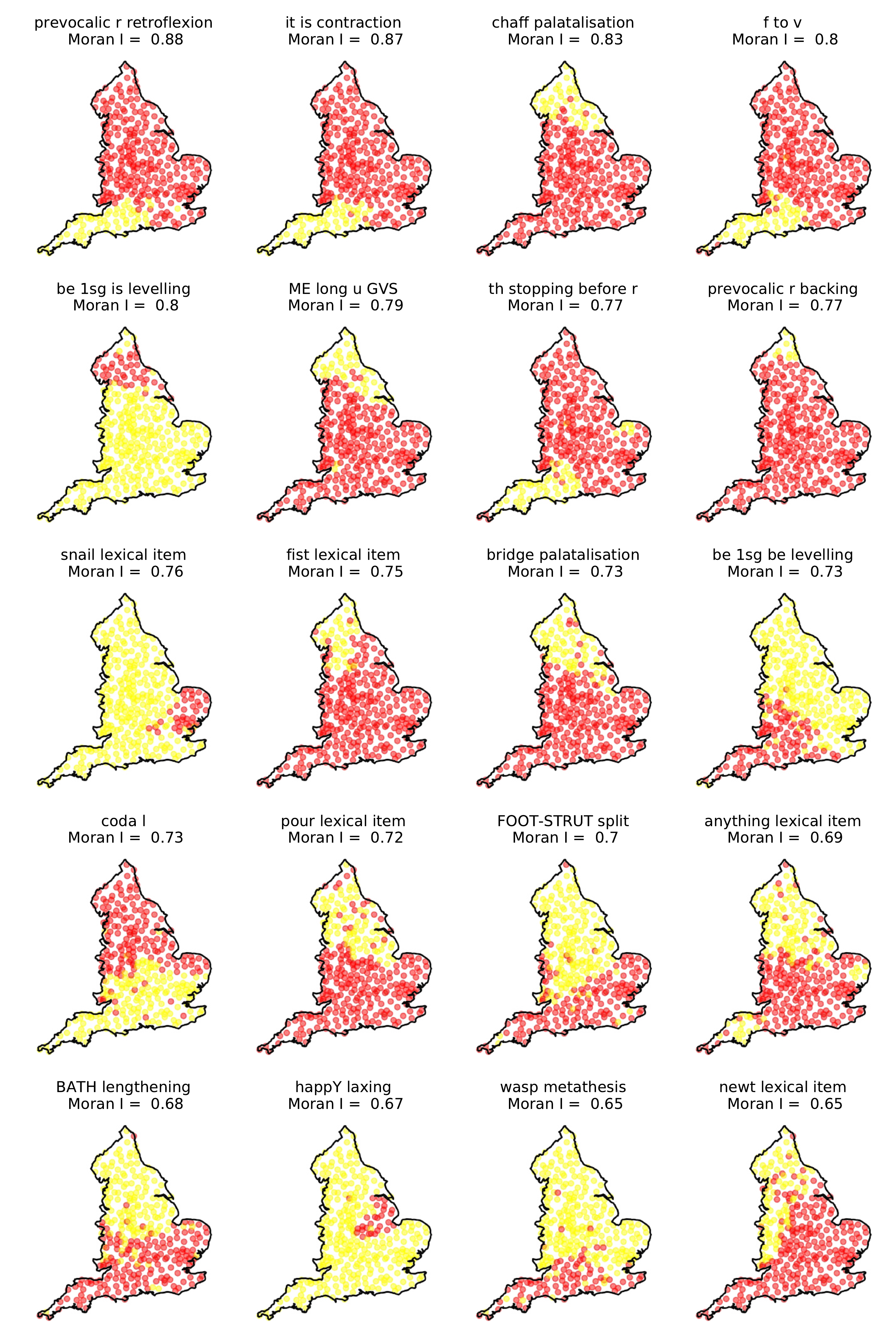}
	\caption{ SED maps of the twenty variables with highest Moran $I$. The title of each map gives the formal linguistic description of the variable, and the Moran $I$ value using size nearest neighbours. }
	\label{fig:high_moran}
\end{figure}

\begin{figure}
	\centering
	\includegraphics[width=\linewidth]{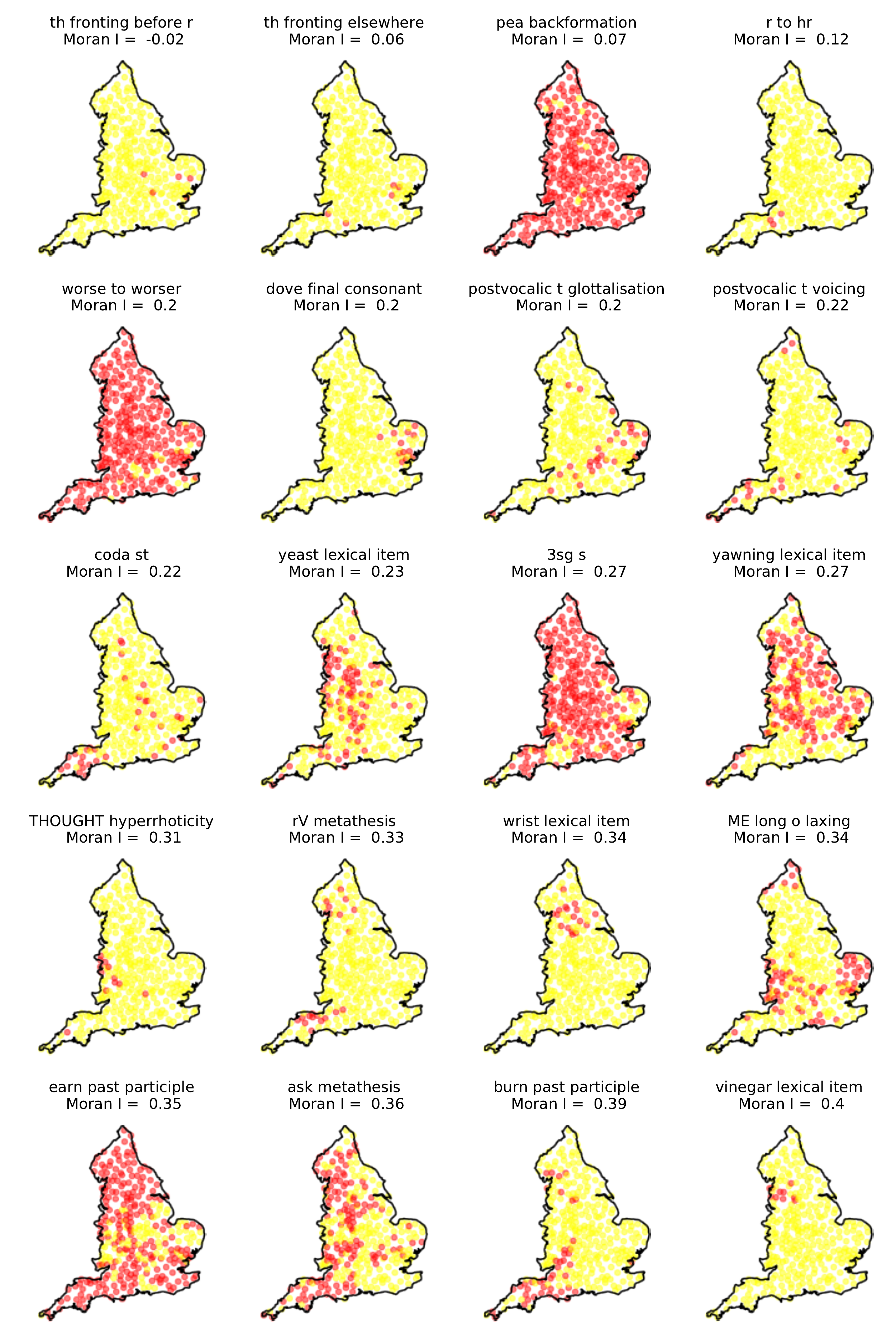}
	\caption{ SED maps of the twenty variables with lowest Moran $I$.  The titles give the formal linguistic description of each variable, and the Moran $I$ value calculated using six nearest neighbours. }
	\label{fig:low_moran}
\end{figure}

For each survey map in our dataset, we wish to infer which of our two copying processes is more likely to have generated it. A visual inspection of the set of maps (Figures \ref{fig:high_moran} and \ref{fig:low_moran}) reveals that while some variables exhibit well defined spatial interfaces, other maps are more spatially ``disordered''. Broadly speaking we expect conformity driven evolution to yield well defined domains with smooth boundaries, and neutral evolution to generate a more complex pattern of spatial boundaries and disorder, if interaction range is short. For neutral evolution with longer interaction ranges we would expect limited spatial order in survey results unless system wide fixation has occurred. Because different parts of the linguistic system evolve by different processes and at different rates, and also carry different social messages, then we do not expect every map to be the result of the same underlying process. We therefore test each language feature individually.  

We have characterised the differences in the spatial distributions generated by the different activation functions using spatial matching probabilities. However, these cannot be used to draw inference from individual maps, because the function $m(\bv{r})$ calculated by averaging over the locations in a single map will strongly depend on the particular distribution realised by that variable. No inference can be drawn about whether matching probabilities for that variable would exhibit exponential or logarithmic decay if we were to re-run its history many times and average $m(\bv{r})$ over the results. To resolve this problem, in section \ref{sect:MFR} we use matching functions to generate approximate multivariate probability distributions over the set of all possible maps, allowing likelihood based inference. However, we first take a much simpler approach based on the extent to which nearby locations are in the same state, known as \textit{spatial autocorrelation}. A simple measure of this, previously used to study regional linguistic variation \cite{gri11}, is Moran's $I$ \cite{mor48}, which, for $N$ locations, is defined
\begin{equation}
I \triangleq \frac{N}{G} \frac{\sum_{ij} G_{ij} (x_i-\bar{x})(x_j - \bar{x})}{\sum_i (x_i - \bar{x})^2}
\end{equation}  
where $x_i$ is the state of the $i$th location and $G_{ij}$ a spatial weight associated with the pair $(i,j)$. The number $G$ is the sum of all spatial weights. In our case the state at location $i$ is the emitted state of the speaker selected for the language survey. Rather than represent this state using $\{\bv{e}_1, \bv{e}_2\}$, we use $x_i \in \{-1,1\}$, so, if the survey contains an equal number of speakers using each variant then $\bar{x}=0$. The spatial weights $G_{ij}$ define what we mean by \textit{nearby}. We take $G_{ij}=1$ if survey location $j$ is one of the six nearest neighbours of $i$ and $G_{ij}=0$ otherwise. The mean and standard deviation of the separation of locations with non-zero weights is then $22.5 \pm 8.4$km. Moran's $I \in [-1,1]$ then measures the extent to which survey locations within this range match their state. Intuitively, if we have large single-variant domains with well defined, smooth interfaces then we expect high $I$ value, because the regions of the system where miss-matches occur are one dimensional and maximally short (due to surface tension), and therefore occupy a small fraction of the total area. If, on the other hand, single variant domains have a complex boundary structure, or variants are otherwise widely dispersed, then we expect a low $I$ value. This intuition is borne out by the results in Figures \ref{fig:high_moran} and \ref{fig:low_moran} which show the variables with, respectively, the highest and lowest $I$ values. 

\begin{figure}
	\centering
	\includegraphics[width=\linewidth]{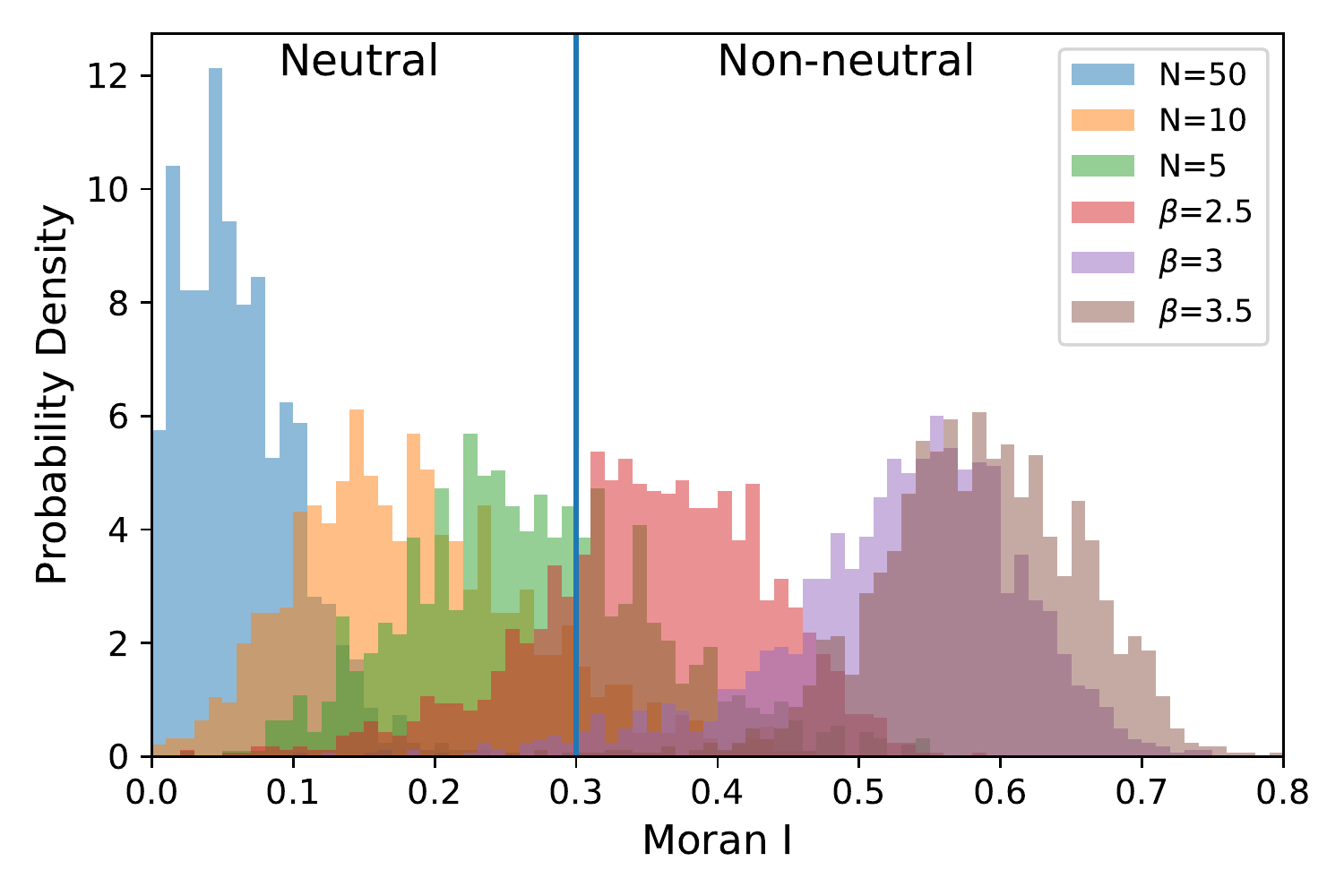}
	\caption{ Distributions of $I$ values (six nearest neighbours) from direct simulation on England using $q=2, \tau_m=2, \tau_s=1$. In the neutral case we have $\sigma_{\text{sim}}=3.16$km, and $N_{\text{sim}} \in \{5,10,50\}$ giving effective interaction ranges $\sigma_{\text{eff}} \in \{70\text{m},100\text{m},220\text{m}\}$ when the true cell population is $N = 10^4$. In the conformity driven model we use $\sigma_{\text{sim}} = 5$km, $N=10^4$ and $\beta \in \{2.5,3,3.5\}$. Vertical line indicates where neutral/non-neutral model cases are more common.  }
	\label{fig:sim_moran}
\end{figure}

To understand the relationship between our evolution models and Moran's $I$ we directly simulate the English dialect domain starting from randomized initial conditions, and extract the emitted states of speakers at each SED survey location at a fixed sequence of time intervals. Using this data we then compute the $I$ value of each sample using the same weights as the survey data, allowing direct comparison to the $I$ values obtained from the SED. The distributions of simulated $I$ values are shown in Figure \ref{fig:sim_moran}. 
In the neutral case we vary the effective interaction range by changing the simulated population per cell, giving  $\sigma_{\text{eff}} \in \{70\text{m},100\text{m},220\text{m}\}$. In the longer range case, spatial variations in external state are small, and most spatial variation is generated by the sampling process; we obtain $I$ values which are close to zero. For the shorter effective interaction ranges, where linguistic subdomains are able to form, we obtain higher $I$ values. Using the non-neutral activation function (\ref{eqn:potts_g}) we have two free parameters: the interaction range and the conformity number $\beta$. Whereas $\beta$ determines the amount of noise in the bulk \cite{dro99}, both $\sigma$ and $\beta$ together determine the width of interfaces (equation (\ref{eqn:omega}). We set the simulated range to $\sigma_{\text{sim}} = 5$km (note: villages recorded in the Domesday book are typically $\approx 2$km separated from their closest neighbour, with remarkable consistency between shires \cite{ber79}). We then simulate the model for $\beta \in \{2.5,3,5\}$, noting that as $\beta \rar 2$, the system approaches a completely disordered state where variants exist in equal proportions in all locations.  

\begin{figure}
	\centering
	\includegraphics[width=\linewidth]{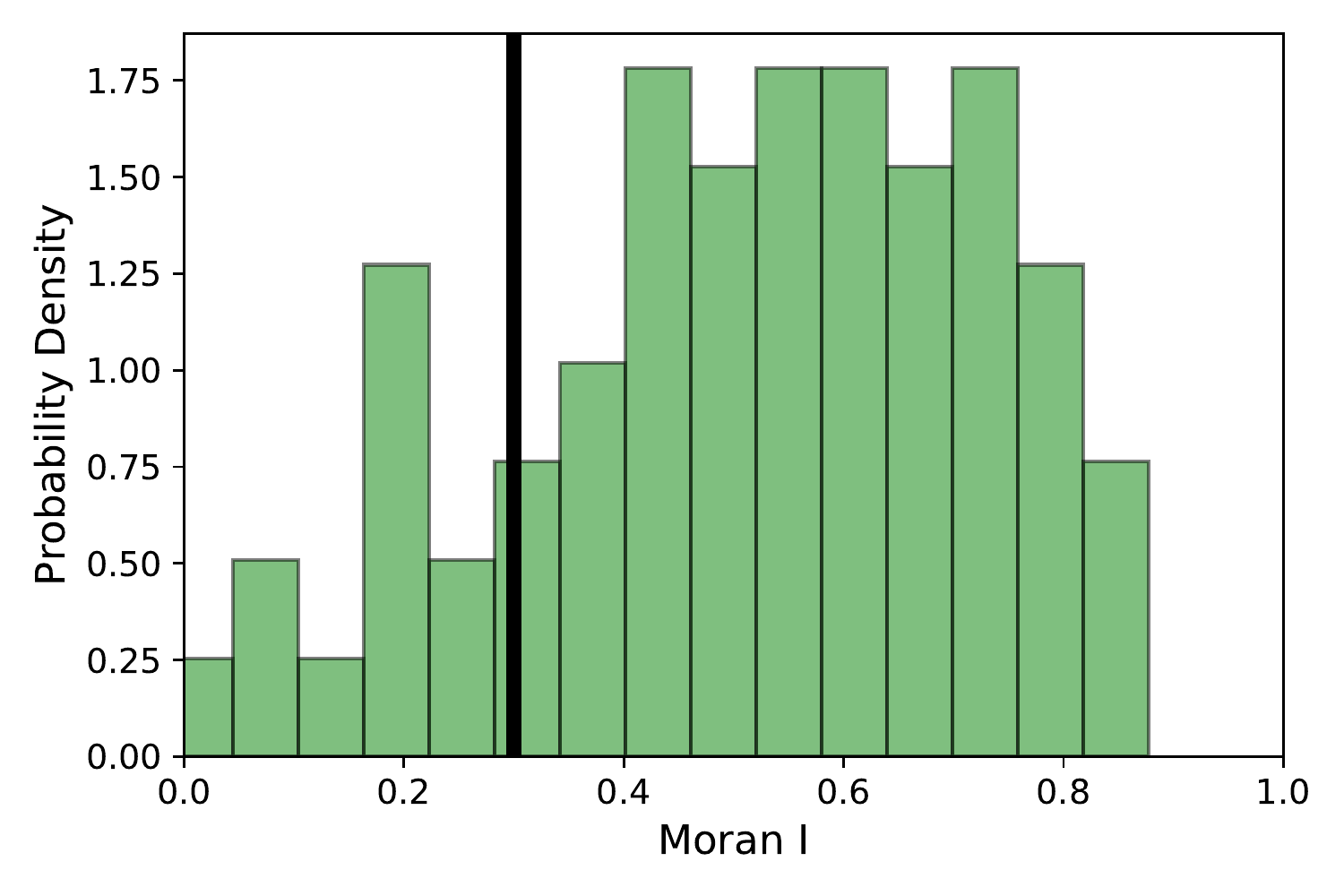}
	\caption{ Distributions of $I$ values (six nearest neighbours) from the SED data.   Vertical line indices where neutral/non-neutral model cases are more common in direction simulations. 82\% of values have $I>0.3$}
	\label{fig:SED_moran}
\end{figure}

Since all of our survey maps contain regions which are linguistically pure, we assume that values of $\beta$ near the disorder transition are not a realistic model of linguistic behaviour. The $I$-distributions obtained from our three $\beta$ values are shown in Figure \ref{fig:sim_moran} and we see that they are much higher than the neutral values. For  $I > 0.3$, the majority of samples are non-neutral. Figure \ref{fig:SED_moran} shows the distribution of $I$ values computed from the SED. Here we see that the majority ($83\%$) of maps have $I$ values which are more likely to have been generated by the three non-neutral models we have tested. From here on we compare conformity driven evolution to the neutral model with lowest cell population $N_{\text{sim}}=5$ on the basis that this model is most likely to be able to match realistic spatial distributions.

Moran's $I$ is an simple and intuitive means to distinguish between different kinds of spatial distribution, and the above analysis suggests that although the neutral model is capable of generating linguistic variant distributions which exhibit the kinds of spatial ordering seen in the SED, the majority of maps are more consistent with conformity driven evolution.   However, Moran's $I$ depends only on matching probabilities at close range when we know that in fact the differences between neutral and non-neutral evolution are manifested in the full $r$-dependence of the matching probability function (Figure \ref{fig:mat}). As an example of why this means that $I$-based inference may  problematic, we note that even if $\beta>2$, so that interfaces exist, it is possible to create any desired level of short range spatial disorder by tuning $\beta$ sufficiently close to $\beta_c=2$. As noted above, such distributions are not attested in the data, but nevertheless have $I$ values typical of the neutral model. We now consider an inference method which removes this issue by accounting for the full $r$ dependence of matching probabilities.

\subsection{Markov Graphical Models}

\label{sect:MFR}

In order to infer what model, and what parameters, are most likely to have generated a given set of data, we require a \textit{statistical model} of that data \cite{was03}. A minimum requirement is that such a model provides a method for generating realisations of the data, given the values of its parameters. Ideally the model will also provide the full probability distribution of the output of a single trial. 
In our case, a single trial corresponds to one SED survey map, and our simulations satisfy the minimum requirement of a statistical model of such maps. However, because the number of possible outcomes of each trial is a high dimensional random vector, it is impossible to attach any form of likelihood to a given map or set of maps - the sample space is too large.  Moreover, as discussed in section \ref{sect:non_neut}, when considering the possible arrangements of non-neutral interfaces, the natural parameter which describes matching probabilities is the density of interfaces in the system (or, equivalently, the average domain size), which cannot be directly controlled in simulations, and may depend on factors exogenous to our simple dynamics.

An alternative statistical model which incorporates the full $r$-dependence of matching probabilities, and gives the probability of any possible map, is the \textit{Markov Random Field} or \textit{Markov Graphical Model} \cite{was03,kol09}. This class of model began with the Ising model, and has since been used in a wide range of fields including computer vision,  spatial data analysis \cite{bes75} and machine learning \cite{kol09}. We have sufficient information about matching probabilities to calibrate a \textit{pairwise model}, in which the probability of map $\bv{x}=(x_1, x_2, \ldots)^T$, where $x_i \in \{-1,1\}$, is
\begin{equation}
P(\bv{x}) = \frac{\exp\left(\frac{1}{2} \bv{x}^T \theta \bv{x}\right)}{\mathcal{Z}(\theta)}
\label{eqn:Px}
\end{equation}    
where $\theta$ is a symmetric matrix of interaction strengths between all possible pairs of sites, and the normalizing constant $\mathcal{Z}$ is the \textit{partition function}. 
The probabilities assigned for different configurations $\bv{x}$ do not depend on the diagonal elements of $\theta$ because $x_i x_i=1$ for all $i$, so these elements contribute a multiplicative constant to the numerator of (\ref{eqn:Px}), which affects the value of the partition function. By convention we set $\theta_{ii}=0$ for all $i$. Samples from (\ref{eqn:Px}) may be obtained via Gibbs sampling \cite{mac03}. Starting from a randomized initial state we propose changes by selecting a single site, $i$, and setting $x_i=1$ with probability \cite{glau63}
\begin{align}
p_i = \frac{1}{2} \left( 1 +  \tanh \sum_j \theta_{ij} x_j\right),
\end{align}  
and $x_i=-1$ with probability $1-p_i$. A distribution $\PP$ is the equilibrium of this update rule if the detailed balance condition is satisfied
\begin{equation}
\PP(x_i=1 \cap \bv{x}_{\setminus i})(1-p_i) = \PP(x_i=-1 \cap \bv{x}_{\setminus i})p_i 
\label{eqn:db}
\end{equation}
where $\bv{x}_{\setminus i}$ denotes the states of all sites excluding $i$. That condition (\ref{eqn:db}) is satisfied by (\ref{eqn:Px}) may be seen by noting that
\begin{equation}
\frac{P(x_i=1 \cap \bv{x}_{\setminus i})}{P(x_i=-1 \cap \bv{x}_{\setminus i})} = e^{2 \sum_j \theta_{ij} x_j} = \frac{p_i}{1-p_i}.
\end{equation}
To determine the interaction matrix we use our exponential and logarithmic matching probabilities  (\ref{eqn:expmat}) and (\ref{eqn:msol}) (with parameters given in Table \ref{tab:param}) to calculate the matching probability between every pair of nodes in the SED, based on their separations. In this way, for each matching curve we obtain a matching probability matrix $M_{ij}$. The equivalent matrix for our statistical model $P(\bv{x})$ is given by
\begin{equation}
\hat{M}_{ij}(\theta) = \EE\left[\frac{1+ x_i x_j }{2}\right]
\end{equation}
which may be estimated by Gibbs sampling. We calibrate our model to the desired matrix by iteratively adjusting the interaction parameters using the descent rule \cite{mac03}
\begin{equation}
\theta_{n+1} = \eta(M-\hat{M}(\theta_n))
\label{eqn:des}
\end{equation}
where $\eta>0$ is a \textit{learning rate}. That is, interaction strengths are incrementally increased or decreased to shift the model matching probabilities toward their targets. The practical (vectorized Python) implementation of the method involves storing many independent realizations of the system (the random vector $\bv{x}$) where each element of each vector is initialized to $\pm 1$ with equal probability, corresponding to $\theta_{ij}=0$ for all $i,j$. After each iteration of (\ref{eqn:des}), each vector is updated (using Gibbs sampling) a sufficient number of times so that the set of vectors $\{\bv{x}\}$ represent a sample from the current model, $\theta_n$. The matching probabilities for this model are then estimated, and used to calculate $\theta_{n+1}$. We note an important difference between our calibrated model, which can in principle have interactions at all separations, and short range Ising-type models. The sub-critical nearest neighbour Ising model, updated using Gibbs sampling, generates domains which grow larger over time leading to a progressively lower interface density. In contrast our model is calibrated so that, at least in the non-neutral case, the interface density stabilizes at a given target value ($\lambda^{-1}$). Examination of calibrated interaction matrices reveals negative interactions at ranges beyond the typical domain size, which limit the expansion of domains once the desired interface density has been achieved.

\begin{figure}
	\centering
	\includegraphics[width=\linewidth]{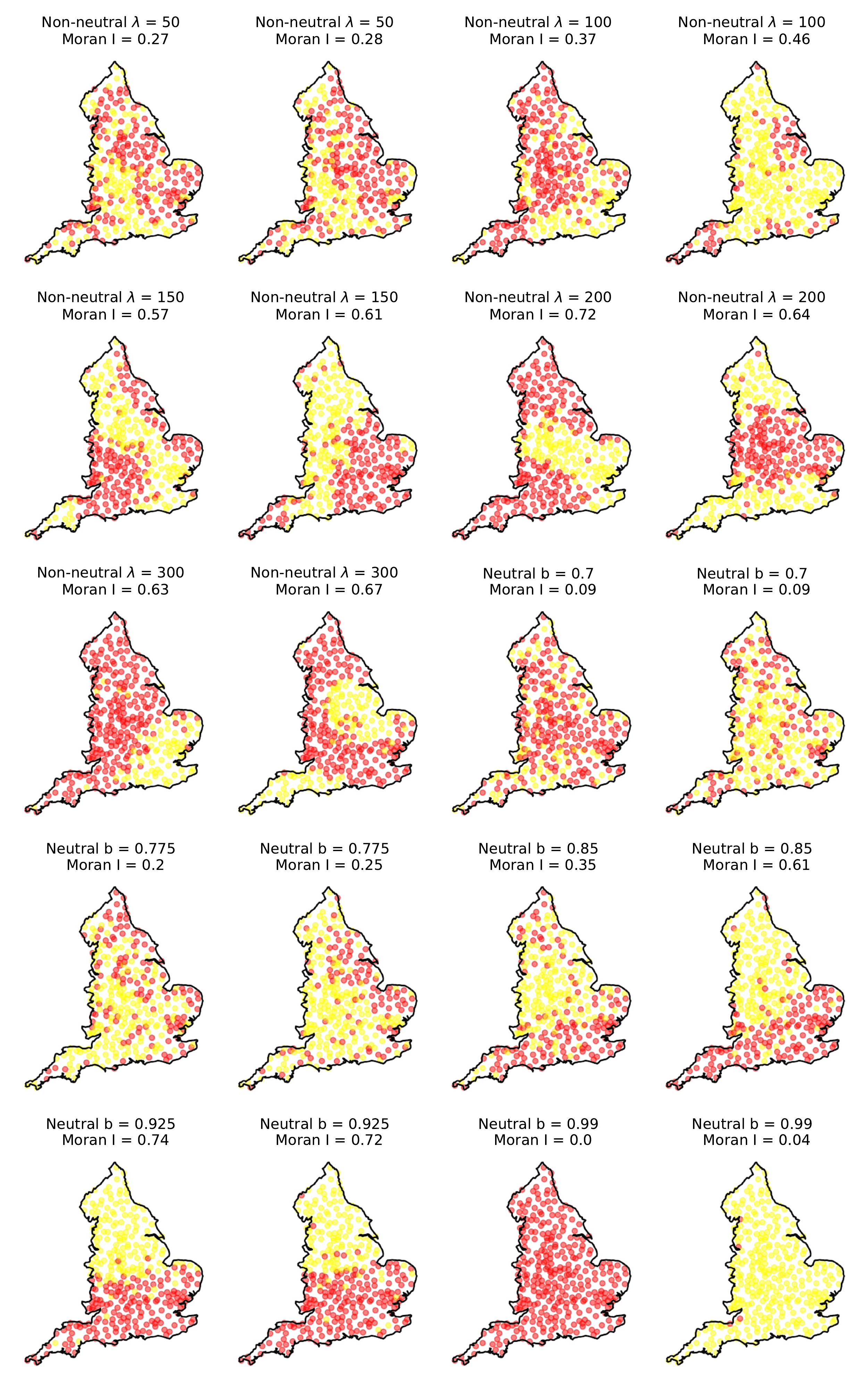}
	\caption{ Samples from Markov Graphical Models calibrated neutral and non-neutral matching curves  in Figure \ref{fig:mat}. }
	\label{fig:MRF_samps}
\end{figure}

\begin{figure}
	\centering
	\includegraphics[width=\linewidth]{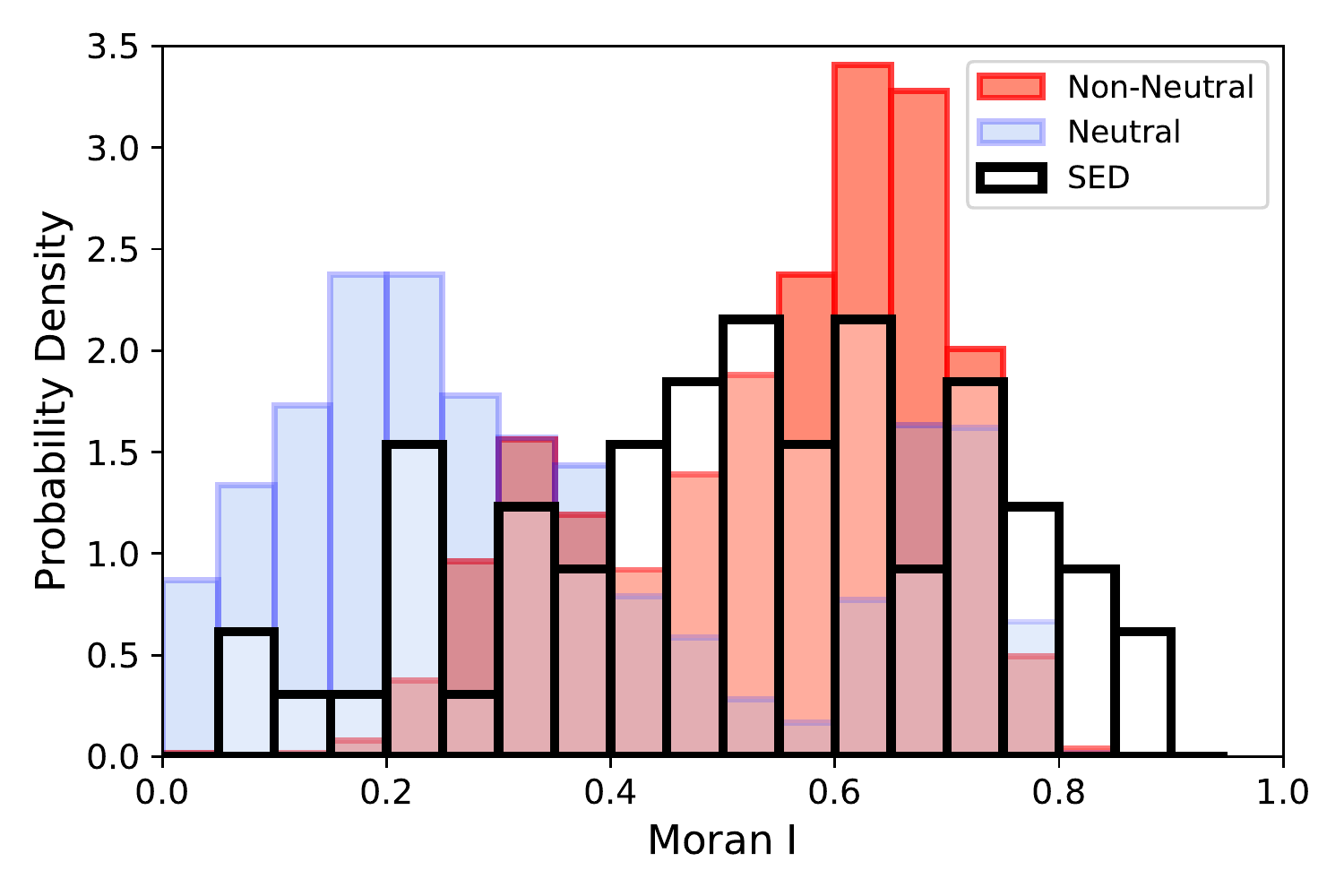}
	\caption{ Distributions of $I$ values (six nearest neighbours) from the SED data, and from Markov Random Field models calibrated to neutral and non-neutral matching functions described by the parameters in Table \ref{tab:param} and plotted in Figure \ref{fig:mat}.  }
	\label{fig:MRF_moran}
\end{figure}

Having calibrated the interaction strengths in this way, we can use our model to generate sample maps consistent with the target matching curves. A set of such maps are shown in Figure \ref{fig:MRF_samps}.
From these examples we see that the calibrated non-neutral model, as expected, generates well defined domains with smooth interfaces. As the density of interfaces declines, Moran's $I$ increases. In the neutral case, although spatial domains appear, they are less well defined in early stage evolution, which is consistent with simulations of the external state shown in Figure \ref{fig:neut_N_5}, and produces Moran $I$ values which are typically lower than those obtained from interface-driven dynamics. By generating a much larger  sample of maps from the calibrated neutral and non-neutral models, we can estimate the empirical distribution of neutral and non-neural $I$ values, as shown in Figure \ref{fig:MRF_moran}, along results for the SED.  
From this we see that the SED and the conformity driven model generate a similar range and distribution of $I$ values, with the neutral model tending to produce maps with lower values. We note that the distribution of neutral $I$ values has a secondary peak around $I \approx 0.7$, which is not reproduced by direct simulation of the model (Figure \ref{fig:sim_moran}). This peak is produced by neutral matching curve 4 in Table \ref{tab:param}, and highlights that fact that our statistical model is only an approximation to the true spatial distributions, for which a closed form probability distribution does not exist.  

Beyond sampling individual maps we can also calculate the log probability of map $\bv{x}$, given the interactions $\theta$ calibrated to a matching curve
\begin{equation}
\ln P(\bv{x}) = \frac{1}{2} \bv{x}^T \theta \bv{x} - \mathcal{Z}(\theta).
\end{equation} 
To evaluate this expression we require an estimate for the partition function $\mathcal{Z}(\theta)$, which cannot be computed exactly due to the intractable sum over all possible states. We adopt the annealed importance sampling method, developed by Neal \cite{nea01} (see appendix \ref{ap:anneal}). For every map we can then estimate its likelihood for every matching curve to which we have calibrated $\theta$. Of the 68 maps in our dataset we find 14 for which a neutral model matching curve has the highest likelihood. Therefore $ \approx 80 \%$ of maps are more likely to be non-neutral, consistent with the $82\%$ result obtained using Moran's $I$. The mean domain size in non-neutral maps is $\bar{\lambda}=182$km with standard deviation 72km.

\begin{figure}
	\centering
	\includegraphics[width=\linewidth]{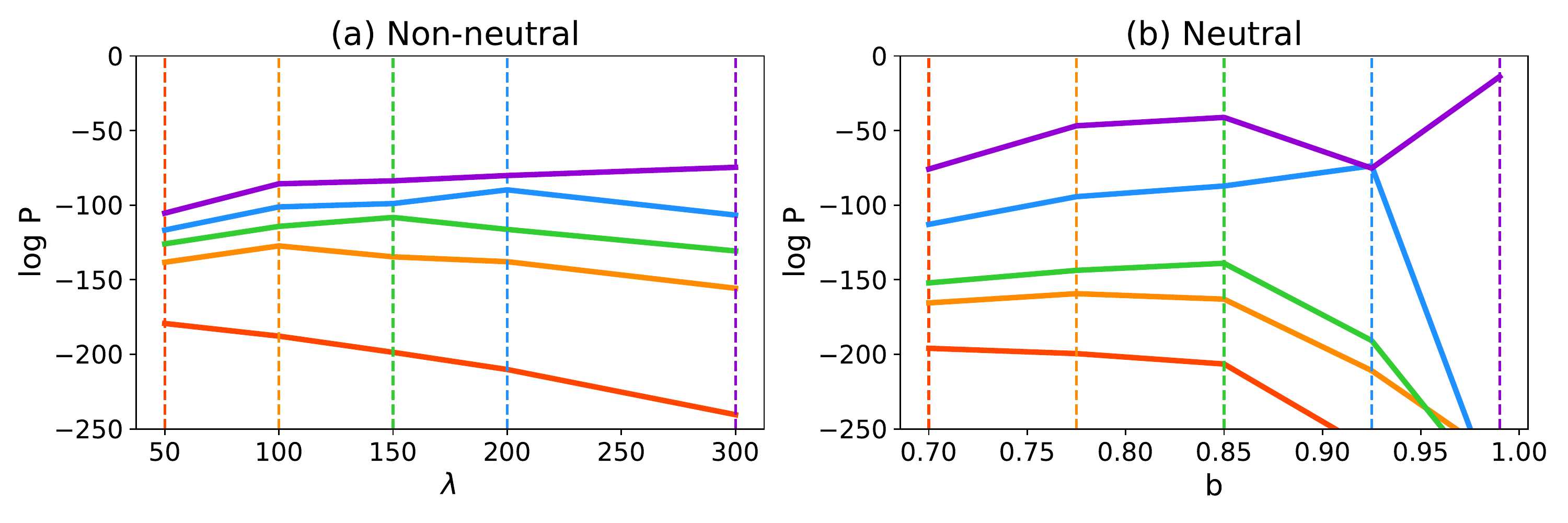}
	\caption{(a) Each curve shows the average log probability (proportional to log likelihood) of 100 samples from a single non-neutral model, according to the five possible non-neutral models (see Table \ref{tab:param} for parameter values). Vertical dashed lines show $\lambda$ values for each model, with curves/ dashed lines of matching colour corresponding to the same $\lambda$ value. The maximum value of each curve occurs at the $\lambda$ value used to generate its samples, as expected. Plot (b) as for plot (a), but using the neutral models with parameter $b$ used to specify model (see Table \ref{tab:param} for $(b,c)$ values).}
	\label{fig:lnPs}
\end{figure}

We test our methodology by generating 100 samples for each calibrated model, and verifying that the  average log probability of these samples is maximised for the model that generated them. The results are displayed in Figure  \ref{fig:lnPs}.
In Figure \ref{fig:lnPs} (a), typical log probabilities increase with $\lambda$ because there are many more ways to cover a map with small domains than there are to cover it with large ones. Likewise, neutral probabilities increase with $b$, which tends to one as fixation is approached and domains grow larger in size. With reference to Figure \ref{fig:lnPs} (b), we note that the neutral log probabilities span a larger range of values than non-neutral values for $\lambda>50$km. This reflects the fact that the neutral model displays a broader range of behaviour. The same model can generate maps with or without domains, and a single realization of the dynamical model can generate a wide variety of different spatial patterns before it reaches fixation. 

\section{Discussion}

Languages are complex structures which exhibit a very wide range of change processes. These processes have been catalogued and studied by linguists for centuries, with the volume and intensity of research rapidly rising in the late twentieth and early twenty first century \cite{lab01,cha98,cam13}. Research into language evolution has become increasingly quantitative \cite{gri11,lab01} and interdisciplinary \cite{bax06,bur17,lor11}, with models often inspired by statistical physics. The analogy between linguistic and genetic evolution is long standing \cite{dar79, bly07}. Linguistic variants are passed from generation to generation like alleles of genes, and the simplest models of genetic evolution are neutral in the sense that every copy of every gene has an equal chance of surviving into the next generation. However, the explanations put forward by linguists for empirically observed changes are rarely consistent with a neutral hypothesis. Nevertheless, mathematical models of language change which assume neutrality have been remarkably successful in describing aspects of language change \cite{bax06,bly07,bly12_2,kau17}. Neutral stochastic processes display a rich variety of behaviour, and it is perhaps for this reason that they are difficult to rule out as evolutionary models of language. 

In this paper we have made an attempt to test the neutral hypothesis against a simple alternative: that speakers exhibit a form of conformity to the majority when learning or copying linguistic variants. We were motivated by  connections between spatial university classes in two dimensional physical models, and the copying behaviour of humans in two dimensional domains. If speakers do not \textit{physically} diffuse to a great extent, then neutral copying should produce maps of linguistic variants which have similar properties to the spatial distributions generated by the voter model (noise driven interfaces and logarithmic matching functions), whereas  conformity driven copying should produce spatial distributions like those of the Ising model (well defined and relatively smooth interfaces and exponentially decaying matching functions). By viewing language communities as Hopfield networks \cite{hop82, hop84} with speaker behaviour defined by an activation function which is either neutral or conformity driven, we have been able to generate fictitious language surveys for the English dialect domain which fall into these two classes. These maps have then been compared to the Survey of English Dialects \cite{ort62}.

We observed that neutral evolution requires speakers to be geographically very isolated (an interaction range of around 100m), or very small in number, in order to produce the significant spatial variations seen in survey data. This would imply that speakers tend to get their linguistic behaviour only from their close neighbours. An alternative explanation is that language communities evolve neutrally but with an \textit{effective population} which is much smaller than the true population. This might occur, for example, in a social network dominated by a small number of very influential individuals for whom $\sum_i \omega_{ij} \gg 1$. We do not rule out either of these possibilities, so we compare our two hypothesis based purely on the spatial distributions that they generate. We have shown first that neutral evolution tends to generate survey maps with low spatial autocorrelation, as measured by Moran's $I$.  Conformity (interface) driven evolution produces higher $I$ values, consistent with the majority ($82\%$) of survey maps.  Second, we constructed statistical models over the space of possible survey maps, derived from theoretical matching curves (logarithmic and exponential) generated by our two activation functions. A likelihood analysis revealed that the majority of maps ($80\%$) were more likely to have been generated by conformity driven evolution, according to our model. 

Interfaces, or ``isoglosses'', are common in linguistic surveys, particularly where the population are not very mobile. In physics, such curvature-driven interfaces arise naturally from Ising type dynamics, although more complex noise-driven versions appear in the voter model. It is interesting to note that only a small sub-population of Ising type agents can switch voter model dynamics to be curvature-driven  \cite{lip17}. We have shown that, at least in the case of the SED, observed spatial variations are more consistent with linguistic conformity than neutral copying. However, a key process in neutral evolution is the spatial diffusion of variants through the population, while speakers themselves remain relatively static.  If we allow the population to diffuse, then even if the underlying copying process is not neutral, linguistic variants will diffuse with their speakers, leading to spatial linguistic distributions with similar properties to those generated by neutral copying in static populations. Exploration of this effect is left for future work.

\subsubsection*{Code and data availability}

All the data and computer code used to generate the results in this paper are available in the publicly accessible GitHub repository \verb!Hopfield_SED!.

\begin{acknowledgements}
The authors are grateful for the Royal Society APEX Award \verb!APX\R1\180117!  which supported this work.
\end{acknowledgements}

\appendix

\section{Normal approximation to multinomial}

\label{ap:nomu}

We review the normal approximation to the multinomial distribution (see \cite{geo12} for details). 
\begin{theorem}
	\label{thm:mult}
	Let $\bv{Y} \sim \text{multinomial}(N,\bv{v})$, and define the standardised form
	\begin{equation}
	\bv{Y}^\ast = \left( \frac{Y_i - N v_i}{\sqrt{N v_i}} \right)_{1 \leq i \leq n}.
	\end{equation}
	Also define the unit vector $ \bv{u} = (\sqrt{v_1}, \ldots, \sqrt{v_n})^T$. Let $\mathsf{O}_{\bv{v}}$ be an orthogonal matrix ($\mathsf{O}_{\bv{v}}^T = \mathsf{O}_{\bv{v}}^{-1}$) whose last column is $\bv{u}$. If $Z_1, Z_2 , \ldots, Z_{n-1}$ are i.i.d. standard normal variates, then
	\begin{equation}
	\bv{Y}^\ast \rightsquigarrow \mathsf{O}_{\bv{v}} \begin{bmatrix}
	Z_1 \\
	Z_2 \\
	\vdots \\
	Z_{n-1} \\
	0
	\end{bmatrix}
	\end{equation}
	as $N \rightarrow \infty$.
\end{theorem}
A proof of this theorem is given in \cite{geo12}. An intuitive understanding may be obtained by noting that $\bv{Y}^\ast$ lies in the $n-1$ dimensional hyperplane
\begin{equation}
H_{\bv{v}} = \left\{ \bv{x} \in \RR^n : \sum_{i=1}^n \sqrt{v_i} x_i = 0 \right\},
\end{equation}
so $\mathsf{O}_{\bv{v}}^T \bv{Y} $ belongs to the hyperplane
\begin{equation}
H = \left\{ \bv{x} \in \RR^n : x_n=0 \right\}, 
\end{equation}
which contains the vector $\bv{Z} \triangleq (Z_1, \ldots, Z_{n-1},0)^T$. The matrix $\mathsf{O}_{\bv{v}}$ therefore rotates $H$ into $H_{\bv{v}}$ ($\bv{Z}$ to $\bv{Y}^\ast$). To practically use this approximation, we require the matrix $\mathsf{O}_{\bv{v}}$, which may be constructed via the Householder transformation. Let $\bv{w}$ be a real unit vector, then the Householder matrix
\begin{equation}
P = \mathsf{I} -2 \bv{w} \otimes \bv{w}
\end{equation} 
is orthogonal. Defining
\begin{equation}
\bv{w} \triangleq \frac{\bv{u}-\bv{e}_n}{|\bv{u}-\bv{e}_n|}
\end{equation} 
we obtain an orthogonal matrix whose last column is $\bv{u}$. For example, when $n=2$ we have
\begin{equation}
\mathsf{O}_{\bv{v}} = \begin{bmatrix}
-\sqrt{v_2} & \sqrt{v_1} \\
\sqrt{v_1} & \sqrt{v_2}
\end{bmatrix}
\end{equation}
and when $n=3$ we have
\begin{equation}
\mathsf{O}_{\bv{v}} = \begin{bmatrix}
\frac{v_2}{1-\sqrt{v_3}} - \sqrt{v_3} &- \frac{\sqrt{v_1 v_2}}{1-\sqrt{v_3}} & \sqrt{v_1} \\
- \frac{\sqrt{v_1 v_2}}{1-\sqrt{v_3}} & 1 - \frac{v_2}{1-\sqrt{v_3}} & \sqrt{v_2} \\
\sqrt{v_1} & \sqrt{v_2} & \sqrt{v_3} 
\end{bmatrix}.
\end{equation}
A more compact statement of theorem \ref{thm:mult} may be made by defining  $\mathsf{E}_n$ to be the diagonal matrix with a 1 in the first $n-1$ diagonal entries and 0 in the $n$th. Letting $\bv{Z} \sim \mathcal{N}(0,\mathsf{E}_n)$ then
\begin{equation}
\bv{Y} \approx N \bv{v} + \sqrt{N} \bv{v}^{\odot \frac{1}{2}} \odot \mathsf{O}_{\bv{v}} \bv{Z}
\end{equation}
where $\odot$ denotes the Hadamard (element-wise) product and $\odot \tfrac{1}{2}$ the Hadamard square root.

\section{Annealed importance sampling}

\label{ap:anneal}

We estimate partition functions by annealed importance sampling \cite{nea01}. Here we explain how the method is  efficiently applied in our case, adapted from the review \cite{sal08}. We have a pairwise exponential measure 
\begin{equation}
\PP(\bv{s}) = \frac{1}{\mathcal{Z}} \exp \left[  E(\bv{s})\right]
\end{equation}
where
\begin{equation}
E(\bv{x}) = \frac{1}{2} \bv{x}^T \theta \bv{x}.
\end{equation}
We wish to calculate the partition function $\mathcal{Z}$.  We begin with a starting measure $\PP_0(\bv{s})$, the model with zero interactions, for which $\mathcal{Z}_0$ is known
\begin{equation}
\mathcal{Z}_0 = n^L.
\end{equation}
We define a sequence of intermediate measures
\begin{equation}
\PP_k(\bv{x}) = \frac{1}{\mathcal{Z}_k} \exp \left[ E_k(\bv{x})\right]
\end{equation}
where $E_k(\bv{x})=k E(\bv{x})/K$ and $k \in \{0,1, \ldots, K\}$. Let $T_k(\bv{x};\bv{x}')$ be a transition probability which leaves measure $k$ invariant in the sense that
\begin{equation}
\sum_{\bv{x}} \PP_k(\bv{x}) T_k(\bv{x}';\bv{x}) = \PP(\bv{x}').
\end{equation} 
In other words, $\PP_k$ the steady state of the transition matrix $T_k$. We then generate sequences of states
\begin{align}
\bv{x}_1 & \sim \PP_0(\bv{x}) \\
\bv{x}_2 & \sim T_1(\bv{x}_2; \bv{x}_1) \\
\ldots \\
\bv{x}_K & \sim T_{K-1}(\bv{x}_K; \bv{x}_{K-1}) 
\end{align}
and for each sequence, $i$, out of $M$, calculate
\begin{align}
\omega^{(i)} &= \prod_{k=1}^K \frac{\exp \left(E_k\left(\bv{x}^{(i)}_k\right)\right)}{\exp \left(E_{k-1}\left(\bv{x}^{(i)}_k\right)\right)} \\
&=\exp \left[ \frac{1}{K} \sum_{k=1}^K E\left(\bv{x}^{(i)}_k\right)  \right].
\end{align}
We then have
\begin{equation}
\ln \mathcal{Z} \approx \ln \mathcal{Z}_0 + \ln \left( \frac{1}{M} \sum_{i=1}^M \omega^{(i)}\right).
\end{equation}
To see why this method works, suppose that $\bv{x}_k \sim \PP_{k-1}$ then
\begin{align}
\EE \left[e^{\frac{E(\bv{x}_k) }{K}} \right] &= \sum_{\bv{x_k}} \PP_{k-1}(\bv{x}_k) \exp \left(\frac{E_k(\bv{x}_k)- E_{k-1}(\bv{x}_k)}{K}  \right) \\
&= \sum_{\bv{x_k}} \PP_{k-1}(\bv{x}_k) \frac{\mathcal{Z}_k \PP_k(\bv{x}_k)}{\mathcal{Z}_{k-1} \PP_{k-1}(\bv{x}_k)} \\
& = \frac{\mathcal{Z}_k}{\mathcal{Z}_{k-1}}
\end{align}
so $\exp(E(\bv{x}_k)/K)$ is an unbiased estimator of $\mathcal{Z}_k/\mathcal{Z}_{k-1}$. If we have an independent sequence $\bv{x}_1, \bv{x}_2, \ldots$ then
\begin{align}
\EE \left[\exp \left(\frac{1}{K} \sum_{k=1}^K E \left(\bv{x}_k\right)  \right) \right] &= \prod_{k=1}^K \frac{\mathcal{Z}_k}{\mathcal{Z}_{k-1}} \\
&=  \frac{\mathcal{Z}_K}{\mathcal{Z}_0}.
\label{eqn:ZK}
\end{align}
As shown by Neal \cite{nea01}, even though the sequence $\bv{x}_1, \bv{x}_2, \ldots$ is not independent, (\ref{eqn:ZK}) still holds, so $\omega^{(i)}$ is an unbiased estimator of $\mathcal{Z}_K/\mathcal{Z}_0$.

\section{Description of variables}

\label{ap:SED}

\subsection{Introduction}
The data used in this study were taken from the SED \textit{Basic materials} \citep{orton1962} rather than from later atlas publications \citep{orton1978,upton1996}. The \textit{Basic materials} presents unmodified transcriptions of question responses rather than defined linguistic variables with discretised variants. Accordingly, here we describe the variables and variants as we have defined them, with references to where in the \textit{Basic materials} the data were taken from. In some instances, especially for lexical variables, data are taken from a single question and the only analysis required is identifying what lexical item(s) each transcription represents; in others, data must be accumulated across multiple questions. In any case, some variants may have to excluded or merged to define a binary variable as described in part IV of the paper.
\par For each variable, we have attempted to give the following information:
\begin{itemize}
	\item the two variants;
	\item reference to where in the SED the data were taken from;
	\item a linguistic description of the variable and the change which produced it;
	\item an identification of which of the two variants represents the innovation and which the conservatism;
	\item where possible, an approximate dating of the change and so a rough idea of how long the variation had existed at the point the SED speakers acquired the language, the 1880s and 90s (this is more often feasible for lexical and morphological variables, where the written record typically provides more direct evidence than for phonetic and phonological variables); it is assumed for the purpose of this estimate that the first attestation of a form in writing cannot be less than 50 years after its innovation in speech;
	\item a description of what variants were merged or excluded to define the binary variable used.
\end{itemize}
\par Non-linguists may wish to consult the glossary at the end of this document for an explanation of some of the specialist terminology used. We use the International Phonetic Alphabet (IPA) \cite{internationalphoneticassociation1999} for transcription throughout. Note that there are differences between the version of the IPA used at the time of publication of the SED (the 1947 chart) and the modern version (the 1999 chart) and we update SED transcriptions to the modern version.

\subsection{Variables}

\begin{center} 1. \textit{\textit{adder} lexical item} \end{center}\par\textit{Variants:}\quad
 \textit{(n)adder, hag-}
\par\textit{Reference:}\quad
 IV.9.4
\par\textit{Background}\quad
This variable describes the lexical item used for the common European viper. The conservative variant is \textit{(n)adder}, found in the OE period as \textit{n\ae dre} (note that Bosworth \& Toller \cite{bosworth1898.1} gloss it only as a general term for snake in this period, whilst the OED suggests it had its more specific meaning already in OE). In written sources \textit{hagworm} is known from the late 15th century according to both the OED and MED (\textit{Catholicon Anglicum} c1475) \citep{lewis.1}; however, it is a Norse loanword (ON \textit{h\k{o}ggormr} `viper') and so probably dates back to the period of the Danelaw. Accordingly, the change in question is the borrowing of \textit{hag-} from ON, and we should assume this variation has existed for at least 1000 years.
\par\textit{Reduction}\quad
As the focus here is on the lexical item, variants of \textit{adder} with and without the metanalytic \textit{n-} and with different reflexes of OE \textit{-d-} were merged; similarly, different formations from \textit{hag-} (\textit{hagworm, hagger, hag}) were merged. The occasional instances of other lexical items (some clearly in error) were excluded.

\begin{center} 2. \textit{\textit{anything} lexical item} \end{center}\par\textit{Variants:}\quad
 \textit{anything, aught}
\par\textit{Reference:}\quad
 V.8.16
\par\textit{Background}\quad
This variable concerns the indefinite pronoun used in the frame \textit{Is \underline{\ \ \ \ \ \ \ \ } left?}, referring to food. Both variants have existed in some form since the OE period (OE \textit{\=awiht, \=\ae nig \th ing}), along with a variety of other indefinite pronouns (\textit{hw\ae t, ahw\ae t}, etc.), but it is not clear that both should be considered fully grammaticalised pronouns at this early point. In dating the variation between them, the question we must answer is at what point this grammaticalisation process was complete. Mitchell \cite{mitchell1985.1} identifies \textit{\=awiht} as a fully grammaticalised pronoun in OE, but notes \textit{\=\ae nig þing} as a common collocation \citep{mitchell1985.2} rather than listing with other pronouns. Bosworth \& Toller \cite{bosworth1898} agree with this implied distinction, in that they give \textit{\=awiht} but not \textit{\=\ae nig þing} its own entry. On the other hand, they do give examples of \textit{\=\ae nig þing} translating Latin \textit{aliquid} without comment (M\ae g \'\ae nig þing gódes beón of Nazareth \textit{a Nazareth potest aliquid boni esse?}; \cite{bosworth1898.2}), and in an investigation into the syntax of a variety of OE quantifiers, Roehrs \& Sapp \cite{roehrs2018} analyse both \textit{\=awiht} and \textit{\=\ae nig þing} as being fully grammaticalised heads. Thus it seems reasonable understand these as already being equivalent pronouns in OE, implying that variation between them has existed for at least 1000 years.
\par\textit{Reduction}\quad
The variant \textit{any} was excluded. The more difficult issue with these data concerns the interpretation of \textipa{[O\super{\:R}:\:t]} and many similar forms recorded in Devon, with a couple of instances in each of Somerset and Cornwall. The SED interprets these as a phonological variant of \textit{aught}. However, given the carrier sentence, it also seems possible that they represent \textit{ort} ``leavings of any description [...] esp. of food'' \citep{markus2019.1} which the EDD does record as occurring in Devon and marginally in Somerset and Cornwall. The problem with the former interpretation is that these are not locations which otherwise exhibit hyperrhoticity; the problem with the latter is that it would be expected to be plural and to appear with a quantifier (i.e.\ *`Are there any orts left?' or similar). The syntactic problems with the \textit{ort} interpretation seem hard to overcome, and so the judgement of the editors of the SED has been followed here and \textipa{[O\super{\:R}:\:t]} has been merged into \textit{aught}. However, it does seem likely that the lexical item \textit{ort} played some role in the history of this form (whether through analogical change or by \textit{ort} and \textit{aught} being reanalysed as a single lexical item).

\begin{center} 3. \textit{\textit{fist} lexical item} \end{center}\par\textit{Variants:}\quad
 \textit{fist, nieve}
\par\textit{Reference:}\quad
 VI.7.4
\par\textit{Background}\quad
This variable describes the lexical item used for a clenched hand. The conservative variant is \textit{fist}, found in the OE period as \textit{f\=yst} \citep{bosworth1898.3}. The OED and MED agree that the earliest written attestation of \textit{nieve} is at the beginning of the fourteenth century (\textit{Havelok} 1300) \citep{lewis.26}; however, it is a Norse loanword (ON \textit{hnefi} `fist') and so probably dates back to the period of the Danelaw. Accordingly, the change in question is the borrowing of \textit{nieve} and we should assume the resulting variation has existed for at least 1000 years.
\par\textit{Reduction}\quad
The phonological variant of \textit{nieve} with coda \textipa{/f/} instead of \textipa{/v/} was merged into \textit{nieve}.

\begin{center} 4. \textit{\textit{frog} lexical item} \end{center}\par\textit{Variants:}\quad
 \textit{frog, paddock}
\par\textit{Reference:}\quad
 IV.9.6
\par\textit{Background}\quad
This variable refers to the lexical item used for the set of amphibians referred to in Standard English as \textit{frogs}. The conservative variant is \textit{frog}, which has existed in this meaning since the OE period \citep{bosworth1898.4}. The innovative variant \textit{paddock} is derived as a diminutive of \textit{pad} `toad'; across the OED and MED, the earliest written attestation is at the beginning of the 14th century (in the compound \textit{padokpipe} c1300, citing Hunt \cite{hunt1989}). Thus we can assume the variation has existed for at least 750 years.
\par\textit{Reduction}\quad
\textipa{[Tr6gs]} understood as a phonological variant of \textit{frog} and so merged with it. An additional variant, \textit{jacky(toad)} (apparently derived from earlier \textit{Jacob} `frog' \citep{markus2019.2}), was excluded on the basis that it is recent, geographically very limited, and entirely embedded in the \textit{frog} domain.

\begin{center} 5. \textit{\textit{hedgehog} lexical item} \end{center}\par\textit{Variants:}\quad
 \textit{hedgehog, urchin}
\par\textit{Reference:}\quad
 IV.5.5
\par\textit{Background}\quad
This variable concerns the lexical item used for the European hedgehog, \textit{erinaceus europaeus}. \textit{Urchin} is a loanword, having been borrowed from Norman French \textit{hirchoun} and first attested in English sources around the turn of the 14th century (\textit{South English legendary}, c1300) \citep{lewis.3}, whilst \textit{hedgehog} is a compound formed within English and first attested in the middle of the 15th century (\textit{Treatise on Fishing}, c1450) \citep{lewis.4}. On this (rather limited) basis we might label \textit{hedgehog} the innovation, but in reality both of these were innovations which competed in replacing earlier \textit{igil}, so this is not a particularly useful framing. For this reason, it is not clear that it would be meaningful to give an age for this variable.
\par\textit{Reduction}\quad
Compounded and modified variants (\textit{prick(l)(y) urchin, pricky black urchin, prick(l)y-back(ed) urchin}) were merged into \textit{urchin}; minor variants entirely embedded in the \textit{hedgehog} domain (\textit{hedgepig, hedgeboar, furzepig}) were excluded.

\begin{center} 6. \textit{\textit{newt} lexical item} \end{center}\par\textit{Variants:}\quad
 \textit{eft} and related variants, \textit{ask} and related variants
\par\textit{Reference:}\quad
 IV.9.8
\par\textit{Background}\quad
Both variants are attested in OE glossing Latin \textit{lacerta} `lizard' \citep{bosworth1898.11}. However, \textit{ask} (OE \textit{\=a\dh exe}) has cognates elsewhere in Germanic whereas \textit{eft} (OE \textit{efete}) is of unknown origin; additionally, according to the OED, attestations of \textit{\=a\dh exe} are found in early OE glosses whereas \textit{efete} is not known until the beginning of the 11th century (OED; \cite{wright1884}). Thus we can take the coining of \textit{eft} to be the innovation (whether it was a loanword or derived from some other lexical item), and the variation to have existed for at least 1000 years.
\par\textit{Reduction}\quad
The raw data contain a great variety of variants. However, many of these are phonological derivations from \textit{eft}, with (\textit{newt, mewt}) and without (\textit{ewt, eff, ebbet, abbet}) metanalytic \textit{n-}; these, along with compounded variants (\textit{water-evet, wet-effet, wet-eff, four-legged evet, four-legged emmet}), were merged into \textit{eft} on the basis that they imply the earlier use of some form of \textit{eft}. Other variants (\textit{askerd, askel, asker}) are suffixed forms from \textit{ask}, phonological derivations of these (\textit{askert, asgel, aster, nasgel}) and compounds (\textit{dry-ask, water-ask}) and all of these were merged into \textit{ask} by the same logic. Unrelated minor variants (\textit{mancreeper, swift, water-swift, waterlizard, padgy-pol, tiddlywink, yellow-belly}) were excluded.

\begin{center} 7. \textit{\textit{owl} lexical item} \end{center}\par\textit{Variants:}\quad
 \textit{owl, howlet}
\par\textit{Reference:}\quad
 IV.7.6
\par\textit{Background}\quad
\textit{owl} is an inherited Germanic word that occurs in OE as \textit{\=ule} (OED; \cite{bosworth1898.5}). The innovation is \textit{howlet}, a loanword from French \textit{hulotte} and first attested in English in the late 15th century (\textit{``Holy berith beris...''} 1475, \textit{Ludus Coventriae} 1475) \citep{lewis.6}. Thus the variation can be taken to be at least 450 years old.
\par\textit{Reduction}\quad
The key feature by which to distinguish these two variants was taken to be the presence of the second syllable, and so variation in the vowel of the first syllable (short \textit{ullet} vs.\ long \textit{howlet}, etc.), the presence of initial /h/ (\textit{howl} vs.\ \textit{owl}, etc.) and rhoticity in the second syllable (\textit{ullet} vs.\ \textit{ullert}) were ignored. Compounded variants were merged with their respective uncompounded variants (\textit{Jenny-owl} and \textit{Meg-owl} with \textit{owl}, \textit{Jenny-howlet} and \textit{Polly-howlet} with \textit{howlet}).

\begin{center} 8. \textit{\textit{pour} lexical item} \end{center}\par\textit{Variants:}\quad
 \textit{pour, teem}
\par\textit{Reference:}\quad
 V.8.8
\par\textit{Background}\quad
This variable refers to the lexical item used for decanting tea from a teapot into a cup. \textit{Teem} is first attested at the beginning of the 15th century (\textit{Cursor Mundi} 1400) \citep{lewis.7}) but as an Old Norse borrowing (cf.\ ON \textit{t\oe ma} `empty') we can assume it dates back much earlier. Thus \textit{pour} is probably the innovation: according to the OED and MED it is probably a loanword from Middle French \textit{purer} and is first attested in the first half of the 14th century (\textit{Amis and Amiloun} c1330) \citep{lewis.8}). Thus we can take the variation to have existed for at least 600 years.
\par\textit{Reduction}\quad
Assorted minor variants were excluded from consideration: \textit{birle, chuck, emp, ent, hale, hell, heave, laden, lade} and \textit{shut}.

\begin{center} 9. \textit{\textit{snail} lexical item} \end{center}\par\textit{Variants:}\quad
 \textit{snail}, forms including \textit{-dod-}
\par\textit{Reference:}\quad
 IV.9.3
\par\textit{Background}\quad
This variable refers to the lexical item given in response to the question: ``What are those slow, slimy things that carry their houses about with them; they come out after rain?'', intended to elicit the name for terrestrial molluscs with spiral shells large enough to retract into (Standard English \textit{snail}). The conservative variant is \textit{snail}, which is an inherited Germanic term known in English from the OE period. The innovation is the use of terms including the element \textit{-dod-}; since the etymology of \textit{-dod-} is unknown, the nature of this innovation (as borrowing vs.\ derivation from some pre-existing element) is uncertain. These are known from the EMoE period according to the OED (\textit{dodman} in John Bale's \textit{King Johan} 1528; \textit{hodmandod} and \textit{dodman} in Bacon 1626 \cite{bacon1626}), so we can assume the variation has existed for at least 400 years.
\par\textit{Reduction}\quad
All compounded variants containing \textit{-dod-} (\textit{hodmedod, hoddy-doddy, dodman}, etc.) were merged as a single variant. Since the etymology of \textit{-dod-} is not known, it is not clear what the precise sequence of derivation was here: *\textit{dod} itself might have been a lexical item meaning snail which spread to this region, making this the innovation and all the compounded forms later derivations; or *\textit{dod} might have had some other meaning (the OED suggests a connection with \textit{dod} `rounded summit'), implying that one of the compounds was the original innovation and the others analogical formations based on it. Either way, however, it seems reasonable to regard the use of terms with \textit{-dod-} as an innovation across this region.

\begin{center} 10. \textit{\textit{upstairs} lexical item} \end{center}\par\textit{Variants:}\quad
 \textit{upover, upstairs}
\par\textit{Reference:}\quad
 V.2.5
\par\textit{Background}\quad
This variable concerns the lexical item used to describe a room in an upper floor. The innovative variant \textit{upover} is not listed in the OED; in the EDD, examples are cited from Devon, but only from 1877 \citep{markus2019.3}. Thus this innovation can be assumed to be very recent.
\par\textit{Reduction}\quad
Minor variants, quite possibly reflecting failures to elicit the relevant term, were excluded: \textit{up a height, up above, up top} and \textit{up aloft}.

\begin{center} 11. \textit{\textit{vinegar} lexical item} \end{center}\par\textit{Variants:}\quad
 \textit{alegar, vinegar}
\par\textit{Reference:}\quad
 V.7.19
\par\textit{Background}\quad
This variable concerns the lexical item used for acetic acid solution used in cooking, elicited as ``that sour liquid you pickle red cabbage in''. Of these two forms, \textit{vinegar} was a loanword from OF \textit{vinaigre}, first attested in this meaning in the first half of the 14th century (\textit{South English Legendary} 1325, \textit{Shoreham Poems} 1350) \citep{lewis.9}. \textit{Alegar} appears to be an analogical formation \textit{ale+eager} on the basis of \textit{vinegar} attested from the end of the 14th century (\textit{Form of Curry} 1399, \textit{Inventories of St. Leonard's Priory} c1422) \citep{lewis.10}. However, since \textit{vinegar} referred specifically to wine vinegar and \textit{alegar} to malt vinegar, the relevant innovation is not the coining of either word but the shift of \textit{alegar} to overlap in meaning with \textit{vinegar} so that the two could be considered variants of a single variable. None of the reference materials consulted here record this meaning, suggesting that this innovation was relatively recent (although in many written contexts it would be very hard to identify the change).

\begin{center} 12. \textit{\textit{wrist} lexical item} \end{center}\par\textit{Variants:}\quad
 \textit{wrist, shackle}
\par\textit{Reference:}\quad
 VI.6.9
\par\textit{Background}\quad
This variable concerns the lexical item used for the end of the arm before the hand. Both words are native Germanic, found already in OE (\textit{wrist, sceacel}) and with cognates in other Germanic languages. However, \textit{shackle} in the meaning `wrist' is a shortening of the compound \textit{shacklebone} `wrist' and so it is the formation of this compound that is the relevant innovation; the first instance recorded in the OED is from the third quarter of the 16th century (\textit{Register of the privy council of Scotland} 1571), so we can assume this variation is at least 350 years old.
\par\textit{Reduction}\quad
Compounded variants with \textit{-wrist} (\textit{armwrist, handwrist}) were merged into \textit{wrist}.

\begin{center} 13. \textit{\textit{yeast} lexical item} \end{center}\par\textit{Variants:}\quad
 \textit{yeast, barm}
\par\textit{Reference:}\quad
 V.6.2
\par\textit{Background}\quad
This variable concerns the lexical item used for the substance added to bread dough to raise it. The two lexical items concerned, \textit{yeast} and \textit{barm}, are both native Germanic words which existed in OE (\textit{gist, beorma}). In Modern Standard English there may be a semantic distinction between the two words whereby \textit{barm} is used to refer to the foam removed from the top of fermenting malt liquors whilst \textit{yeast} refers to the fungus that causes fermentation; historically, however, both terms had the former meaning (OED; \cite{bosworth1898.6,lewis.11}). Thus we should understand the variation between these two variants as having existed for at least 1000 years.

\begin{center} 14. \textit{Participial adjective from \textit{burn}} \end{center}\par\textit{Variants:}\quad
 \textit{burnt, burned}
\par\textit{Reference:}\quad
 V.6.7
\par\textit{Background}\quad
This is a morpholexical variable: it is part of a wider pattern of morphological variation between (regular) /d/ and (irregular) \textipa{/t/} suffixes for forming the preterite, but the choice is lexically controlled for the individual speaker and the variation is independent across lexical items. The verb \textit{burn} is descended from two OE verbs, strong class III \textit{beornan} and weak class 1 \textit{b\ae rnan}, which merged during the ME period, along with admixture with parts of the two corresponding ON verbs (strong intransitive \textit{brenna} and weak transitive \textit{brenna}) (OED; \cite{lewis.12}); the MoE past tense forms are clearly only related to the weak formations. The innovative \textipa{/t/} variant does not seem to occur in OE (it is not mentioned in Bosworth \& Toller \cite{bosworth1898.7} and there are no occurrences in the Old English Web Corpus \cite{healey2009}). The first ME occurrences according to the MED \citep{lewis.13} and LAEME \citep{laing2013} are in the late 13th and early 14th centuries (Cambridge Trinity B.14.39 (=LAEME text \#246) after 1253, \textit{Genesis \& Exodus} 1325), so we can assume the variation is at least 750 years old.
\par\textit{Reduction}\quad
Excluded responses with unrelated verbs (\textit{kizzened, swinged}).

\begin{center} 15. \textit{Past participle of \textit{earn}} \end{center}\par\textit{Variants:}\quad
 \textit{earnt, earned}
\par\textit{Reference:}\quad
 VIII.1.26
\par\textit{Background}\quad
This is a morpholexical variable: it is part of a wider pattern of morphological variation between (regular) /d/ and (irregular) \textipa{/t/} suffixes for forming the preterite, but the choice is lexically controlled for the individual speaker and the variation is independent across lexical items. The verb \textit{earn} is descended from a weak class 2 OE verb \textit{earnian}; as such, the regular variant (\textit{earned}) is conservative and innovative \textit{earnt} must be by analogy with another verb. The irregular variant appears first in the written record in the 18th century (early attestations include: Anon. 1730 \cite{anon.1730}; Smith 1737 \cite{smith1737}; Ward 1758 \cite{ward1758}; Nugent's 1763 translation of Rousseau \cite{rousseau1763}), so we can assume the variation is at least 300 years old.

\begin{center} 16. \textit{Past participle of \textit{get}: presence of \textit{-en}} \end{center}\par\textit{Variants:}\quad
 -Ø, \textit{-en}
\par\textit{Reference:}\quad
 IX.6.4
\par\textit{Background}\quad
This is part of broader variation in the morphology of English past participles, but the variation occurs at the level of individual lexical items and so should be classed as morpholexical. The OED notes that in varieties with both variants \textit{got} and \textit{gotten} there is often a semantic distinction where \textit{have gotten} refers to the process of obtaining something whilst \textit{have got} refers to simple possession (as might be expected for an ongoing grammaticalisation process). The framing sentence used in the SED is: \textit{You say to a friend: Shall I give you one of these pups? But he answers: No thanks, we \underline{\ \ \ \ \ } one.} This perhaps leaves open space for either interpretation, but simple possession seems more likely. The verb \textit{get} is primarily descended from the ON class V strong verb \textit{geta}, perhaps with some influence from its OE cognate \textit{gietan} (OED). The -Ø variant is the innovation. Forms without the final /n/ (i.e.\ \textit{-e}) are found as a result of general final /n/ loss in ME as early as the late 14th century (`gote' in \textit{Wycliffe Bible} 1382) but since the MoE vowel is short these are unlikely to be precursors of the -Ø form. The MED lists \textit{g(h)et} and \textit{gat} as possible forms of the past participle, but the only examples given are one instance of \textit{gat} at the beginning of the 15th century (\textit{Cleanness} (Nero A.10) c1400) which context renders ambiguous between a past participle and a simple preterite, and one of \textit{geth} in the mid-15th century (\textit{Paston letters} 3.2 1454) \citep{lewis.14}). Thus, the earliest we can confidently date the innovation to is the early 15th century, rendering it at least 600 years old.
\par\textit{Reduction}\quad
Variation in the stem vowel was ignored, so that \textit{getten, gitten} and \textit{gotten} were merged as one variant and \textit{(a)got, gat} and \textit{got} as the other.

\begin{center} 17. \textit{Preterite of \textit{grow}} \end{center}\par\textit{Variants:}\quad
 \textit{growed, grew}
\par\textit{Reference:}\quad
 IX.3.9
\par\textit{Background}\quad
This is a morpholexical variable: it is part of larger patterns of morphological variation between weak and strong preterites, but the choice of variant is lexical and not correlated across different verbs and locations. The strong form \textit{grew} is the conservative variant and attested from the OE period, whereas the weak form \textit{growed} is an analogical formation first attested in ME from the latter half of the 14th century (\textit{William of Pallern} 1375, \textit{Wycliffe Bible} 1382) according to citations in the MED \citep{lewis.27}; accordingly we can assume the variation is at least 650 years old.
\par\textit{Reduction}\quad
The SED records an occasional third variant, \textit{did grow}, in localities such as So4 and Ha6. However, these are likely to reflect examples of habitual \textit{do} and not truly a variant of the simple preterite; accordingly, these were excluded.

\begin{center} 18. \textit{Backformation of sg.\ \textit{pea}} \end{center}\par\textit{Variants:}\quad
 sg.\ \textit{pea} pl \textit{peas}, sg.\ \textit{pease} pl \textit{pease}
\par\textit{Reference:}\quad
 V.7.13
\par\textit{Background}\quad
This morpholexical variable concerns variation in the paradigm of \textit{pea(se)}. The noun is originally weak, with plural in \textit{-n} (OE \textit{pise} : \textit{pisan}, ME \textit{pese} : \textit{pesen}). When weak plurals were lost in ME, it gained a strong plural in a handful of dialects (e.g. \textit{peses} in Piers Plowman C 9.307, c1400) but in most varieties the result were identical singular and plural forms \textit{pease} : \textit{pease}. The final \textipa{/z/} of the stem was later reinterpreted as plural -s and so a singular form \textit{pea} was backformed from it. The earliest attestations of this form in the OED are from the mid- to late 17th century \citep{boyle1666,plot1677,sinclair1683}, so we can assume the variation is at least 400 years old.

\begin{center} 19. \textit{\textit{worse}: formation of \textit{worser}} \end{center}\par\textit{Variants:}\quad
 \textit{worse, worser}
\par\textit{Reference:}\quad
 VI.12.3
\par\textit{Background}\quad
This morpholexical variable refers to the formation of the comparative of \textit{bad} as \textit{worse} vs.\ \textit{worser}. This word probably did originally have a distinct comparative suffix (Gothic has \textit{wairs\underline{iza}}, on the basis of which Magnússon reconstructs PG *werz-\underline{izan}- \citep{magnusson1989}) but by the OE period this is no longer synchronically recognisable (OE \textit{wiersa}) making \textit{worse} the conservatism. The suffix \textit{-er} is then added to this by analogy with regular comparatives in some varieties in the late ME / EMoE period. The earliest attestations in the OED are from the end of the 15th century and beginning of the 16th (including \textit{De proprietatibus rerum} 1495, Mirk 1508 \cite{mirk1508}) so we can assume the variation dates back at least 550 years.
\par\textit{Reduction}\quad
The unrelated lexical variant \textit{waur} was excluded.

\begin{center} 20. \textit{\textit{worse}: lexical item} \end{center}\par\textit{Variants:}\quad
 \textit{worse, waur}
\par\textit{Reference:}\quad
 VI.12.3
\par\textit{Background}\quad
This variable refers to the lexical item used suppletively as the comparative of \textit{bad}: either native \textit{worse(r)} ($<$ OE \textit{wiersa}) or the borrowing \textit{waur} ($<$ ON \textit{verri}). The earliest attestation of the loanword in citations in the OED and MED is in the late 12th century (\textit{Ormulum} 1175) \cite{lewis.28} implying that the variation has existed for at least 850 years, but, as for any ON loanword, identifying a terminus post quem in this way is likely to give an underestimate of the age of the variation: we can assume that this was borrowed during the period of the Danelaw, suggesting an age of at least 1000 years.
\par\textit{Reduction}\quad
The variant \textit{worser} was merged with \textit{worse}, since \textit{worser} is transparently a later derivation from \textit{worse}.

\begin{center} 21. \textit{Possessive pronouns} \end{center}\par\textit{Variants:}\quad
 \textit{-s, -(e)n}
\par\textit{Reference:}\quad
 IX.8.5
\par\textit{Background}\quad
This is a morphological variable referring to the formation of the possessive personal pronouns. The histories of the individual person-number forms should initially be considered separately.
\par For the 3sg.\ feminine, both variants are ME analogical constructions based on the 3sg.\ feminine personal pronoun \textit{hire} and the possessive pronouns \textit{m\=in/þ\=in} in the case of \textit{hiren}, and gen.sg.\ \textit{-es} in the case of \textit{hires}. According to citations in the MED \cite{lewis.29}, \textit{hiren} is attested as early as the first half of the 13th century (\textit{Ancrene Riwle} c1230) whereas \textit{hires} is not attested until the fourth quarter of the 14th century (\textit{Wycliffe Bible} 1382), suggesting that we should see \textit{hires} as the innovation; this is consistent with the fact that the expansion of gen.sg.\ \textit{-es} was not complete until the end of the ME period \citep{allen2008}.
\par For the 3sg.\ masculine, the \textit{-s} variant is the original form and found regularly since the OE period \citep{bosworth1898.8}. The \textit{-n} variant \textit{hisen}, like \textit{hiren}, is an analogical formation based on the 1sg./2sg.\ possessive pronouns, but is not attested until the mid-15th century according to the MED (\textit{Laud Troy Book} c1425, \textit{Letters pertaining to the Guilds of Coventry} 1440) \citep{lewis.15}.
\par For the 1pl, both variants are analogical formations and instances of both are found from the late 14th century (\textit{ourn} in \textit{Wycliffe Bible} 1382, \textit{oures} in the \textit{Pardoner's Tale} 1390) \citep{lewis.16}.
\par For the 3pl, as with the 3sg.\ feminine and the 1pl, both variants are analogical formations. At least one instance of the \textit{-s} variant is found as early as the late 12th century (\textit{Ormulum} c1175) but it starts to appear regularly only from the late 14th (\textit{Wycliffe Bible} 1382, \textit{Cursor Mundi} 1400) \citep{lewis.17}. The \textit{-n} variant appears to be substantially later: the MED notes just one example, in the mid-15th century (\textit{Treatise on the Ten Commandments} c1425) and suggests this may be a secondary analogical formation based on the 3sg.\ feminine instead of being by analogy with the 1sg./2sg.\ \citep{lewis.18}.
\par We can see that the two variants have somewhat different histories and exact dating for the different persons/numbers. However, the innovation of these systems, in which the possessive pronouns are all formed with \textit{-s} or are all formed with \textit{-n}, can be given termini post quem by looking at the latest forms to appear: the mid-15th century for the \textit{-n} variant, the late 14th century for the \textit{-s} variant. Thus the variation is at least 650 years old.
\par\textit{Reduction}\quad
The 3sg.\ feminine, 3sg.\ masculine, 1pl and 3pl are treated together, so that \textit{hers}, \textit{his}, \textit{yours} and \textit{theirs} are merged as -s and \textit{hern, hisn, ourn} and \textit{theirn} are merged as \textit{-n}. Double marked variants (\textit{hersn, ourns} etc.) could equally be derived from earlier \textit{-s} or \textit{-n} variants, and so were excluded. Zero marked variants (\textit{her, our} etc.) probably reflect a misunderstanding of the question and so were also excluded.
\par 1sg.\ \textit{mine} and 2sg.\ \textit{thine} do not participate in this system (they always have \textit{-n}), and so are not included. 2sg.\ \textit{yours/yourn} is not recorded in many localities where \textit{thine} is still used, and so is not included; additionally, as the use of the historical plural in the singular is more recent than the innovation of this variable and (in recent decades) reflects spread from Standard English, it is not clear that we would expect it to be part of the same system. 2pl \textit{yours/yourn} has a substantially different distribution to the other person/number combinations, presumably again reflecting this interaction with Standard English and with the 2sg., and so was not included.

\begin{center} 22. \textit{Verbal 3sg.\ \textit{-s}} \end{center}\par\textit{Variants:}\quad
 \textit{-s}, -Ø
\par\textit{Reference:}\quad
 VI.5.5 (\textit{speaks}), VI.13.3 (\textit{aches, hurts}), VI.14.2 (\textit{suits}), VI.14.14 (\textit{wears}), VIII.1.9 (\textit{looks, favours, resembles}), VIII.6.2 (\textit{begins, breaks, closes, comes, finishes, leaves, opens, shuts, starts}), IX.3.6 (\textit{makes})
\par\textit{Background}\quad
This morphological variable concerns the form of the verb used with a 3sg.\ pronoun subject. There has been variation in subject-verb agreement at least since the OE period, with Northumbrian texts such as the 10th century Lindesfarne Gospels showing variable \textit{-es} for all persons/numbers alongside more conservative forms (see e.g. \cite{brunner1965.1}). This \textit{-es} ending spread from northern to southern English varieties throughout the ME period and became restricted to the 3sg., rising in frequency dramatically in London English in the late 16th century \cite{nevalainen2000.1,miller2002}. The zero ending may be the result of analogical levelling across the paradigm or may have its origin in subjunctive zero endings. Either way, it existed at a very low frequency in many EMoE varieties, but became particularly established in East Anglia from the 16th century; it was also found particularly in parts of the south west of England, perhaps as a result of the changes involving positive declarative \textit{do} (for these points and further, see Wright \cite{wright2015}).
\par As can be seen from this brief account, it is not straightforward to identify a conservative and an innovative variant here. Both endings have existed for a very long period of time, but their functions and the roles they play in the larger inflectional system have shifted. We first see them occurring in systems that look broadly like the MoE systems (with levelled -Ø in all person/number combinations on the one hand, or with \textit{-s} distinguishing the 3sg.\ from all other cells in the other) in the south of England at roughly the same period of EMoE. Accordingly, it seems reasonable to think of this variation as around 500 years old.
\par\textit{Reduction}\quad
Instances of habitual \textit{do}+verb were excluded from consideration.

\begin{center} 23. \textit{1sg.\ present of \textit{be}: levelling to \textit{be}} \end{center}\par\textit{Variants:}\quad
 levelled to \textit{be}, not levelled to \textit{be}
\par\textit{Reference:}\quad
 IX.7.1
\par\textit{Background}\quad
The verb \textit{to be} in Standard English differs from all other verbs in showing a pattern of subject-verb agreement that distinguishes more than just 3sg.\ vs.\ other. In traditional dialects, this system is simplified in a large variety of different ways. In this variable, we look just at the levelling of 1sg.\ (Standard English \textit{am}) to \textit{be}, but for most speakers that reflects a system with levelling to \textit{be} in all person/number combinations. In OE there were two verbs meaning `be': \textit{wesan} and \textit{b\=eon}, of which \textit{wesan} was unmarked whilst \textit{b\=eon} was typically used for the gnomic present, the future, or the iterative present/future, with many exceptions \citep{mitchell1985.3}. It seems likely that MoE dialectal systems with levelling to \textit{be} date back to the collapse of the \textit{wesan:b\=eon} system, rather than being a later development: the innovation, under this understanding, is levelling of \textit{wesan} forms to \textit{b\=eon} as the semantic distinction between them was lost.
\par LAEME offers some evidence for \textit{be} forms used in the 1sg.\ with future meaning in the Midlands \citep{laing2013.1}, but has no map for 1sg.\ \textit{be} forms used in other contexts. There is no evidence of this form in southern ME in LALME \citep{benskin2013.1}; however, the MED lists examples in the indicative from the middle of the 15th century (\textit{King Ponthus} 1450, \textit{Pilgrimage of the Life of Man} 1500) \citep{laing2013.3}. By contrast, LAEME, LALME and the MED offer copious evidence for \textit{be-} forms in the 2sg., 3sg.\ and pl (indeed, \textit{be-} forms are universal in the pl in southern ME), suggesting that spread to the 1sg.\ was the last stage of this levelling process. Together with the fact that a system with complete \textit{be-}levelling must have existed by the end of the 16th century since it was part of the input to Caribbean Englishes, this suggests that we can date this innovation to some time in the 15th century, making the variation at least 550 years old.
\par\textit{Reduction}\quad
The variants \textit{be} and \textit{bin} were merged as showing \textit{be}-levelling; the variants \textit{am}, \textit{are} and \textit{is} were merged as not showing \textit{be}-levelling.

\begin{center} 24. \textit{1sg.\ present of \textit{be}: levelling to \textit{is}} \end{center}\par\textit{Variants:}\quad
 levelled to \textit{is}, not levelled to \textit{is}
\par\textit{Reference:}\quad
 IX.7.1
\par\textit{Background}\quad
The verb \textit{to be} in Standard English differs from all other verbs in showing a pattern of subject-verb agreement that distinguishes more than 3sg.\ vs.\ other. In traditional dialects, this system was simplified in a large variety of different ways. In this variable, we look just at the levelling of 1sg.\ (Standard English \textit{am}) to \textit{is}. Historically, this variant might have been related to the pattern known as the `North Subject Rule' by which present tense verbs took the 3sg.\ \textit{-s} form in all contexts except where they were directly adjacent to a personal pronoun; the NSR could have generated 1sg.\ \textit{is} when the subject was not adjacent to the verb which might later have been extended to other contexts, and was associated with a similar spatial region to that we see in the SED for this variable. However, this has not been investigated in detail. There are no tokens of 1sg.\ \textit{is} in LALME \citep{benskin2013.2}, but LALME only has data for southern England for this variable. LAEME does not map this variable specifically; exploring the tag dictionary we find one text with relevant examples, 1sg.\ $<$es$>$ in hand C of the 14th century \textit{Cursor Mundi} \citep{laing2013.2}, but these reflect just two tokens in this long text which otherwise uses $<$am$>$. In a study of early evidence for the NSR, de Haas finds that NSR with full verbs dates from as early as the 10th century \citep{dehaas2011.1}; but this study excluded \textit{to be}, meaning it offers no direct evidence for this variable \citep{dehaas2011.2}. We have not been able to identify any occurrences in the Parsed Corpus of Early English Correspondence \citep{nevalainen2006}, which covers the period 1410-1681. Overall, then, all we can say about the age of this variant is that it likely has its origins in NSR which is of OE or EME age and that it may have existed in some form since the ME period, but that we do not have clear enough evidence to offer a specific date.
\par\textit{Reduction}\quad
The variants \textit{am, are, be} and \textit{bin} were merged as not showing \textit{is}-levelling.

\begin{center} 25. \textit{\textit{it is} contraction} \end{center}\par\textit{Variants:}\quad
 \textit{'tis, it's}
\par\textit{Reference:}\quad
 V.7.3
\par
\begin{figure}[!]
	\includegraphics[width=9 cm]{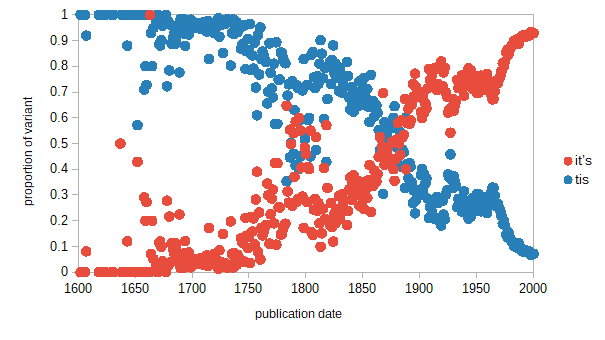}
	\caption{Relative rate of \textit{'tis} and \textit{it's} in the Google Books corpus by date}\label{fig:figure1}
\end{figure}

\textit{Background}\quad
This morphological variable concerns the contracted form of the 3sg.\ inanimate pronoun, \textit{it}, plus the 3sg.\ of \textit{to be}, \textit{is}, in phrase-internal position. The verb \textit{to be} has exhibited contractions with various pronouns and negative adverbs since the OE period, but contractions of \textit{(h)it}+\textit{is} in particular seem to go back to the late ME period. The OED lists examples of \textit{tis} as early as the late 13th century (\textit{Ancrene Riwle} 1289) but without syncope (i.e.\ \textit{hit tis}); in both the OED and MED the first cited example with syncope of the vowel of \textit{is} is from the latter half of the 15th century (\textit{Mankind} c1475) \citep{lewis.19}. There are no examples with syncope of the vowel of \textit{it} in the MED, and the earliest cited instance of \textit{it's} in the OED is in the mid-16th century \citep{horace1566}, suggesting a point of innovation at some time in the 16th century. These dates suggest that \textit{tis} was the conservativism, well-established by the time that \textit{it's} was innovated; the relative trajectory of the two variants in printed materials in Google Books \citep{michel2011} supports this, cf.\ Figure~\ref{fig:figure1}. Thus we can infer that this variation has existed for around 350 years.

\par\textit{Reduction}\quad
Responses with no vowel (i.e.\ [ts]) were excluded, as they are ambiguous between \textit{it's} and \textit{tis}.

\begin{center} 26. \textit{Coda /l/ velarisation} \end{center}\par\textit{Variants:}\quad
 clear \textipa{[l]}, dark or vocalised \textipa{[\textbarl \ U]}
\par\textit{Reference:}\quad
 IV.7.6 (\textit{owl}), V.9.7 (\textit{shelf}), VII.3.7 (\textit{fall}), VII.6.10 (\textit{dull})
\par\textit{Background}\quad
This variable concerns velarisation (and potentially subsequent vocalisation) of /l/ in coda position. There is evidence for this sound change throughout the history of English (and, indeed, its reconstructed prehistory, if vocalisation of Proto-Indo-European syllabic liquids is taken into account). Sporadic instances from the EME period give MoE forms like \textit{which}, \textit{such} and \textit{as}; systematic occurrence in the frame [V+back]\_[C+labial, C+back] starting in the north of England from the 15th century onwards gives MoE forms like \textit{yolk}, \textit{half} and \textit{folk} \citep{minkova2014.1}; systematic occurrence after back vowels regardless of following context gives Modern Scots forms like \textit{a'} `all', \textit{pou} `pull' and \textit{fou} `full' \citep{grant}. However, these earlier instances of the sound change are excluded from consideration here, and only the most recent occurrence is examined, which applies to all coda /l/ regardless of preceding vowel (and, although these are not included here, also syllabic /l/) and is associated with the south-east of England. Since this latest sound change rarely affects the orthography, it is difficult to date from written sources.
\par\textit{Reduction}\quad
Vocalised realisations, given that they are universally back, must have proceded via dark \textipa{[\textbarl]}, and so are merged with it.

\begin{center} 27. \textit{Coda /st/ simplification} \end{center}\par\textit{Variants:}\quad
 [s], [st]
\par\textit{Reference:}\quad
 VI.6.9 (\textit{wrist}), VI.7.4 (\textit{fist}), VII.6.6 (\textit{frost})
\par\textit{Background}\quad
This variable concerns the simplification of coda /st/ clusters to [s]. The existence of this process as a fast speech process, but not a regular sound change, is a universal of MoE varieties and can be found throughout much of the history of English. For example, looking at the superlative suffix \textit{-est}, LALME shows just six points with simplification ($<$-es$>$, $<$-ys$>$) and these are scattered evenly across the map, suggesting spelling errors or sporadic sound change rather than a regular sound change \citep{benskin2013.3}. However, the SED offers evidence for a more consistent regular sound change in certain regions, in particular Devon, east Cornwall and West Somerset. It is hard to date this later change specifically.

\begin{center} 28. \textit{FACE vowel diphthongisation} \end{center}\par\textit{Variants:}\quad
 fronting diphthong, other
\par\textit{Reference:}\quad
 V.1.5 (\textit{gable}), VI.13.3 (\textit{aches}), IX.3.6 (\textit{make, makes, made})
\par\textit{Background}\quad
This is a phonetic variable, referring to the MoE realisation of ME \textipa{/a:/}. Changes affecting ME \textipa{/a:/} are extremely complex, with a great diversity of different reflexes at the time of the SED. Anderson \cite{anderson1987.1} identifies four major groups, reflecting four sound changes:\renewcommand{\theenumi}{\alph{enumi}}
\begin{enumerate}
    \item rising diphthongs, reflecting the development of a stressed, high front onset;
    \item centring diphthongs, reflecting the development of a schwa offglide;
    \item long monophthongs, reflecting either no changes beyond the raising that affected all ME \textipa{/a:/}, or only further raising;
    \item upgliding diphthongs, reflecting the development of a high front second element.
\end{enumerate}
Of these, Anderson reasons on the basis of information from Ellis that (d) is a recent development, spreading rapidly from London only in the 19th century (Anderson \cite{anderson1987.2} citing Ellis \cite{ellis1889}), during which time other variants have been recessive. Accepting this argument, we examine here the distribution of this latest change.
\par\textit{Reduction}\quad
All of the (d) variants \textipa{[aI, \ae I, EI, Ei, eI, ei]} were merged as showing the change in question. All of the (a) \textipa{[ia, Ia, I@, ja:, je:, jE]}, (b) \textipa{[ea, e@, E@]} and (c) \textipa{[E, E:, e, e:, i:, i]} variants were merged as `other'.

\begin{center} 29. \textit{\textit{h}-dropping} \end{center}\par\textit{Variants:}\quad
 [h], [Ø]
\par\textit{Reference:}\quad
 IV.5.10 (\textit{hare}), IV.10.9 (\textit{holly}), V.1.1 (\textit{houses}), VI.9.1 (\textit{hip, haunch, huck}), VI.10.9 (\textit{height}), VII.6.6 (\textit{hoar-frost})
\par\textit{Background}\quad
The variable refers to the non-realisation of etymological onset /h/. There has been at least sporadic loss of onset /h/ in English since the OE period, but this became a more established part of the phonology of many varieties during the ME period, partly as a result of Romance contact \citep{minkova2014.2}. This history is relatively complex and the evidence equivocal, with different behaviours of native words and loanwords, prescriptive pressures and sociolinguistic effects, and influence from the orthography; however, this is not important for our purposes.

\begin{center} 30. \textit{happY lowering} \end{center}\par\textit{Variants:}\quad
 lowered or laxed vowel, high vowel
\par\textit{Reference:}\quad
 VIII.1.16 (\textit{ready}) 
\par\textit{Background}\quad
This is a phonetic variable, dealing with realisations of the unstressed, word-final front vowel labelled as the `happY' vowel in Wells' lexical sets \citep{wells1982}. We see two sound changes reflected in the SED data: one which lowers the vowel to \textipa{[e]} or \textipa{[9]}, and one which tenses it to [i]. Since the distributions of these reflexes do not overlap, they are both included here as separate variables.
\par\textit{Reduction}\quad
The high variants \textipa{[I i]} were merged as not showing lowering, the mid variants \textipa{[e 9]} were merged as showing lowering.

\begin{center} 31. \textit{happY tensing} \end{center}
\par\textit{Variants:}\quad
 tensed vowel, lax or lowered vowel
\par\textit{Reference:}\quad
 VIII.1.16 (\textit{ready}) 
\par\textit{Background}\quad
See happY lowering (30).
\par\textit{Reduction}\quad
The lax and lowered variants \textipa{[I e 9]} were merged as not showing tensing.

\begin{center} 32. \textit{ME \textipa{/o:/}: EMoE fronting} \end{center}\par\textit{Variants:}\quad
 fronted \textipa{[Y, Y:]}, other
\par\textit{Reference:}\quad
 V.1.2 (\textit{roof}), V.2.4 (\textit{rooms}), V.2.14 (\textit{broom}), VI.5.6 (\textit{tooth})
\par\textit{Background}\quad
The SED shows a huge diversity of reflexes of ME \textipa{/o:/}, reflecting a series of overlapping sound changes. We examine three sets of reflexes reflecting three changes whose outputs are phonetically distinct and whose spatial distributions are relatively well separated: the fronting of ME \textipa{/o:/} to \textipa{[\o:]} (or similar) in the north-west of England, giving MoE reflexes with front falling diphthongs (\textipa{[I@]} etc.) \cite{anderson1987.3}; the shortening of long \textipa{/u:/} to \textipa{[U]} \cite{anderson1987.5}; and the fronting of raised \textipa{/u:/} to \textipa{[Y:]} \cite{anderson1987.4}. The last of these is a phonetic variable; the other two are phonological, in the sense that they result in mergers with other phonemic vowels. The first had happened already in the ME period, whereas the latter two are later. See ME \textipa{/o:/}: shortening (62) and ME \textipa{/o:/}: ME fronting (63) for the other two changes.
\par\textit{Reduction}\quad
Front rounded monophthongs \textipa{[Y, Y:]} were merged as showing the change, all other variants \textipa{[EU, aU, 6U, ou, OU, @U, i@, I:@, IU, I@, Iu, I7, j7, jU, u:, Uu:, Uu, UI, Iu:, @u:, 0, U, 2, 2:, 7]} were merged as not showing the change.

\begin{center} 33. \textit{ME \textipa{/u:/}: MOUTH monophthongisation} \end{center}\par\textit{Variants:}\quad
 MOUTH monophthongisation, no MOUTH monophthongisation
\par\textit{Reference:}\quad
 IV.5.2 (\textit{mouse}), V.1.1(a) (\textit{houses}), VI.14.14 (\textit{trousers}), VII.1.16 (\textit{thousand})
\par\textit{Background}\quad
See ME \textipa{/u:/}: Great Vowel Shift (64).
\par\textit{Reduction}\quad
Long low or mid monophthongs \textipa{[A:, a:, \ae , \ae :, E:]} and low or mid front vowels with a schwa or low offglide \textipa{[a:@, \ae :a, \ae :@, \ae a, \ae @, Ea, E@, ea]} were merged as showing the change; all other reflexes \textipa{[AI, aI, @u, @u:, @U, @U:, 2U, 5u, EU, Eu, EU:, Eu:, Ew, eu:, eU, Uu:, u:, u, a:U, au, au:, aU, Au, AU, \ae u, \ae U, \ae :U, \ae 7, \oe 7]} were merged as not showing the change.

\begin{center} 34. \textit{Onset \textipa{/f/} voicing} \end{center}\par\textit{Variants:}\quad
 \textipa{[f]}, \textipa{[v]}
\par\textit{Reference:}\quad
 IV.9.6 (\textit{frogs}), IV.10.11 (\textit{furze}), V.3.1 (\textit{fire})
\par\textit{Background}\quad
This variable concerns the realisation of etymological \textipa{/f/} as \textipa{[v]} in initial position in native Germanic words. This is part of a larger set of changes by which all initial non-back fricatives could be voiced: \textipa{/f T s S/} $>$ \textipa{[v D z Z]}. The chronology and evidence is somewhat different for the different fricatives, however: \textipa{\textipa{/S/}} $>$ \textipa{[Z]} is much more inconsistent than the others and should probably be assumed to be a later, separate change; \textipa{/f/} $>$ \textipa{[v]} is best evidenced in medieval sources (but possibly only because the orthography had ways of representing \textipa{[v]}, viz.\ $<$u v w$>$) and shows a relatively high rate of application in potential items in the SED (66.3\%); \textipa{/s/} $>$ \textipa{\textipa{[z]}} is less well evidenced in medieval sources, but this is expected given that EME orthography did not have a widely accepted ortheme for \textipa{/z/}, and it shows a similarly high consistency in the SED (70.3\%); \textipa{/T/} $>$ \textipa{[D]} is so poorly evidenced in medieval sources that it is hard to meaningfully trace its distribution, but it is the most consistent in the SED (81.1\%) \citep{lass.1}. The first direct spelling evidence for this change is an instance of $<$u$>$ for \textipa{/f/} in a document from around 950 \citep{fisiak1985.1}. The orthographic evidence then mounts through the EME period; the text which shows the change most consistently is a Kentish text from the first half of the 14th century (\textit{Ayenbite of Inwit} 1340) \citep{lass.2}. There is some disagreement about whether the changes at different places of articulation should be seen as separate sound changes or whether it might be possible to unify them as a single change (the latter is argued for by Fisiak \cite{fisiak1985}); however, in either case, it seems clear that later retreat of these isoglosses has proceeded somewhat differently for the different phonemes involved, justifying our treating \textipa{/f/} $>$ \textipa{[v]} and \textipa{/T/} $>$ \textipa{[D]} here as  separate changes.
\par There are broadly two possible positions on the dating of this change. Either the orthographic evidence is taken as offering us evidence for timing the of the change (at least indirectly and at a delay), implying that it took place somewhere around the EME period; or the timing of the orthographic changes are seen as entirely unrelated to the timing of the sound change, in which case the sound change might have happened much earlier in the OE period. The former `traditional' view is put forward in Brunner \cite{brunner1965}, Berndt \cite{berndt1960} and Pinsker \cite{pinsker1974}, among others. The latter view is argued tentatively by Fisiak \cite{fisiak1985}, more directly by Bennet \cite{bennet1955} and Lass \cite{lass}. This latter view has the advantage that it allows the sound change to be identified as the same sound change that voiced initial fricatives in varieties of Dutch and Low German. We accept this view, implying an extremely early date: that this variation existed in the speech community since before the migration of Germanic speakers to the British Isles, and thus the data here reflect a geospatial distribution that has been evolving since the migration period.

\begin{center} 35. \textit{Onset \textipa{/T/} voicing} \end{center}\par\textit{Variants:}\quad
 \textipa{[T]}, \textipa{[D]}
\par\textit{Reference:}\quad
 V.8.16 (\textit{anything}), V.10.2 (\textit{thread}), V.10.9 (\textit{thimble}), VI.6.3 (\textit{throat}), VII.6.15 (\textit{thawing}), VIII.1.3 (\textit{three}), VIII.7.7 (\textit{throwing})
\par\textit{Background}\quad
See onset \textipa{/f/} voicing (34).
\par\textit{Reduction}\quad
Stopped and fronted variants were merged with dental fricatives according to their voicing (i.e.\ \textipa{[f t \textsubbridge{t} \textsubbridge{t}\super h T]} were merged as \textipa{[T]} and \textipa{[v d \:d D]} as \textipa{[D]}). This is on the basis of the assumption that the change in voicing preceded other changes affecting this phoneme.

\begin{center} 36. \textit{Postvocalic \textipa{/t/} glottalisation} \end{center}\par\textit{Variants:}\quad
 \textipa{[t]}, \textipa{[P]}
\par\textit{Reference:}\quad
 V.8.7 (\textit{kettle}), VI.5.11 (\textit{eat, ate, eaten}), VII.3.7 (\textit{autumn}), VIII.1.4 (\textit{daughter}), VIII.1.11 (\textit{brought}), VIII.1.12 (\textit{aunt}), IX.3.8 (\textit{caught}), IX.6.4 (\textit{got, gotten})
\par\textit{Background}\quad
This variable refers to the realisation of postvocalic \textipa{/t/} as a glottal stop \textipa{[P]}. This sound change is characteristic of British English much more than colonial varieties, suggesting a late date of innovation. However, since there are no explicit contemporary commentaries before the second half of the 19th century, dating it specifically is difficult \citep{minkova2014.3}. Here, we take it that there is no reason to date it earlier than the beginning of the 19th century.
\par\textit{Reduction}\quad
Glottalised variants were treated together, whether or not they were debuccalised (i.e.\ \textipa{[P]} and \textipa{[\t*{tP}]} are merged as \textipa{[P]}); voicing and aspiration are ignored, meaning that \textipa{[t, t\super h, d]} are merged as \textipa{[t]}.

\begin{center} 37. \textit{Postvocalic \textipa{/t/} voicing} \end{center}\par\textit{Variants:}\quad
 \textipa{[t]}, \textipa{[d]}
\par\textit{Reference:}\quad
 V.8.7 (\textit{kettle}), VI.5.11 (\textit{eat, ate, eaten}), VII.3.7 (\textit{autumn}), VIII.1.4 (\textit{daughter}), VIII.1.11 (\textit{brought}), IX.3.8 (\textit{caught}), IX.6.4 (\textit{got, gotten})
\par\textit{Background}\quad
This variable refers to the realisation of postvocalic \textipa{/t/} as a voiced stop \textipa{[d]}. This sound change is found in North American English varieties as well as British English varieties, suggesting an early date of innovation.
\par\textit{Reduction}\quad
All voiceless variants \textipa{[P, \t*{tP}, t, t\super h]} are merged as not showing the change.

\begin{center} 38. \textit{Prevocalic \textipa{/r/} backing} \end{center}\par\textit{Variants:}\quad
 uvular \textipa{[K]}, coronal \textipa{[r R \*r \:R]}
\par\textit{Reference:}\quad
 II.9.1 (\textit{grass}), IV.5.8 (\textit{squirrel}), IV.11.1 (\textit{blackberries, brambles}), V.5.3 (\textit{cream}), V.3.11 (\textit{draught}), VII.4.8 (\textit{Christmas}), IX.3.9 (\textit{grow, grew, growed})
\par\textit{Background}\quad
This phonetic variable refers to the place of articulation of (prevocalic) \textipa{/r/} as uvular or coronal. Following Minkova \cite{minkova2014.4}, we assume that the historically prior realisation was an alveolar or dental trill \textipa{[r]} and so the innovation here is the backing of this phoneme to a uvular trill or approximant \textipa{[\;R K]}. The earliest written reference to this sound change dates from 1724 \cite{jones1724}, cf.\ \cite{wales2006}, so we assume that this variable has existed for at least 200 years.
\par\textit{Reduction}\quad
All of the coronal realisations \textipa{[r R \*r \:R]} were merged as a single variant.

\begin{center} 39. \textit{Prevocalic \textipa{/r/} retroflexion} \end{center}\par\textit{Variants:}\quad
 retroflex \textipa{\textipa{[\;R]}}, dental/alveolar \textipa{[r R \*r]}
\par\textit{Reference:}\quad
 II.9.1 (\textit{grass}), IV.5.8 (\textit{squirrel}), IV.11.1 (\textit{blackberries, brambles}), V.5.3 (\textit{cream}), V.3.11 (\textit{draught}), VII.4.8 (\textit{Christmas}), IX.3.9 (\textit{grow, grew, growed})
\par\textit{Background}\quad
This phonetic variable refers to the realisation of preconsonantal \textipa{/r/} as retroflex \textipa{[\:R]} vs.\ dental/alveolar \textipa{[r R \*r]}. It is generally agreed that the earliest realisation of this sound was a dental/alveolar trill (although see Minkova \cite{minkova2014.4} for references to the argument that the uvular variant is historically prior). Tristram \cite{tristram1995} argues that the next stage of the sequence of changes was a shift to retroflex place of articulation, that this happened already in West Saxon, and spread during the OE period to the limit of the Danelaw. However, here we instead accept the more parsimonious account that lenition from trill to approximant preceded changes in place \cite{erickson2002,tristram1995.1}, citing \cite{scheiders1991,trautmann1880}, and so the retroflex variant reflects the output of a much more recent sound change. Dating this change, however, is extremely difficult, as it would be expected to leave no orthographic evidence.
\par\textit{Reduction}\quad
All of the dental/alveolar realisations \textipa{[r R \*r]} were merged as a single variant. The uvular realisation \textipa{[K]} was excluded from consideration since, strictly speaking, it is impossible to tell whether the coronal variant which existed before \textipa{/r/} backing was dental, alveolar, postalveolar or retroflex; however, sinec these back realisations existed in a delimited area fully embedded within the non-retroflex domain, this has no effect on the overall distribution of this change.

\begin{center} 40. \textit{\textipa{/r/} $>$ \textipa{[hr]}} \end{center}\par\textit{Variants:}\quad
 aspirated \textipa{[h\:R]}, non-aspirated \textipa{/r/}
\par\textit{Reference:}\quad
 V.1.2 (\textit{roof}), V.1.15 (\textit{rubbish}), V.2.4 (\textit{rooms}), V.10.7 (\textit{red}), VI.6.9 (\textit{wrist}), VI.7.15 (\textit{reach, reached})
\par\textit{Background}\quad
A small area in Somerset shows consistent aspiration of word-initial \textipa{/r/}. This does not result in any mergers or splits, and so is classed as a phonetic variable. We know of no evidence by which to date this change.
\par\textit{Reduction}\quad
Place and manner of articulation of \textipa{/r/} were ignored, so that \textipa{[\:R \*r R r \;R]} were all treated together as not showing aspiration.

\begin{center} 41. \textit{Rhoticity} \end{center}\par\textit{Variants:}\quad
 rhoticity, no rhoticity
\par\textit{Reference:}\quad
 II.5.1 (\textit{corn}), II.5.1 (\textit{barley}), V.3.1 (\textit{fire}), V.6.2 (\textit{barm}), V.6.7 (\textit{burnt/burned})
\par\textit{Background}\quad
This phonological variable refers to the non-realisation of etymological nonprevocalic \textipa{/r/}, excluding the sequence \textipa{/rs/}. Sporadic loss of coda \textipa{/r/}, especially preceding coronals, is attested from the OE period, but this is regarded as a separate sound change with different phonological consequences (contra Minkova \cite{minkova2014.6}); reflexes of that earlier change are seen in variation in etymological \textipa{/rs/} clusters in the SED. Evidence for the general nonprevocalic \textipa{/r/} loss which we examine here becomes clear from the mid-17th \citep{mcmahon2000} or around the turn of the 18th century \citep{minkova2014.7}, so we assume that this variable has existed in this form for around 250 years.
\par\textit{Reduction}\quad
All consonantal realisations of postvocalic \textipa{/r/}, including r-colouring of the preceding vowel, were merged as \textipa{/r/} (rhoticity); all others were classed as Ø (loss of rhoticity).

\begin{center} 42. \textit{th-fronting before \textipa{/r/}} \end{center}\par\textit{Variants:}\quad
 fronted, non-fronted
\par\textit{Reference:}\quad
 V.10.2 (\textit{thread}), VI.6.3 (\textit{throat}), VII.1.3 (\textit{three}), VIII.7.7 (\textit{throwing})
\par\textit{Background}\quad
Th-fronting refers to the sound change \textipa{/T/} $>$ \textipa{/f/}; this is a phonological variable, since it results in merger with existing \textipa{/f/}. We have two datasets representing th-fronting, on the basis that its distribution preceding \textipa{/r/} appears to be quite different to its distribution in other positions, suggesting that it represents a different change.
\par\textit{Reduction}\quad
Voicing was ignored, so that \textipa{[f v]} were merged as showing fronting, and \textipa{[d, D, \:d, t, \textsubbridge{t}, \textsubbridge{t}\super h, T]} as not showing fronting.

\begin{center} 43. \textit{th-fronting elsewhere} \end{center}\par\textit{Variants:}\quad
 fronted, non-fronted
\par\textit{Reference:}\quad
 IV.3.11 (\textit{path}), V.7.20 (\textit{broth}), V.8.16 (\textit{anything}), V.10.9 (\textit{thimble}), VI.5.6 (\textit{tooth, teeth}), VII.6.15 (\textit{thawing}), VII.2.11 (\textit{both})
\par\textit{Background}\quad
See th-fronting before \textipa{/r/} (43).

\begin{center} 44. \textit{th-stopping before \textipa{/r/}} \end{center}\par\textit{Variants:}\quad
 stops, fricatives
\par\textit{Reference:}\quad
 V.10.2 (\textit{thread}), VI.6.3 (\textit{throat}), VII.1.3 (\textit{three}), VIII.7.7 (\textit{throwing})
\par\textit{Background}\quad
This variable refers to the change that changes the fricative \textipa{/T/} into a stop. This is a phonological change, since it results in merger with existing coronal stop phonemes. It seems to apply regardless of other changes which affect voicing and place of articulation, so we find dental, alveolar and retroflex, voiced and voiceless variants as a result. It is likely that it postdates initial fricative voicing, so this indicates that \textipa{[D]} was affected by this change just as \textipa{[T]} was. The changes in place of articulation among coronal places (retroflexion before \textipa{/r/}, retraction to alveolar) do not apply to fricative variants, and so must postdate th-stopping. Th-fronting does not apply to the stopped variants, and so must also postdate th-stopping.
\par\textit{Reduction}\quad
Voicing was ignored, so that \textipa{[d, \:d, t, \textsubbridge{t}, \textsubbridge{t}\super h]} were merged as showing stopping and \textipa{[D, T, v, f]} as not showing stopping.

\begin{center} 45. \textit{THOUGHT hyperrhoticity} \end{center}\par\textit{Variants:}\quad
 rhoticity, no rhoticity
\par\textit{Reference:}\quad
 II.5.2 (\textit{straw}), IV.8.8 (\textit{straw}), V.8.9 (\textit{draw}), VII.3.7 (\textit{autumn}), VIII.1.4 (\textit{daughter}), VIII.1.11 (\textit{brought}), IX.3.8 (\textit{caught}), IX.4.6 (\textit{ought}), IX.4.7 (\textit{ought}), IX.4.9 (\textit{ought})
\par\textit{Background}\quad
This variable refers to the reanalysis of some long low vowels as representing underlying \textipa{/}V\textipa{r/} sequences, with the result that some varieties show a rhotic realisation in vowels which have no etymological *\textit{r}. Here we look only at this phenomenon in the THOUGHT vowel (using the label from Wells' lexical sets \cite{wells1982}); this can result in merger with existing \textipa{/}V\textipa{r/} sequences, and so is classified as a phonological variable.
\par\textit{Reduction}\quad
Vowel quality was ignored, so that \textipa{[a\:R:, A\:R:, O\*r:, O\:R:, U@\*r]} were merged as showing hyperrhoticity, and \textipa{[a:, \ae :, \ae :@, \ae U, aU, A:, AU, 6, 6:, 6u:, 6U, e:, e:I, @:, @O:, @OU, EI, EU, I@, o:, o:@, o:U, o@, oU, O, O:, O:@, O:U, O@, OU, u:, U, U@, Uu, 2U]} as not showing it.

\begin{center} 46. \textit{ME \textipa{/Ou/} monophthongisation} \end{center}\par\textit{Variants:}\quad
 monophthongised, non-monophthongised
\par\textit{Reference:}\quad
 VIII.1.11 (\textit{brought}), IX.4.6 (\textit{ought}), IX.4.7 (\textit{ought}), IX.4.9 (\textit{ought})
\par\textit{Background}\quad
This variable concerns the monophthongisation of ME \textipa{/Ou/}; this change did not take place in the north of England. It is classed as a phonological variable, since it results in merger with other sources of \textipa{/O:/} (forming the THOUGHT set). 
\par\textit{Reduction}\quad
Among non-monophthongised realisations, the quality of the initial element was ignored, so that \textipa{[\ae U, aU, AU, 6U, EU, oU, OU, 2U]} were merged as non-monophthongised; various later sound changes (including later diphthongisations) were ignored, so that \textipa{[a:, \ae :, a\:R:, A:, A\:R:, 6:, e:, EI, o@, O, O:, O:@, O@, O\*r:, O\:R:, U, U@]} were merged as showing monophthongisation.

\begin{center} 47. \textit{\textit{wasp} metathesis} \end{center}\par\textit{Variants:}\quad
 \textit{wasp, waps}
\par\textit{Reference:}\quad
 IV.8.7
\par\textit{Background}\quad
This is a phonolexical variable in the sense that it refers to a sound change which affected a single word: metathesis in the coda consonant cluster of \textit{waps}. Both forms are attested from the OE period: Bosworth \& Toller list forms \textit{waefs, w\ae ps, weaps} and \textit{w\ae sp} \citep{bosworth1898.9}, of which the \textit{-fs-} forms appear to be earlier; there are just four instances of the \textit{-sp-} forms in the Old English Web Corpus \cite{healey2009}, and all are 10th century or later. The impression that the \textit{-sp-} forms are the innovation is confirmed by comparison with cognates outside English, such as OHG \textit{wefsa}. Thus the variation has existed since at least the 10th century.
\par\textit{Reduction}\quad
Other variants were unrelated lexical items and relatively rare, and accordingly were excluded.

\begin{center} 48. \textit{\textit{ask} metathesis} \end{center}\par\textit{Variants:}\quad
 \textit{ask, aks}
\par\textit{Reference:}\quad
 IX.2.4
\par\textit{Background}\quad
This variable deals with variation in the form of the stem of the verb \textit{ask}. The conservative form has \textipa{/sk/}, as demonstrated by cognates elsewhere in West Germanic (Old Frisian \textit{\=askia}, Old Saxon \textit{\=eskon}, Old High German \textit{eisc\=on}). The metathesised form has \textipa{/ks/} and is abundantly attested from the OE period \cite{bosworth1898.10,lewis.20}; the OED suggests that in OE the metathesised form is particularly characteristic of West Saxon, consistent with the fact that the metathesised form is more common in the south and the Midlands during the ME period \cite{laing2013.4,benskin2013.4}. 
\par\textit{Reduction}\quad
Realisations with a postalveolar fricative (\textipa{[aS]} and similar) were merged into the non-metathesised variant on the basis the development \textipa{/sk/} $>$ \textipa{/S/} is characteristic of some varieties of OE whilst *\textipa{/ks/} $>$ \textipa{/S/} does not occur, and so a postalveolar fricative implies that metathesis never took place. Realisations with an alveolar fricative but no stop (\textipa{[as]} and similar) were excluded on the basis that it is impossible to tell whether these represent a reduced form of \textipa{/ask/} or \textipa{/aks/}.

\begin{center} 49. \textit{\textit{birch} palatalisation} \end{center}\par\textit{Variants:}\quad
 palatalised \textit{birch}, non-palatalised \textit{birk}
\par\textit{Reference:}\quad
 IV.10.1
\par\textit{Background}\quad
Velar stops *\textipa{/k g/} underwent palatalisation before the OE period, probably first becoming *\textipa{[c]} and *\textipa{[J \textbardotlessj]}, later undergoing lenition to \textipa{[\textteshlig]} and \textipa{[j \textdyoghlig]}. There appears to have been variation between palatalised and velar consonants in many words in the OE period, although OE orthography did not distinguish the palatalised and velar consonants, making it difficult to assess the situation precisely; certainly, by the ME period when the orthography begins to make these distinctions clearly, such variation is widespread, with non-palatalised reflexes in northern and Danelaw areas, and palatalised reflexes from the south and the Midlands. We follow Ringe \& Taylor \cite{ringe2006a.1} in assuming that this variation did not reflect dialectal variation in the application of the palatalisation rule, but borrowing of cognate forms without palatalisation from ON. Thus although we see apparently similar phonological variation across multiple words, this must have spread lexically; certainly changes in distribution in the following centuries has progressed differently for different stems. For these reasons, we treat palatalisation in \textit{birch} (49), \textit{bridge} (50), \textit{chaff} (51) and \textit{reach} (55) as separate variables.

\begin{center} 50. \textit{\textit{bridge} palatalisation} \end{center}\par\textit{Variants:}\quad
 palatalised \textit{bridge}, non-palatalised \textit{brig}
\par\textit{Reference:}\quad
 IV.1.2
\par\textit{Background}\quad
See \textit{birch} palatalisation (49).

\begin{center} 51. \textit{\textit{chaff} palatalisation} \end{center}\par\textit{Variants:}\quad
 palatalised \textit{chaff}, non-palatalised \textit{caff}
\par\textit{Reference:}\quad
 II.8.5, III.5.3
\par\textit{Background}\quad
See \textit{birch} palatalisation (49).

\begin{center} 52. \textit{\textit{dove} final consonant} \end{center}\par\textit{Variants:}\quad
 \textit{dove, doe}
\par\textit{Reference:}\quad
 IV.7.4
\par\textit{Background}\quad
This word has shown variation in the presence of the final consonant since the ME period; the conservative form has the consonant, the innovative form does not. The MED cites forms which probably lack final \textipa{/f/} or \textipa{/v/} from the first half of the 15th century onwards (\textit{The Fire of Love} 1435, \textit{The Book of Margery Kempe}, Book 1 1438) \citep{lewis.21}.
\par\textit{Reduction}\quad
Compounded variants were merged with their respective uncompounded variants: \textit{ringdoe} with \textit{doe, ringdove} and \textit{turtledove} with \textit{dove}.

\begin{center} 53. \textit{\textit{partridge} final consonant} \end{center}\par\textit{Variants:}\quad
 voiced \textit{partridge}, voiceless \textit{partrich}
\par\textit{Reference:}\quad
 IV.7.8.(a)
\par\textit{Background}\quad
This word is a borrowing from French (Anglo-Norman \textit{pardriz, partreiz}, Old French \textit{perdriz}), first occurring in ME with final /\textteshlig/ reflecting French \textipa{/\texttslig/}. The innovation is thus the voicing of the final consonant. This sound change is not restricted to this lexical item alone, but the results are not fully regular, and the variation in \textit{partridge} does not appear to pattern with other lexical items with coda \textipa{/\textteshlig$\sim$\textdyoghlig/} variation in the SED, so it is here treated as a phonolexical variable. On the basis of citations in the OED and MED, the voiced variant seems to have occurred from the mid-15th century (\textit{Terms of Association (1)} 1450, \textit{Sir Gawain and the Carl of Carlisle} 1475) \citep{lewis.22}.

\begin{center} 54. \textit{\textit{porridge} final consonant} \end{center}\par\textit{Variants:}\quad
 voiced \textit{porridge}, voiceless \textit{porrich}
\par\textit{Reference:}\quad
 V.7.3
\par\textit{Background}\quad
This variable concerns the voicing of the coda consonant of the final syllable of \textit{porridge}. This word, originally also with different medial consonant, is a loanword from French (cf.\ Old French \textit{potage}); the conservative form is thus voiced \textipa{/\textdyoghlig/}, and the innovation is devoicing to \textipa{/\textteshlig/}. The MED offers just one citation for the voiceless form, tentatively dated in the mid-15th century (\textit{Herbal (misidentified as a ME version of the De Viribus Herbarum of `Macer Floridus')} 1425(?)) \citep{lewis.23}; the earliest attestation of the voiceless form in the OED is not until much later \citep{learmont1791}, but a voiceless final consonant is recorded for another related form, \textit{poddish}, in the 16th century \citep{tyndale1528}. Thus we accept the dating of the innovation implied by the mid-15th century attestation.
\par\textit{Reduction}\quad
The two voiceless realisations, \textipa{/S/} and /\textteshlig/, were merged on the basis that it is reasonable to assume that \textipa{/S/} went through an earlier stage as /\textteshlig/.

\begin{center} 55. \textit{\textit{reach} palatalisation} \end{center}\par\textit{Variants:}\quad
 palatalised \textit{reach}, non-palatalised \textit{reak}
\par\textit{Reference:}\quad
 VI.7.15
\par\textit{Background}\quad
See \textit{birch} palatalisation (49).

\begin{center} 56. \textit{\textipa{/kw/} $>$ \textipa{/tw/}} \end{center}\par\textit{Variants:}\quad
 \textipa{/kw/}, \textipa{/tw/}
\par\textit{Reference:}\quad
 V.2.11 (\textit{quilt}), VI.7.9 (\textit{quick})
\par\textit{Background}\quad
This is a phonological variable, concerning the merger of \textipa{/kw/} into \textipa{/tw/}. This sound change is centred on the north-west of England.
\par\textit{Reduction}\quad
The variant \textipa{\textipa{[t]}} was merged into \textipa{/tw/}. Whether this just reflects a fast-speech phenomenon, or truly an additional sound change, it presupposes the application of \textipa{/kw/} $>$ \textipa{/tw/}.

\begin{center} 57. \textit{BATH lengthening} \end{center}\par\textit{Variants:}\quad
 short, long
\par\textit{Reference:}\quad
 I.7.7 (\textit{shaft}), II.1.3 (\textit{pasture}), II.8.5 (\textit{chaff}), II.9.1 (\textit{grass}), III.5.3 (\textit{chaff}), III.5.4 (\textit{grass}), IV.3.11 (\textit{path}), IX.2.4 (\textit{ask})
\par\textit{Background}\quad
This variable concerns lengthening in ME \textipa{/a/} preceding voiceless fricatives. This is generally referred to as the TRAP-BATH split, but this term is used to refer to both lengthening (whether contrastive or not) and later backing \textipa{[a]} $>$ \textipa{[a:]} $>$ \textipa{[A:]}. Here we refer to it as BATH lengthening to make it clear that we are dealing only with the changes in length and not quality, since the change in length is the earlier change. This is a phonological change since it results in a phonemic split in the \textipa{/a/} vowel (unambiguously so in varieties which also undergo later backing, but, we would argue, also in varieties which do not; cf.\ \cite{blaxter2019a}; contra \cite{hughes2012,kester1979}).
\par\textit{Reduction}\quad
All long variants \textipa{[\ae :, a:, A:]} are merged as long; \textipa{[A]} is also treated as long, as, since only the long vowel underwent backing, it must reflect an earlier lengthened form. All other short variants \textipa{[\ae , a]} are merged as short. 

\begin{center} 58. \textit{CARD-CORD merger} \end{center}\par\textit{Variants:}\quad
 merger, no merger
\par\textit{Reference:}\quad
 II.5.1 (\textit{corn, barley}), V.6.2 (\textit{barm})
\par\textit{Background}\quad
This variable refers to the merger of the START and NORTH lexical sets (to use the terminology of Wells \cite{wells1982}), which takes place by loss of rounding of the NORTH vowel. It is a phonological change. This was assessed by comparing words with the NORTH vowel and words with the START vowel and identifying where a speaker had the same vowel for both.

\begin{center} 59. \textit{CLOTH lengthening} \end{center}\par\textit{Variants:}\quad
 lengthening, no lengthening
\par\textit{Reference:}\quad
 V.7.20 (\textit{broth}), VII.4.7 (\textit{on}), VII.6.6 (\textit{frost})
\par\textit{Background}\quad
The low back unrounded vowel \textipa{/6/} underwent lengthening in certain contexts in many varieties of English, with the result that this set merged with the THOUGHT set or, in a few cases, with the BATH set. This was assessed by comparing words with in the CLOTH set with words in the LOT, BATH, TRAP and THOUGHT sets and identifying which pair the speaker had the same vowel for. Note that this change is often referred to as the LOT-CLOTH split; we refer to it as CLOTH lengthening here to indicate that we treat speakers with same vowel in CLOTH and BATH together with speakers with the same vowel in CLOTH and THOUGHT.

\begin{center} 60. \textit{FOOT-STRUT split} \end{center}\par\textit{Variants:}\quad
 split, no split
\par\textit{Reference:}\quad
 IV.4.4 (\textit{cut}), IV.6.14 (\textit{ducks}), IV.6.21 (\textit{pluck}), IV.7.4 (\textit{doves}), IV.7.5 (\textit{gull}), IV.9.2 (\textit{slugs}), IV.12.4 (\textit{stump}), V.1.15 (\textit{rubbish}), V.2.5 (\textit{upstairs}), V.8.4 (\textit{some}), V.9.12 (\textit{up}), VI.5.7 (\textit{double}), VII.6.10 (\textit{dull})
\par\textit{Background}\quad
ME \textipa{/u/} underwent loss of rounding and lowering in certain lexical items in many varieties of English, resulting in a split into the STRUT set (which undergoes the change) and the FOOT set (which does not). This is a phonological variable.
\par\textit{Reduction}\quad
The lowered, unrounded variants \textipa{[@, 2]} were treated together as evidencing the split. The front rounded variant \textipa{[\oe :]} and other minor variants, mostly reflecting lexical variation in which phoneme was found in particular words rather than variation in realisation of the phoneme, were excluded. The back unrounded variant \textipa{[7]} represents a problem. It occurs in two regions: Norfolk and Northumberland. Norfolk is mostly within the split area and Northumberland mostly within the non-split area, although both are adjacent to isoglosses (since Scottish English has the split, although with a rather different history). In Norfolk, speakers who use \textipa{[7]} consistently also use \textipa{[2]} for STRUT words and never use \textipa{[7]} for FOOT words, suggesting \textipa{[7]} should be treated as a variant of \textipa{/2/} and so as evidence for the split. In Northumberland, speakers who use \textipa{[7]} consistently also use \textipa{[U]} and sometimes also use \textipa{[7]} for FOOT words, suggesting that \textipa{[7]} is a variant of \textipa{/U/} and so evidence against the split. For this reason, \textipa{[7]} was excluded from consideration.

\begin{center} 61. \textit{Intrusive \textit{r}} \end{center}\par\textit{Variants:}\quad
 intrusive \textit{r}, no intrusive \textit{r}
\par\textit{Reference:}\quad
 I.7.16 (\textit{sawing-horse}), VII.6.15 (\textit{thawing})
\par\textit{Background}\quad
This variable describes the excrescence of a non-etymological \textipa{/r/} to break up the hiatus between a non-high vowel and a following vowel. This is generally understood to be a consequence of the loss of rhoticity and so cannot be dated any later than that sound change \citep{minkova2014.8}; the earliest evidence for the sound change is from Sheridan (\cite{sheridan1762} cf.\ \citep{soskuthy2013.1}), so we cannot assume an age for this variable of greater than 150 years.
\par\textit{Reduction}\quad
All consonantal realisations of \textipa{/r/} were merged as showing intrusive \textit{r}, regardless of place of articulation. Instances of intrusive \textit{l} were excluded from consideration.

\begin{center} 62. \textit{ME \textipa{/o:/}: shortening} \end{center}\par\textit{Variants:}\quad
 lax/short \textipa{[U, 0, 2, 2:, 7, Y]}, other
\par\textit{Reference:}\quad
 V.1.2 (\textit{roof}), V.2.4 (\textit{rooms}), V.2.14 (\textit{broom}), VI.5.6 (\textit{tooth})
\par\textit{Background}\quad
See ME \textipa{/o:/}: EMoE fronting (32).
\par\textit{Reduction}\quad
Shortened monophthongs \textipa{[U, 0, 2, 2:, 7, Y]} were merged as showing the change. Short \textipa{[0]} was included in this category on the assumption that it was a variant realisation of \textipa{/U/}. Long \textipa{[2:]} was also included in this category on the basis that it must be a lengthened realisation of \textipa{/2/} (and thus had previously undergone shortening), in order for it to undergo the FOOT-STRUT split. All other variants \textipa{[Y, Y:, EU, aU, 6U, ou, OU, @U, u:, Uu:, Uu, UI, Iu:, @u:, i@, I:@, IU, I@, Iu, I7, j7, jU]} were merged as not showing the change.

\begin{center} 63. \textit{ME \textipa{/o:/}: ME fronting} \end{center}\par\textit{Variants:}\quad
 fronted \textipa{[i@, I:@, IU, I@, Iu, I7, j7, jU]}, other
\par\textit{Reference:}\quad
 V.1.2 (\textit{roof}), V.2.4 (\textit{rooms}), V.2.14 (\textit{broom}), VI.5.6 (\textit{tooth})
\par\textit{Background}\quad
See ME \textipa{/o:/}: EMoE fronting (32).
\par\textit{Reduction}\quad
Diphthongs with a close front first element \textipa{[i@, I:@, IU, I@, Iu, I7, j7, jU]} were merged as showing the change, all other variants \textipa{[Y, Y:, EU, aU, 6U, ou, OU, @U, u:, Uu:, Uu, UI, Iu:, @u:, 0, U, 2, 2:, 7]} were merged as not showing the change.

\begin{center} 64. \textit{ME \textipa{/u:/}: Great Vowel Shift} \end{center}\par\textit{Variants:}\quad
 GVS applied to \textipa{/u:/}, GVS did not apply to \textipa{/u:/}
\par\textit{Reference:}\quad
 IV.5.2 (\textit{mouse}), V.1.1(a) (\textit{houses}), VI.14.14 (\textit{trousers}), VII.1.16 (\textit{thousand})
\par\textit{Background}\quad
The SED shows a great variety of reflexes of ME \textipa{/u:/}. Two sound changes have been selected here as separate variables on the basis that their reflexes are clearly distinct, and occur in well-separated regions. The first is the application of the Great Vowel Shift: diphthongisation of \textipa{/u:/}, presumably first to \textipa{[@u]} or something similar \cite{anderson1987.7}; this is a phonological variable that dates back to the EMoE period. The second is the monophthongisation of the MOUTH vowel (i.e.\ the reflex of ME \textipa{/u:/} in those varieties in which it did undergo the GVS) to a long, low vowel (sometimes with subsequent rediphthongisation with an offglide) \cite{anderson1987.6}; this is a phonetic variable, since the varieties in question did not undergo the TRAP-BATH split and so had no other phonemic long low vowel (for this latter variable, see ME \textipa{/u:/}: MOUTH monophthongisation (33)).
\par\textit{Reduction}\quad
High back vowels \textipa{[Uu:, u:, u]} were merged as not showing the GVS; all other reflexes \textipa{[A:, ea, E:, Ea, E@, a:, a:@, \ae , \ae :, \ae :a, \ae :@, \ae a, \ae @, AI, aI, @u, @u:, @U, @U:, 2U, 5u, EU, Eu, EUU, Eu:, Ew, eu:, eU, a:U, au, au:, aU, Au, AU, \ae u, \ae U, \ae :U, \ae 7, \oe 7]} were merged as showing the GVS.

\begin{center} 65. \textit{\textit{ng}-coalescence} \end{center}\par\textit{Variants:}\quad
 \textipa{[N]}, \textipa{[Ng]}
\par\textit{Reference:}\quad
 I.7.3 (\textit{string}), VI.5.4 (\textit{tongue}),  IX.2.12 (\textit{among})
\par\textit{Background}\quad
This variable concerns the realisation of ME \textipa{/ng/} at morpheme boundaries: in some varieties the non-nasal stop element is deleted \textipa{[Ng]} $>$ \textipa{[N]}, whereas other varieties retain \textipa{[Ng]}. This is a phonological change, since the result is that in varieties with the change, \textipa{[N]} can contrast with \textipa{[n]} and so represents a separate (if marginal) phoneme \textipa{/N/}.

\begin{center} 66. \textit{/rV/ metathesis} \end{center}\par\textit{Variants:}\quad
 metathesis \textipa{/}V\textipa{r/}, no metathesis \textipa{/r}V{/}
\par\textit{Reference:}\quad
 IV.1.2 (\textit{bridge}), V.2.14 (\textit{brush}), V.9.11 (\textit{brush}), V.10.7 (\textit{red}), VII.4.8 (\textit{Christmas})
\par\textit{Background}\quad
This variable concerns metathesis in \textipa{/r}V\textipa{/} sequences so that \textit{bridge} is realised \textipa{[b@\super{\:R}:\:dZ]}, \textit{brush} \textipa{[b@\super{\:R}:S]}, etc. Metatheses involving \textipa{/r/} have long been a feature of the phonology of all English varieties and many alternations between \textipa{/}(C)V\textipa{r}C\textipa{/} and \textipa{/}(C)\textipa{r}VC\textipa{/} are dated to the OE period \cite{ringe2006a.2,minkova2014.9}. However, for the words in question, we appear to be dealing with a specific, much later and more locally delimited sound change. Of these words, \textit{brush} has only existed in English since the ME period, and for \textit{bridge}, \textit{brush} and \textit{red}, the OED and MED record no metathesised forms in ME \citep{lewis.24}; for \textit{Christmas} a metathesised form is recorded (\textit{Churchwardens' Accounts of the Parish of St. Mary, Thame} 1442) \citep{lewis.25}, however, since this is not in the same region as the metathesis we see in these data, it is reasonable to assume that it is an independent sound change. The EDD records \textit{bursh} in Somerset \cite{elworthy1888}. Thus, we take this to be a recent sound change.
\par\textit{Reduction}\quad
Occasional realisations with a fully deleted \textipa{/r/} were excluded on the basis that it is impossible to tell whether these represent metathesis followed by loss of rhoticity, or simplification of an onset \textipa{/}C\textipa{r/} cluster.

\begin{center} 67. \textit{First equative conjunction} \end{center}\par\textit{Variants:}\quad
 \textit{so, as}
\par\textit{Reference:}\quad
 VIII.1.22
\par\textit{Background}\quad
This is a syntactic variable referring to the first conjunction used in the construction which expresses that an adjective has identical degree for two referents (e.g. \textit{\underline{as} good as} vs.\ \textit{\underline{so} good as}). Of the two conjunctions, \textit{as} descends from OE \textit{ealsw\=a} and \textit{so} from OE \textit{sw\=a}; both of these are found in equative constructions already in OE (cf.\ examples cited in the OED and Bosworth \& Toller \cite{bosworth1898.12,bosworth1921.1}), so we can assume the variation is at least 1000 years old (note, however, that the \emph{second} conjunction in this construction also differed in OE). \textit{Sw\=a} is far more common in this construction in OE and etymologically \textit{ealsw\=a} is a compounded variant of \textit{sw\=a}, so we can assume that \textit{ealsw\=a} is the innovation.

\section{Linguistic abbreviations and glossary}
\begin{longtable}{p{0.15\textwidth} p{0.32\textwidth}}
\textbf{Abbreviation} & \textbf{} \\[6pt]
EDD & English Dialect Dictionary \cite{markus2019} \\[6pt]
EME & Early Middle English \\[6pt]
EMoE & Early Modern English \\[6pt]
LAEME & A Linguistic Atlas of Early Middle English \cite{laing2013} \\[6pt]
LALME & A Linguistic Atlas of Late Medieval English \cite{benskin2013} \\[6pt]
ME & Middle English \\[6pt]
MED & Middle English Dictionary \cite{lewis} \\[6pt]
MoE & Modern English \\[6pt]
OE & Old English \\[6pt]
OED & Oxford English Dictionary \\[6pt]
ON & Old Norse \\[6pt]
PG & Proto-Germanic \\[6pt]
\end{longtable}

\begin{longtable}{p{0.15\textwidth} p{0.32\textwidth}}
\textbf{Term} & \textbf{Gloss} \\[6pt]
1sg., 2sg., 3sg., 1pl, 2pl, 3pl & first person singular, second person singular, etc. \\[6pt]
analogy & process of language change whereby the form of one morpheme or word influences the form of another \\[6pt]
coda & the last part of the syllable, typically the zero or more consonants which follow the vowel \\[6pt]
comparative & the form of an adjective which expresses higher degree for one referent than another \\[6pt]
conjunction & a word that has the function of linking other words or phrases such as \textit{and}, \textit{or}, \textit{as}, etc. \\[6pt]
conservatism & the historically prior variant \\[6pt]
conservative & (of a variant) historically prior \\[6pt]
consonant cluster & a sequence of adjacent consonants without intervening vowel \\[6pt]
construction & an arrangement of multiple words to express a given grammatical function \\[6pt]
degree & the relational extent to which an adjective applies to a referent (typically distinguishing positive vs.\ comparative vs.\ superlative) \\[6pt]
function words vs.\ content words & function words are those which have little lexical meaning of their own but instead express grammatical relationships among other words in the sentence; content words are those words which do have lexical meaning \\[6pt]
innovation & (1) the historically more recent variant; (2) the process by which a new variant is introduced to the language \\[6pt]
innovative & (of a variant) historically more recent \\[6pt]
isogloss & boundary between multiple spatial domains in which different variants are dominant \\[6pt]
levelling & (1) the spread of a single form through a paradigm so that cells which were previously morphologically distinct are no longer so; (2) the spread of a single form across communities so that there is no geospatial variation where such variation previously existed \\[6pt]
lexical & having to do with individual words \\[6pt]
lexical item & a word and all its morphological forms (such as: \textit{speak}, including \textit{speaks}, \textit{speaking}, \textit{spoke}, \textit{spoken}) \\[6pt]
metathesis & a sound change by which two phonemes in a word exchange positions \\[6pt]
morpheme & meaningful units smaller than words, such as prefixes, suffixes and stems \\[6pt]
morpholexical (and phonolexical, etc.) & -lexical as a suffix here indicates variation which applies only to a single word and does not reflect a wider pattern of variation in the language; for example, `morpholexical' refers to variables having to do with the formation (morpho-) of a specific word from its constituent parts where this is not part of a larger pattern involving other words \\[6pt]
morphology, morphological & having to do with the formation of words from morphemes \\[6pt]
nonprevocalic & not preceding a vowel (i.e.\ preceding a consonant or a word-boundary) \\[6pt]
onset & the first part of the syllable, typically the zero or more consonants which precede the vowel \\[6pt]
phoneme & a unit of sound which can distinguish words \\[6pt]
phonetic & having to do with the realisation of particular sounds where this does not affect the structure of the overall system in which those sounds are placed \\[6pt]
phonology, phonological & having to do with the system of contastive sounds (phonemes) used by a language to construct morphemes \\[6pt]
preterite & past tense \\[6pt]
prevocalic & preceding a vowel \\[6pt]
sound change & changes in pronunciation \\[6pt]
suppletion & a morphological property where cells in a single paradigm are supplied by unrelated stems \\[6pt]
syntax, syntactic & having to do with the formation of sentences from words (covering word order, choice of function words, etc.) \\[6pt]
(linguistic) variable & a linguistic context in which there is variation in form with no corresponding variation in function/meaning (``two ways of saying the same thing'' \citep{labov1972}) \\[6pt]
variant & one possible form of a given variable 
\end{longtable}

\bibliographystyle{unsrt}
\bibliography{SED_refs_plus_sup}

\end{document}